\documentclass[preprint,3p,times,authoryear]{elsarticle}

\usepackage{subcaption}
\usepackage{booktabs}
\usepackage{multirow}
\usepackage{siunitx}
\usepackage{tabularx}
\usepackage[symbol]{footmisc}
\usepackage{ragged2e} % 用于 \RaggedRight
\usepackage{varwidth}
\usepackage{bm}
\usepackage{graphicx}
\usepackage{epstopdf}

\usepackage{caption} 

\usepackage{longtable}
\usepackage{array}      % for >{\raggedright\arraybackslash}X if needed
\usepackage{ltablex}    % optional, if you want X columns with longtable

\usepackage{makecell}
\keepXColumns

\usepackage{amssymb}
\usepackage{amsmath}
\usepackage{paralist}
\usepackage{enumitem}

\usepackage[colorlinks=true, citecolor=darkgreen]{hyperref}
\usepackage{cleveref}
\journal{Journal of Symbolic Computation}

\usepackage{graphicx}
\usepackage{hyperref}
\usepackage{xcolor}

\def \R {\mathbb{R}}
\def \C {\mathbb{C}}
\def \Q {\mathbb{Q}}

\def \mvar {\mathrm{mvar}}
\def \res {\mathrm{res}}
\def \discrim {\mathrm{discrim}}
\def \lc {\mathrm{lc}}

\newtheorem{example}{Example}
\newtheorem{remark}{Remark}
\newtheorem{definition}{Definition}
\newtheorem{notation}{Notation}

\begin{document}

\begin{frontmatter}



\title{Breaking the Data Barrier in Learning Symbolic Computation: A Case Study on Variable Ordering Suggestion for Cylindrical Algebraic Decomposition\footnote[1]{This work is supported by the National Key Research Project of China under Grant No. 2023YFA1009402, the National Natural Science Foundation of China under Grant No. 12101267 and Chongqing Talents Plan Youth Top-Notch Project (2021000263). Rui-Juan Jing (rjing@ujs.edu.cn) and Yuegang Zhao (ygzhao@stmail.ujs.edu.cn) contributed equally to the work. Corresponding author: Changbo Chen (chenchangbo@cigit.ac.cn).}} %% Article title

\author[JSU]{Rui-Juan Jing$^\dagger$}
\author[JSU]{Yuegang Zhao$^\dagger$}

\author[CIGIT,UCAS]{Changbo Chen}
\affiliation[JSU]{organization={School of Mathematical Sciences, Jiangsu University},
            %addressline={},
            city={Zhenjiang},
            %postcode={},
            state={Jiangsu},
            country={China}}

\affiliation[CIGIT]{organization={Chongqing Institute of Green and Intelligent Technology, Chinese Academy of Sciences},
            %addressline={},
            city={Liangjiang New Area},
            %postcode={},
            state={Chongqing},
            country={China}}
\affiliation[UCAS]{organization={Chongqing School, University of Chinese Academy of Sciences},
            city={Liangjiang New Area},
            %postcode={},
          state={Chongqing},
            country={China},
            }

%% Abstract
\begin{abstract}

Symbolic computation, powered by modern computer algebra systems, 
has important applications in mathematical reasoning through exact deep computations. 
The efficiency of symbolic computation is largely constrained by such deep computations in high dimension. 
This creates a fundamental barrier on labelled data acquisition if leveraging supervised deep learning to 
accelerate symbolic computation. 
Cylindrical algebraic decomposition (CAD) is a pillar symbolic computation method for reasoning with first-order logic formulas over reals with many applications in formal verification and automatic theorem proving. Variable orderings have a huge impact on its efficiency.
Impeded by the difficulty to acquire abundant labelled data, 
existing learning-based approaches are only competitive with the best expert-based heuristics.
 In this work, we address this problem by designing a series of intimately connected tasks for which a large amount of annotated data can be easily obtained. We pre-train a Transformer model with these data and then fine-tune it on the datasets for CAD ordering. Experiments on publicly available CAD ordering datasets show that on average the orderings predicted by the new model are significantly better than those suggested by the best heuristic methods.

\end{abstract}

%% Keywords

\end{frontmatter}

\section{Introduction}

The advent of deep learning and subsequent emergence of large language models (LLMs)
have revolutionized the area of image and language processing 
and have been deeply impacting many areas in engineering and science. 
Despite the decades advances in artificial intelligence (AI), 
efficient and robust reasoning remains a challenging task~\citep{Melanie2025}.
For now, the logic and computation oriented approaches remain the dominant ones for trustworthy reasoning, powered respectively by proof assistants and computer algebra systems. 
Some efforts have been made to leverage the synergy between the learning-oriented and logic/computation-oriented approaches~\citep{DBLP:conf/acl/Lu00WC23}. However, it remains a challenge to combine them towards more efficient and robust reasoning~\citep{Xia_Li_Liu_Wu_Liu_2025}.

Cylindrical algebraic decomposition (CAD)~\citep{col75} is one of the main general approaches for performing quantifier elimination and
reasoning with first-order logic formulas over reals~\citep{Tarski98}.
Despite the worst  time complexity being doubly exponential~\citep{DBLP:conf/issac/BrownD07}, 
its practical efficiency has been incrementally improved through numerous theoretical and algorithmic advances 
as well as various implementation efforts in independent packages or  computer algebra systems, such as in QEPCAD~\citep{DBLP:journals/jsc/CollinsH91,brown2003qepcad}, Mathematica~\citep{DBLP:journals/jsc/Strzeboski00}, Redlog~\citep{seidl2003generic},  Maple~\citep{DBLP:conf/icms/ChenM14a}, etc.
The theoretical/algorithmic improvements have been achieved in different ways, 
including incremental improvements to Collins' original projection-lifting framework via 
improved projection operator~\citep{Hon90,Laz94,McC98,Bro01,MPP19}, improved lifting~\citep{DBLP:journals/jsc/Strzebonski06,IWANE201343}, input formula aware partial cylindrical algebraic decomposition~\citep{DBLP:journals/jsc/CollinsH91,DBLP:journals/jsc/BradfordDEMW16}, as well as more geometric ways through triangular decompositions~\citep{DBLP:conf/issac/ChenMXY09} and Gr\"obner bases~\citep{chen2025geometric}.

In particular, it has become an indispensable part of satisfiability modulo theory (SMT) for non-linear arithmetic~\citep{Jovanovic2012} and has been integrated into various SMT solvers~\citep{moura2008z3,dutertre2014yices,barbosa2022cvc5,Corzilius15}, which in turn unleashed the potential of CAD on addressing problems of larger size as a local approach~\citep{brown2015constructing,DBLP:conf/cav/LiXZ23,nalbach2024levelwise}.
There is no doubt that well-designed heuristics have played an important role 
in boosting the efficiency of SAT/SMT solvers. 
Recently, machine learning techniques have started to be integrated~\citep{Luzhengyang2024} into them.

For cylindrical algebraic decomposition,  
as the variable ordering can significantly affect its efficiency, 
several heuristics have been proposed~\citep{Brown04,DBLP:conf/issac/DolzmannSS04,DBLP:journals/cca/ChenDLMXXX11,DBLP:journals/jsc/LiXZZ23,DBLP:conf/casc/RioE22,DBLP:journals/jsc/PickeringAEC24}.
Currently, the heuristics perform much better than random choices, but 
are still far away from the optimal ordering~\citep{DBLP:conf/icms/ChenZC20,DBLP:conf/casc/ChenJQYZ24}.
The huge room for improvement have attracted several attempts to utilize machine learning
to suggest better variable orderings, reviewed in Section~\ref{sec:related}.
These learning-oriented models have demonstrated the potential to surpass the heuristic methods designed by experts , but the lead is minor in general. 
In fact, the Transformer model~\citep{transformer2017}, initially invented for natural language processing, has been used in~\citep{DBLP:conf/casc/ChenJQYZ24} for the CAD ordering selection problem. However, it does not noticeably surpass other models or the heuristic methods.

On the other hand, Transformer has found applications in learning some other tasks related to symbolic computation. 
Existing work mainly fall into two groups. 
The first group of work focus on end-to-end learning a simple output from a simple input for complex tasks, 
such as symbolic integration~\citep{DBLP:conf/iclr/LampleC20}, computing Lyapunov functions~\citep{DBLP:journals/corr/abs-2410-08304}, computing Gr\"obner bases~\citep{DBLP:journals/corr/abs-2311-12904,malhou2026hatsolver}, etc.
The second group of work focus on understanding the limitation and power of Transformer architecture 
on learning basic arithmetic operations like integer addition and multiplication~\citep{quirke2024understanding,DBLP:conf/iclr/LeeSL0P24,McLeish2024,cho2025accum}. 
Surprisingly, these ``basic tasks'' are not necessarily easier to learn than ``complex tasks''.
In this work, we are interested to explore the power of Transformer to learn better strategies to 
replace existing heuristic ones used in the computation of complex tasks.
For the above mentioned work, we note that what is in common for successful learning is that:
(i) a huge/large amount of labelled 
data are available, by either a forward or a backward process, for training Transformer models; 
(ii) both input and output must be relatively simple expressions.

In the case of selecting variable orderings for CAD, we do not know an economic way 
to synthesize the data.
At present, to create the correct label for the input system, one has to construct CADs 
for all possible orderings and pick the best one according to either the minimal time or the least number of cells. 
Hence, regards CAD ordering, even both the input system 
and the output ordering are simple expressions, the ordering selection task itself should be treated as a complex one
as it involves conducting a complex computation task.
The complex nature of this task also implies the data scarcity. 
Indeed, in~\citep{DBLP:conf/casc/ChenJQYZ24}, building a dataset of size 20K costs 
more than 140 CPU days. 
This makes CAD ordering a good case study for leveraging 
deep learning for complicated symbolic computation
in the circumstance of data scarcity.

The proposed approach is to leverage the pre-training and fine-tuning paradigm to address the  data scarcity problem 
for CAD ordering selection. 
More precisely, our main contribution in this work is three-fold:
\begin{itemize}
    \item We propose several simple pre-training tasks intimately connected to CAD. By generating large random datasets for them, we pre-train Transformer models for these tasks
          to achieve reasonable performance. 
     \item We  design a pre-training \& fine-tuning framework to leverage the pre-trained models to help learning the variable ordering selection task. We explain our choices through intensive ablation analysis.
    \item  We enhance a public four variables random dataset~\citep{DBLP:conf/casc/ChenJQYZ24}
    for CAD variable ordering selection (the new labelled data cost about 36 days to generate) and make the new dataset publicly available at 
    \begin{center}
        \url{https://doi.org/10.57760/sciencedb.31989}
    \end{center}
    We conduct experiments on public random datasets for three and four variables as well as a real dataset originated from benchmarks for satisfiability modulo theory (SMT) solvers. Employing the learned models to handle systems with more than four variables is also briefly discussed.
\end{itemize}

\section{Related work on learning the optimal variable ordering for CAD}
\label{sec:related}

In this section, we briefly review existing works on variable ordering selection 
for cylindrical algebraic decomposition and explain how they motivate the present work.

The first work~\citep{DBLP:journals/cca/HuangEWDP14} applies machine learning to choose the best heuristic method
from several candidates. 
The learned one outperforms each individual, but the performance is also limited by the best heuristic. 
The work~\citep{DBLP:conf/mkm/EnglandF19} and~\citep{ZhuChen2020a} propose classification models to directly choose the best variable ordering respectively 
according to the running time and number of cells by expanding features used in heuristic methods
in different ways.
As the number of possible variable orderings is $n!$ for $n$ variables,
direct classification models would be infeasible for large $n$ in terms of data acquisition and classification. 

To overcome this problem, the work~\citep{DBLP:conf/icms/ChenZC20} proposes an iterative method IVO with quadratic complexity to explore the variable ordering space to find orderings performing much 
better than the heuristic one. 
Like the earlier iterative approach {\sf sotd}~\citep{DBLP:conf/issac/DolzmannSS04}
and the latter proposed heuristic {\sf mods}~\citep{DBLP:conf/casc/RioE22}, 
it involves heavy CAD computation and cannot predict the variable ordering on the fly. 
In the same paper, they propose an approach  PVO learning  to predict only the leading variable in the ordering, 
which is then used to rotate an initial variable ordering suggested by a heuristic method. 
This approach relies on the same graph features proposed in~\citep{ZhuChen2020a}.  

Later, it is realized in both~\citep{jia2023suggesting} and~\citep{Jing2024} that
the CAD ordering selection could be modelled as a Markov decision process and the
reinforcement learning approaches naturally applies.
Instead of making a $n!$-classification to select an ordering, it suffices to take 
a sequence of $n-1$ actions and each action only needs to make at most $n-i$ different choices at step $i$, $i=0,\ldots,n-1$. 
The work~\citep{jia2023suggesting} is only trained on random datasets with $3$ variables
and its generalization ability on  datasets with more variables is limited. 
The work~\citep{Jing2024} achieves significant speedup for systems up to $16$ variables
but is limited to systems with the same support. 
The limitation of these learning based methods further motivates the work~\citep{DBLP:journals/jsc/PickeringAEC24}
to leverage explainable AI to propose better heuristics (see independent evaluations in ~\citep{DBLP:conf/casc/ChenJQYZ24}).

In summary, we have the following observations on the variable ordering problem for CAD:
(i) the potential for accelerating CAD computation by choosing an optimal variable ordering is huge
and such potential is only exploited in a moderate level; (ii) the existing approaches that further exploit the deep features of projected polynomials,  not matter heuristic or learning oriented,
could outperform the ones that utilize only the shallow features
of input systems; (iii) the ones that further exploit the deep features of projected polynomials
cannot suggest the variable ordering on the fly as they all have to explicitly compute projected polynomials; 
(iv) it is difficult, if not impossible, to generate a huge dataset for the CAD ordering selection problem.
Although data augmentation by permuting variables 
can increase the dataset size for free, 
it brings limited benefit on the model's performance~\citep{Hester2023,DBLP:conf/scsquare/RioE23}.
Based on these observations, we view CAD ordering as a complex task featured with data scarcity and propose to apply 
the pre-training and fine-tuning paradigm to solve this problem.

%%%%%%%%%%%%%%%%%%%%%%%%%%%%%%%%%%%%%%%%%%%%%%%%%%%%%%%%%%%%%%%

\section{Cylindrical algebraic decomposition}
\label{sec:background}
In this section, we briefly recall the notion of cylindrical algebraic decomposition. 

Let $F\subset\Q[x_1,\ldots,x_n]$ and let $x_1<\cdots<x_n$ be a given variable ordering.  
An $F$-invariant CAD, w.r.t. the ordering $x_1<\cdots<x_n$, decomposes $\R^n$ into finitely many connected semi-algebraic sets, called cells, 
such that these cells form a cylindrically arranged partition  of $\R^n$, namely the projections of any two cells 
onto any $\R^i$, $i=1,\ldots, n-1$ are either identical or disjoint.
Moreover, above each cell, the polynomials in $F$ enjoy some invariant property,
such as sign invariant or order invariant.

%Let $F\subset\Q[x_1,\ldots,x_n]$ be a finite set of nonzero polynomials. 
Let $F_n := \{f\mid f\in F, \mvar(f)=x_n\}$ and $F_o=F\setminus F_n\subset\Q[x_1,\ldots,x_{n-1}]$.
Here $\mvar(f)$ denotes the largest variable appearing in $f$ according to a given variable ordering. 
Let $F_s$ be a squarefree basis of $F_n$. 
The original idea of Collins for computing an $F$-invariant CAD is to find a projection operator which maps $F_s$ into another
finite set of polynomials $F_{p}\subset\Q[x_1,\ldots,x_{n-1}]$, 
such that an $F$-invariant CAD ${\cal C}_n$ of $\R^n$ can be constructed from an $F_p\cup F_o$-invariant CAD ${\cal C}_{n-1}$ of $\R^{n-1}$
by building a stack of cells of $\R^n$ above each cell of ${\cal C}_{n-1}$.

In this work, we are based on the Maple built-in command (since Maple 2020) {\sf CylindricalAlgebraicDecompose}~\citep{DBLP:conf/icms/ChenM14a}  inside the RegularChains library (denoted by RC-CAD for short) for computing CADs.
This command implements the triangular decomposition based algorithms for computing a CAD~\citep{DBLP:conf/issac/ChenMXY09,CMM14b}. 
More precisely, given  a system of polynomial equations and inequalities,  
say $sys := \{f_1=\cdots=f_m=0, g_1>0, \ldots, g_s>0\}$, it first computes a special triangular decomposition, 
called cylindrical decomposition ${\cal D}$ of the zero set of $sys := \{f_1=\cdots=f_m=0, g_1\neq 0, \ldots, g_s\neq 0\}$ in  $\C^n$ such that the cells of ${\cal D}$
are cylindrically arranged. 
If a set of polynomials $F$ is given, 
it computes a cylindrical decomposition ${\cal D}$ of $\C^n$ such that above 
each cell of ${\cal D}$, each polynomial in $F$ either vanishes at all points of the cell 
or vanishes at no points of the cell.
To obtain a cylindrical algebraic decomposition,
it further refines each cell in the cylindrical decomposition into a disjoint union of cylindrically arranged connected semi-algebraic sets via real root isolation of regular chains.

%{
Although the projection-lifting scheme and the triangular decomposition scheme compute CAD following different philosophy,  they share many common operations 
in the specific algorithms implementing the philosophy, 
especially when the input is a pure set of polynomials or has no equational constraints. 
In particular, for most of the algorithms implementing both the two schemes, if no equational constraints are present, they all need to perform at least the following projection operation. 

\begin{definition}
[Projection]Let $X=\left\{  x_{1},\ldots,x_{n}\right\}  $ and let
$F\subset\mathbb{Q}\left[  X\right]$ be finite. The $\emph{projection}$ of
$F\ $with respect to $x_{n}$, denoted by $\operatorname*{pf}\left(
F,x_{n}\right)  $,  is defined as follows. 
Let  $G$ be the set of all irreducible factors of $F$,  $G_{n}=\left\{  g\mid g\in G,  \ \ \deg_{x_{n}}(g)>0\right\}$,
$G_{0}=G\backslash G_{n}$ and $E_{n-1}= 
\left\{  \ \operatorname*{lc}\left(  g,x_{n}\right)  \ |\ g\in G_{n}%
\ \right\}  \cup 
\left\{  \ \operatorname*{discrim}\left(  g,x_{n}\right)  \ |\ g\in
G_{n}\ \right\}  \cup 
\left\{  \ \operatorname*{res}_{x_{n}}\left(  g,h\right)  \ |\ g,h\in
G_{n},g\neq h\ \right\}$, 
where $\lc(g, x_n)$, $\discrim(g, x_n)$, and $\res_{x_{n}}\left(  g,h\right)$
denote respectively the leading coefficient of $g$, the discriminant of $g$, and the resultant of $g$
and $h$ w.r.t. the variable $x_n$.
%\end{compactitem}
\noindent Then $\operatorname*{pf}\left(  F,x_{n}\right)  =G_{0}%
\ \cup\ E_{n-1}\ \backslash\ \mathbb{Q}$.
\end{definition}

%}

\begin{example}
Let $f := y^2(y^2-b^2)-x^2(x^2-a^2)$.
For fixed values of $a$ and $b$, its zero set in $\R^2$ defines a Devil's curve.
Let $a=4/5$, $b=1$, $g=25f$, and $F := \{g\}$.
Then an $F$-sign invariant CAD under the order $x>y$ (resp. $y>x$)
is illustrated by the left (resp. right) subfigure of  Figure ~\ref{fig:Devil}.
The number of CAD cells in the left (resp. right) subfigure is 89 (resp. 49). 
\begin{figure}[ht!]
    \centering
    \includegraphics[width=0.4\textwidth]{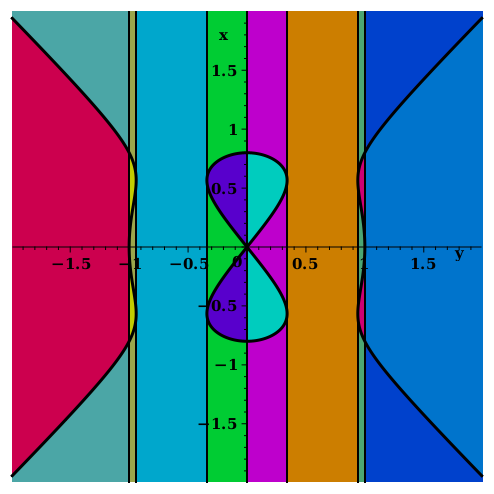}
    \includegraphics[width=0.4\textwidth]{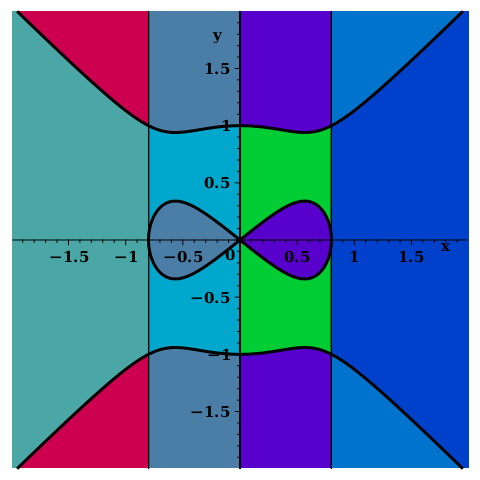}
    \caption{CADs of the Devil's curve under the ordering \texorpdfstring{$x>y$}{x>y} and \texorpdfstring{$y>x$}{y>x}.}

    \label{fig:Devil}
\end{figure}
\end{example}

%%%%%%%%%%%%%%%%%%%%%%%%%%%%%%%%%%%%%%%%%%%%%%%%%%%%%%%%%%%%%%%

\section{Method}
\label{sec:method}

In this section, we introduce a Transformer-based framework to tackle the problem of variable ordering selection for CAD by leveraging a pre-training and fine-tuning paradigm.

Pre-training and fine-tuning is a learning paradigm popularized by the encoder-only Transformer model BERT~\citep{devlin-etal-2019-bert}. 
% \reworked{It is a particular way of transfer learning~\citep{wang2023introduction}
% and is suitable for addressing the problem of learning complex tasks facing  data scarcity. }
% {
It represents a specific form of transfer learning~\citep{wang2023introduction}, 
well suited for tackling complex learning tasks under limited data availability.
% }
%Now, it has become the standard learning paradigm for training LLMs and adapting LLMs for downstream tasks. 
Originally used for natural language processing, the core idea is to  pre-train a model to ``understand" the input language or have a good representation of the input language by utilizing the contextual information. 
%For instance, the contextual information can be obtained by masking some words in a sentence or masking some sentences in a paragraph. In this way, the mask scheme provides a free way to annotate data and a latent representation of the input language can be learned in a self-supervised way. 
In general, pre-training can be used in the situation where huge amount of labelled data are accessible for training a model to learn a common representation/embedding which might be useful for many downstream tasks. Fine-tuning, on the other hand, can inherit such a common representation  and may keep refining it for a particular downstream task.

In our context, the main task is to train 
a model to effectively predict the variable ordering for CAD. 
However, we only have limited labelled data for this task. 
To leverage the power of the pre-training and fine-tuning paradigm, 
we have to create suitable pre-training tasks such that on one hand these tasks are closely related to the main task and on the other hand collecting 
a large amount of data for these pre-training tasks should be relatively easy. 
Section~\ref{subsec:frame} provides an overview of the framework.
The pre-training tasks are detailed in Section~\ref{subsec:pretask}.
Some of these tasks are inspired by the features used in heuristic methods for variable ordering selection, 
which are recalled in  Section~\ref{subsec:feature}. 
The process for generating the labelled dataset for each task is detailed in Section~\ref{subsec:data}. 
These data have to be tokenized to be used with the Transformer models.
The tokenization scheme is elaborated in Section~\ref{subsec:encode}.

%\subsection{\reworked{The pre-training and fine-tuning framework}
%{
\subsection{The framework for CAD variable ordering selection}
%}
\label{subsec:frame}

\begin{figure}[ht!]
    \centering
    \includegraphics[width=0.8\textwidth]{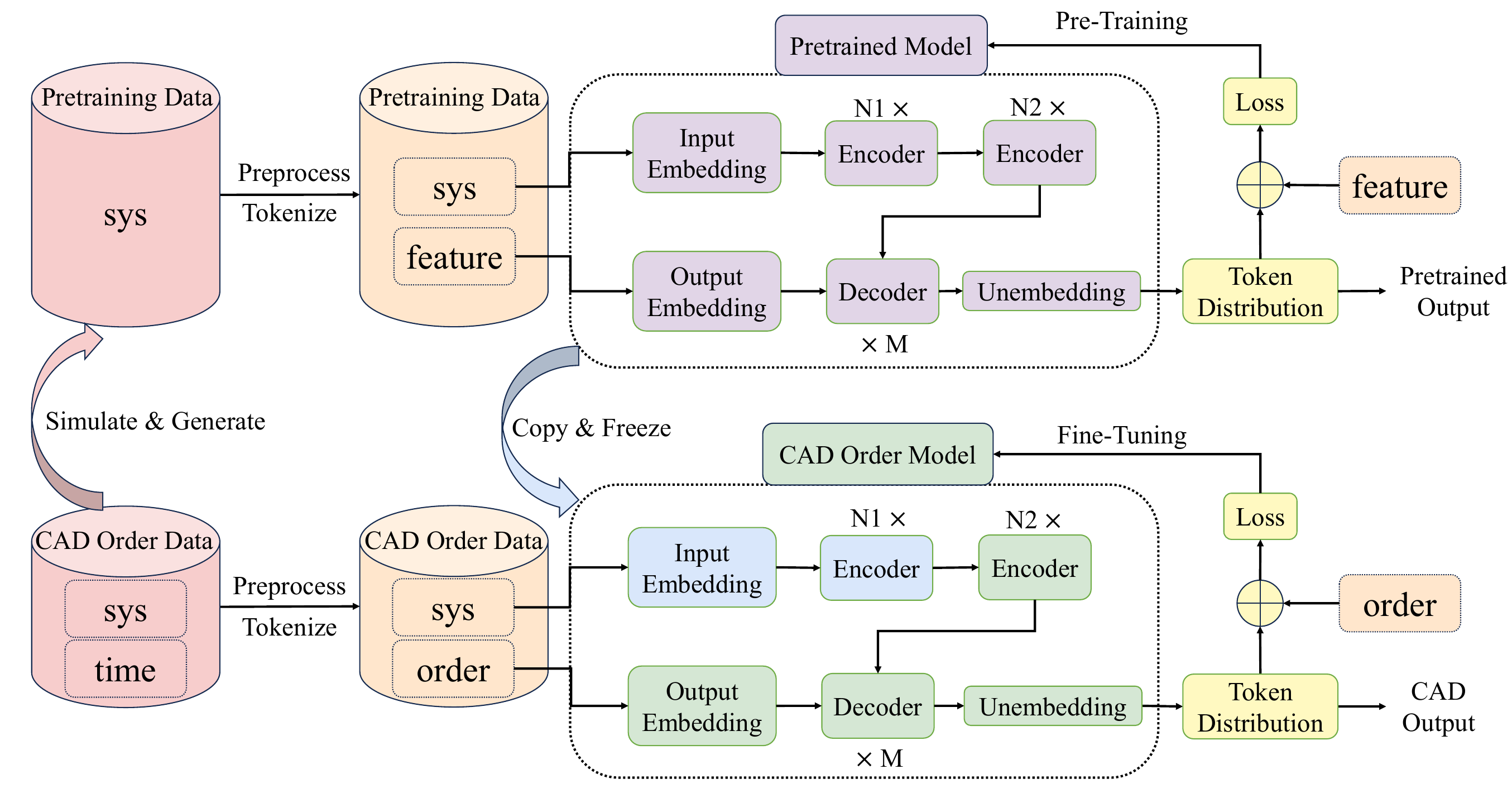}
    \caption{An overview of the proposed framework for CAD variable ordering selection.}
    \label{fig:frame}
\end{figure}

An overview of the proposed
%\reworked{pre-training and fine-tuning framework}
%{
pre-training and fine-tuning framework for CAD variable ordering selection
%}
is illustrated by Figure~\ref{fig:frame}. 
The entire process begins with generating a large number of random systems 
that reflect the structural characteristics of those in the CAD order dataset, 
including the number of constraints, the number of variables, and the exponents of variables. 
Each generated system is then preprocessed to extract a feature vector, which serves as its label according to a designed pre-training task. 
The resulting labeled dataset is subsequently tokenized and used to train and evaluate a Transformer model. 
The model obtained through this pre-training procedure is referred to as the \emph{pre-trained model}.
%}

After pre-training, we fine-tune the model on the CAD order dataset. 
During fine-tuning, we freeze the input embedding layers and some early encoder layers (highlighted in light blue in Figure~\ref{fig:frame}). 
These layers are expected to encode task-independent structural properties of input systems, as well as basic computational behaviors that facilitate learning effective variable orderings for CAD construction. 
The remaining unfrozen layers are then trained using the CAD order dataset, enabling the model to specialize in CAD variable ordering prediction. 
This fine-tuning strategy allows the model to leverage the general representations learned during pre-training while adapting to the specific patterns of the downstream task.
%}

\subsection{Important features relevant to the variable ordering }
\label{subsec:feature}

%{ 

\begin{notation}
Let $F\subset\mathbb{Q}\left[  X\right]$, finite. 
Let $\mathcal{T}_{f}$ denote the set of all terms appearing in $f$.
For $T\in\mathcal{T}_{f}$ and $v\in X$, let

\begin{compactitem}
\item ${\sf d}\left(  T,v\right)  =\deg_{v}(T)$

\item ${\sf a}\left(  T,v\right)  =\left\{
\begin{array}
[c]{ll}%
1 & \text{if\ }v\ \text{appears in }T\\
0 & \text{else}%
\end{array}
\right.  $

\item ${\sf e}\left(  T,v\right)  =\left\{
\begin{array}
[c]{ll}%
\deg (T) & \text{if\ }v\ \text{appears in }T\\
0 & \text{else}%
\end{array}
\right.  $
\end{compactitem}

For $U,V\in\left\{  \max,\operatorname*{sum},\operatorname*{avg}%
\right\}  $ and $o\in\left\{  {\sf d},{\sf a}%
,{\sf e}\right\}  $, let $U\_V\_o\left(  F,v\right)
\ =\ \underset{f\in F}{U}\ \underset{T\in T_{f}}{V}\ o\left(  T,v\right)  $.

\end{notation}

In Table 5 of~\citep{PICKERING2024102276}, 
with the SHAP (SHapley Additive exPlanations) method, $27$
different feature operators are ranked based 
on the impact of the features they produce on the output
of machine learning models. 
Since the reduction operation ${avg}$ introduces rational numbers, causing extra difficulty for learning with Transformer, 
we remove the operators involving ${avg}$ and leave the rest to maintain their original order.
Moreover, since ${\rm max}\_{\rm max}\_{\sf a}$ would always be $1$ if a variable appears in the input polynomials, 
we replace it with ${\rm sum}\_{\rm max}\_{\sf a}$. 
This results the following $11$ sorted feature operators: 
% \begin{table}[h]
% \centering
% \begin{tabular}{l}
\begin{equation}
\label{eq:feature}
\begin{array}{l}
\emph{{\rm sum}\_{\rm max}\_{\sf d}}, \emph{{\rm sum}\_{\rm max}\_{\sf e}}, \emph{{\rm sum}\_{\rm sum}\_{\sf d}}, \emph{{\rm sum}\_{\rm sum}\_{\sf e}},
\emph{{\rm sum}\_{\rm sum}\_{\sf a}}, \emph{{\rm max}\_{\rm max}\_{\sf d}},\\
\emph{{\rm max}\_{\rm sum}\_{\sf e}}, \emph{{\rm max}\_{\rm max}\_{\sf e}},
\emph{{\rm sum}\_{\rm max}\_{\sf a}}, \emph{{\rm max}\_{\rm sum}\_{\sf d}}, \emph{{\rm max}\_{\rm sum}\_{\sf a}}.
\end{array}
\end{equation}
In the above feature operators, the correspondence between the symbols here and the ones in~\citep{PICKERING2024102276} is: 
${\sf d}$=``v", ${\sf a}$=``sg\_v" and ${\sf e}$=``sv".
% \end{tabular}
% \end{table}
The operators listed in Eq.~(\ref{eq:feature})
are exactly the elements of the following set 
\[
R=\left\{  U\_V\_o\ |\ U,V\in\left\{  \max,\operatorname*{{\rm sum}}\right\}
,\ o\in\left\{  {\sf d},{\sf a},{\sf e}\right\}
\right\}  \ \backslash\ \left\{  {\rm max}\_{\rm max}\_{\sf a}\right\}.
\]
Note that $\#R$\thinspace$=$ $11=2\times2\times3-1$. 
We will order the elements of $R$ as Eq.~(\ref{eq:feature}).

\begin{definition}[Feature matrix]
The \emph{feature matrix} of $F$, denoted by $\operatorname*{fm}(F)$, is the
$\#R\times n$ matrix whose entries are given by
$\operatorname*{fm}(F)_{ij}=R_{i}\left(  F,x_{j}\right)$.

\end{definition}
%}

\subsection{Pre-training tasks}
\label{subsec:pretask}

%{
We denote by  $\texttt{task\_c}$ the task of predicting an optimal variable ordering for CAD.
The precise meaning of ``optimal" will be made clear in Section~\ref{subsec:data}.
Roughly speaking, an optimal variable ordering is the one that minimizes the running time of a CAD solver under all possible variable orderings.
As generating a large amount of labelled data for this task is very challenging, 
we would define a series of pre-training tasks for which it is much easier to obtain abundant labelled data. On the other hand, these pre-training tasks should be relevant to $\texttt{task\_c}$.
Let us elaborate this point a bit more.
A naive and the only certain way we know so far to find an optimal variable ordering 
is to compute CADs under all possible variable orderings and then choose the one 
minimizing the CAD running time. Thus it is natural to require a pre-training task to extract useful information from some computation steps in constructing a CAD. 
This motivates us to define six pre-training tasks in Definition~\ref{def:pre-training}.

\begin{definition}
[Pre-training features and pre-training tasks]
\label{def:pre-training}
Let $F\subset\mathbb{Q}\left[  x_{1},\ldots,x_{n}\right]  $ and $v\in X$.
Write $q=\prod\limits_{f\in F}f$ and $g=\operatorname*{gsfp}\left(  q\right)$, where
$\operatorname*{gsfp}$ stands for the greatest square free part.
Let $r_{v}=\operatorname*{res}_{v}\left(  g,\frac{\partial g}{\partial
v}\right)  $ and  $sr_{v}=\operatorname*{gsfp}\left(  r_{v}\right)  $.
We define

\begin{compactitem}
\item $\texttt{feature\_f}=\operatorname*{fm}\left(  F\right)  $,

\item $\texttt{feature\_p}=\operatorname*{fm}\left(  \underset{1\leq i\leq n}{\cup
}\operatorname*{pf}\left(  F,x_{i}\right)  \ \right)  $,
\item $\texttt{feature\_s}=\operatorname*{fm}\left(  \underset{1\leq i\leq n}{\cup
}\operatorname*{pf}\left(  \left\{  g\right\}  ,x_{i}\right)  \right)  $,

\item $\texttt{feature\_e}=\left\{  \left[  \left(  \deg_v(T)\ |\ v\in X\right)
\ |\ T\in\mathcal{T}_{f}\setminus\Q \right]  :f\in F\right\}  $,

\item $\texttt{feature\_m}=\left[  \left(  \deg_{v}(T)\ |\ v\in X\right)  \ |\ T\in
\mathcal{T}_{q}\setminus\Q\right]  $,

\item $\texttt{feature\_r}=\left[  \left(  \deg (r_{v}) |\ v\in X\right),\left( \deg (sr_{v})|\ v\in X\right),\left( \#\mathcal{T}_{r_{v}}|\ v\in X\right), \left( \#\mathcal{T}_{sr_{v}}|\ v\in X\right) \right]  $.
\end{compactitem}
For $i\in\{f, p, s, e, m, r\}$, 
denote by \texttt{task\_i} the pre-training task for predicting the feature $feature\_i$.
\end{definition}

In Definition~\ref{def:pre-training}, $\texttt{feature\_f}$ contains only information of the input polynomials and has been shown to be effective in both heuristic and learning-oriented 
approaches for selecting the variable ordering of CAD~\citep{DBLP:conf/scsquare/FlorescuE19,PICKERING2024102276}. 
The features $\texttt{feature\_p}$, $\texttt{feature\_s}$ and $\texttt{feature\_r}$
include important information on polynomials in the projection, such as the maximum degree of a variable appearing in them, the number of polynomials containing a variable, the total number of terms including/excluding a variable, etc.
The features $\texttt{feature\_e}$ and $\texttt{feature\_m}$
provide the exponents of the input polynomials
and the exponents of their product.
% \fbox{$\texttt{feature\_e}$ discards constant terms, what about $feature_m$?}
%}

Next, we use Example~\ref{exam:features} to illustrate the pre-training features and tasks.

\begin{example}
\label{exam:features}
Let $f_1=-6x_1^3x_2-4x_1x_2x_3^2+2x_2^2x_3+1$ and $f_2=-5x_3^4+x_3^3-7$ be two polynomials in variables  $x_1,x_2,x_3$. 
Let $F := \{f_1, f_2\}$ and $sys := \{f_1=0, f_2=0\}$.
The system $sys$ and the timings for computing a CAD for it under different variable orderings are taken from the dataset DQ-3 in~\citep{DBLP:conf/casc/ChenJQYZ24}.

{  \noindent{\texttt{task\_c}.}}
The six variable orderings, 
%arranged in lexicographic order, are  
$\{[x_1, x_2, x_3], [x_1, x_3, x_2], [x_2, x_1, x_3], [x_2, x_3, x_1], [x_3, x_1, x_2], [x_3, x_2, x_1]\},$
with the corresponding CAD running times (in seconds)  
$\{0.035, 0.080, 0.034, 0.048, \textit{timeout}, \textit{timeout}\}$, 
where the time limit was set as $900$ seconds.  
The absolute optimal ordering, i.e. the one yielding the shortest running time, is  
\[
\texttt{best\_order} \boldsymbol{=} [\texttt{$x_2$},\; \texttt{$x_1$},\; \texttt{$x_3$}].
\]

{  \noindent{\texttt{task\_e}.}}
We have ${\cal T}_{f_1} =\{-6x_1^3x_2, -4x_1x_2x_3^2, +2x_2^2x_3,+1\}$ and ${\cal T}_{f_2} = \{-5x_3^4,+x_3^3,-7\}$, discarding constant terms, yielding the exponent vectors:
\[
\texttt{feature\_e} \boldsymbol{=} \{\,[\texttt{(3\ 1\ 0)};\; \texttt{(1\ 1\ 2)};\; \texttt{(0\ 2\ 1)}],\; [\texttt{(0\ 0\ 4)};\; \texttt{(0\ 0\ 3)}]\,\}.
\]

{  \noindent{\texttt{task\_f}.}}
% As ${\sf d}(F,x_1) := \{\{3, 1, 0, 0\}, \{0,0,0\}\}$, ${\sf d}(F, x_2)=\{\{1,1,2,0\}, \{0,0,0\}\}$, 
% and  ${\sf d}(F, x_3)=\{\{0,2,1,0\}, \{4,3,0\}\}$, we have  $\emph{sum\_max\_{\sf d}}(F, x_1)=3$, 
% $\emph{sum\_max\_{\sf d}}(F, x_2)=2$, and $\emph{sum\_max\_{\sf d}}(F, x_3)=6$, yielding 
% the first feature vector $(3\ 2\ 6)$. 
We have
\[
\begin{aligned}
\texttt{feature\_f} \boldsymbol{=} [&\texttt{(3\ 2\ 6)};\; \texttt{(4\ 4\ 8)};\; \texttt{(4\ 4\ 10)};\; \texttt{(8\ 11\ 14)};\; \texttt{(2\ 3\ 4)};\; \texttt{(3\ 2\ 4)}; \\&
\texttt{(8\ 11\ 7)};\; \texttt{(4\ 4\ 4)};\; \texttt{(1\ 1\ 2)};\; \texttt{(4\ 4\ 7)};\; \texttt{(2\ 3\ 2)}].
\end{aligned}
\]

{  \noindent{\texttt{task\_m}.}}
The exponent vectors of the product of $f_1$ and $f_2$ are:
\[
\begin{aligned}
\texttt{feature\_m} \boldsymbol{=} [&\texttt{(3\ 1\ 4)};\; \texttt{(1\ 1\ 6)};\; \texttt{(3\ 1\ 3)};\; \texttt{(1\ 1\ 5)};\; \texttt{(0\ 2\ 5)};\; \texttt{(0\ 2\ 4)}; \\ &
\texttt{(3\ 1\ 0)};\; \texttt{(1\ 1\ 2)};\; \texttt{(0\ 0\ 4)};\; \texttt{(0\ 2\ 1)};\; \texttt{(0\ 0\ 3)}].
\end{aligned}
\]

{  \noindent{\texttt{task\_p}.}}
%\reworked{
As $\cup_{i=1}^3 pf(F,x_i) = \{5\,{{x_3}}^{4}-{{x_3}}^{3}+7,6\,{x_2},-1536\,{{x_2}}^{4}{ {x_3}}^{6}-3888\,{{x_2}}^{6}{{x_3}}^{2}-3888\,{{x_2}}^{4}{ x_3}-972\,{{x_2}}^{2},5\,{{x_3}}^{4}-{{x_3}}^{3}+7,-2\,{ x_3},36\,{{x_1}}^{6}+48\,{{x_1}}^{4}{{x_3}}^{2}+16\,{
%{\it x1}
x_1
}^{2}{{x_3}}^{4}-8\,{x_3},4\,{x_1}\,{x_2},-96\,{{x_1 }}^{4}{{x_2}}^{2}+4\,{{x_2}}^{4}+16\,{x_1}\,{x_2},32400\,{ {x_1}}^{12}{{x_2}}^{4}+864\,{{x_1}}^{10}{{x_2}}^{4}-2160\, {{x_1}}^{9}{{x_2}}^{5}-21600\,{{x_1}}^{9}{{x_2}}^{3}+40320 \,{{x_1}}^{8}{{x_2}}^{4}-432\,{{x_1}}^{7}{{x_2}}^{3}+1080 \,{{x_1}}^{6}{{x_2}}^{4}+4032\,{{x_1}}^{5}{{x_2}}^{5}- 13440\,{{x_1}}^{4}{{x_2}}^{6}+5400\,{{x_1}}^{6}{{x_2}}^{2} -13440\,{{x_1}}^{5}{{x_2}}^{3}+12544\,{{x_1}}^{4}{{x_2}}^{4} -224\,{x_1}\,{{x_2}}^{7}+560\,{{x_2}}^{8}+72\,{{x_1}}^{4}{{x_2}}^{2}-180\,{{x_1}}^{3}{{x_2}}^{3}-672\,{{x_1}}^{2}{ {x_2}}^{4}+2240\,{x_1}\,{{x_2}}^{5}-600\,{{x_1}}^{3}{x_2}+1120\,{{x_1}}^{2}{{x_2}}^{2}-4\,{x_1}\,{x_2}+10\,{{x_2}}^{2}+25\}$, we have
$$
\begin{aligned}
\texttt{feature\_p} \boldsymbol{=} [&\texttt{(23\ 20\ 19)};\; \texttt{(30\ 35\ 25)};\; \texttt{(120\ 107\ 31)};\; \texttt{(202\ 224\ 51)};
\texttt{(26\ 31\ 11)};\; \texttt{(12\ 8\ 6)};\;\\
&\texttt{(174\ 184\ 23)};\; \texttt{(16\ 16\ 10)};\texttt{(4\ 5\ 5)};\; \texttt{(102\ 82\ 9)};\; \texttt{(20\ 22\ 3)}].
\end{aligned}
$$

{ \noindent{\texttt{task\_r}.}}
Let $g$ be the greatest squarefree part of the product of polynomials in $F$.
Let $v=x_1$, 
we have $r_v=\operatorname*{res}_{v}\left(  g,\frac{\partial g}{\partial
v}\right)=72\,{{x_2}}^{3} \left( 128\,{{x_2}}^{2}{{x_3}}^{6}+324\,{{
x_2}}^{4}{{x_3}}^{2}+324\,{{x_2}}^{2}{x_3}+81 \right) 
 \left( 5\,{{x_3}}^{4}-{{x_3}}^{3}+7 \right) ^{5}$. 
Thus the total degree of $r_v$ is $31$, 
and the total degree of the squarefree part $sr_v$  of $r_v$ is $13$. 
The result can be obtained in a similar way, yielding:  
$$
\texttt{feature\_r} \boldsymbol{=} [\texttt{(31\ 19\ 40)};\; \texttt{(13\ 11\ 23)};\; \texttt{(60\ 40\ 144)};\; \texttt{(12\ 12\ 45)}].
$$

{ \noindent {\texttt{task\_s}.}}
Computing $\cup_{i=1}^3 pf(\{g\},x_i)$  brings 
six polynomials  $\{6x_2(5x_3^4-x_3^3+7),-960000\,{{x_2}}^{4}{{x_3}}^{22}+\cdots, -2x_3(5x_3^4-x_3^3+7),900\,{{x_1}}^{6}{{x_3}}^{8}+\cdots, 20x_1x_2,-1105994585041920000\,{{x_1}}^{28}{{x_2}}^{10}-\cdots\}$, yielding:  
$$ 
\begin{aligned}
\texttt{feature\_s} \boldsymbol{=} [&\texttt{(35\ 28\ 43)};\; \texttt{(54\ 71\ 50)};\; \texttt{(1506\ 1643\ 703)};\; \texttt{(3010\ 3616\ 961)};
\texttt{(157\ 196\ 75)};\;\\ & \texttt{(28\ 20\ 22)};\;\texttt{(2816\ 2880\ 722)};\; \texttt{(38\ 38\ 26)};  \; \texttt{(3\ 4\ 4)};\; \texttt{(1433\ 1447\ 532)};\; \texttt{(138\ 144\ 47)}].
\end{aligned}
$$

\end{example}

\subsection{Preprocessing the data}
\label{subsec:data}
In this section, we detail the preprocessing pipeline applied to the raw datasets used in this work, which include both the CAD data and the pre-training data. The goal is to articulate how these two types of data are transformed from their initial raw forms into structured input–output pairs suitable for model learning.

For raw CAD data, each instance is represented as an input–output pair $(sys, time)$, where $sys$ denotes an $n$-variable polynomial system and $time$ records the running timings of the CAD solver under all $n!$ possible variable orderings. 
Let  $t^\ast$ denote the minimum running time.
% The objective of preprocessing is to transform each instance into $(sys, best\_order)$, where $best\_order$ is the variable ordering that minimizes the runtime, referred to as the absolute optimal ordering, with corresponding runtime $t^\ast$. 
Extensive empirical evidence indicates that certain orderings, while not attaining the exact minimum, yield running timings that are nearly indistinguishable from $t^\ast$. To capture these cases, we introduce a tolerance parameter $\tau = 0.03$ and define the set of relative optimal orderings as
$
\mathcal{O}_{\mathrm{rel}} = \{\, order \;|\; t(order) \leq (1+\tau)\,t^\ast \,\},
$
%By definition, the absolute optimal ordering necessarily belongs to $\mathcal{O}_{\mathrm{rel}}$.
{where \(t({order})\) denotes the running time of the CAD solver under the given variable ordering \({order}\).}
For convenience, unless otherwise stated, the term ``optimal'' is used to refer to any ordering within $\mathcal{O}_{\mathrm{rel}}$, thereby subsuming the absolute optimum as a special case. 
For instance, in Example~\ref{exam:features}, with $\tau = 0.03$, both $[x_2, x_1, x_3]$ and $[x_1, x_2, x_3]$ are considered to be optimal.
The objective of preprocessing raw CAD data is to transform each instance into pairs of the form $(sys, optimal\_order)$.

The raw pre-training data consist of only  the input polynomial systems. 
The objective of preprocessing in this setting is to construct input–output pairs of the form $(sys, feature)$, where $feature$ denotes the task-specific feature vector described in Definition~\ref{def:pre-training}.
%{
The preprocessing step also includes a data screening procedure, 
detailed in Section~\ref{subsec:CAD-pretrain-dataset}.
%}

Next, we clarify two subtle considerations regarding the CAD data preprocessing:  
$(i)$ For both the training and validation datasets, the \emph{absolute optimal ordering} is used as the target label. It may happen that multiple orderings yield the identical minimum running time.To accommodate this, we apply a multi-label strategy: if a system $sys$ admits two absolute optimal orderings $order_1$ and $order_2$, 
we record two separate samples $(sys, order_1)$ and $(sys, order_2)$. 
$(ii)$ Since testing-validation and testing datasets are only used for prediction, no multi-label augmentation preprocessing on the data is applied. 
Instead, we assess the quality of evaluation under two complementary metrics: \emph{absolute accuracy} (probability of predicting an absolute optimal ordering) and \emph{relative accuracy} (probability of predicting a relative optimal ordering).

\subsection{The tokenization scheme}
\label{subsec:encode}

%The transformer model requires sequence representations as input. 
To encode a CAD problem into a sequential format, we construct a specialized token set tailored for polynomial systems, encompassing both input and output sequence components.
Table~\ref{tab:tokens} presents the complete token vocabulary,  organized into seven categories, used for $n$-dimensional CAD variable ordering selection problems in our study.
\begin{table}[!ht]
\caption{The vocabulary of tokens.}
\label{tab:tokens}
\centering
\begin{tabular}{ll|ll}
\toprule
\text{Category} & \text{Tokens} & \text{Category} & \text{Tokens}\\
\midrule
special\_tokens & \texttt{$\langle s\rangle$}  \texttt{$\langle/s\rangle$}  \texttt{$\langle \text{pad}\rangle$} & separators & \texttt{$\langle \text{sep}\rangle$} \texttt{;}  \texttt{,} \\
variables  $x_1 \ldots x_{n}$ & \texttt{x1}  $\ldots$  \texttt{xn} & digits\_of\_exponents & \texttt{0}--\texttt{9} \\
arithmetic\_operators &  \texttt{+}  \texttt{-}  \texttt{*}  \texttt{$\wedge$} & digits\_of\_coefficients & \texttt{c0}--\texttt{c9} \\
relational\_operators & \texttt{=}  \texttt{>}  \texttt{$\geq$}  \texttt{<}  \texttt{$\leq$}  \texttt{$\neq$} &  digits\_of\_feature\_vectors &  \texttt{0}--\texttt{9}\\
\bottomrule
\end{tabular}
\end{table}

In the table, three special control symbols are introduced to facilitate sequence modeling: $\langle s\rangle$ marks the start of a sequence, $\langle/s\rangle$ denotes its termination, and \texttt{$\langle \text{pad}\rangle$} serves as a padding placeholder for sequence alignment during batch processing. 
% For an $n$-dimensional polynomial system, variables are indexed systematically from $x_1$ to $x_{n}$ following Maple’s polynomial generation rules. To represent fundamental operations in polynomial expressions, we include the standard arithmetic operators $+$, $-$, $*$, and $\wedge$, while relational operators such as $=$, $>$, $\geq$, $<$, $\leq$, and $\neq$ are incorporated to capture the constraints and logical relationships inherent in polynomial systems. 
To further ensure structural clarity, three types of separators are defined: the comma “\texttt{,}” separates (features of) different polynomials in the same polynomial system, the semicolon “\texttt{;}” distinguishes different types of output features in pre-training tasks, and \texttt{$\langle \text{sep}\rangle$} separates the $n$ values associated with a single feature, thereby supporting a hierarchical and unambiguous representation.

For both exponents and feature vectors, we encode each digit of them separately. 
In contrast, the digits in the coefficients are prefixed with the letter ``$c$" yielding $c_0$–$c_9$,
This strategy is adopted for two reasons: $(i)$ the feature vectors are constructed based on the exponents of input/computed polynomials;
$(ii)$ exponents play a much more important role than coefficients in variable ordering selection, and it is desirable to distinguish their digits
explicitly. 

Furthermore, to enable the pre-training task to more effectively capture the structural features of polynomial systems, a completion procedure is applied to each polynomial term. Specifically, if a variable is absent, it is explicitly represented with an exponent of zero; similarly, variables of degree one are explicitly represented with an exponent of one. For example, in a system with three variables $\{x_1,x_2,x_3\}$, the term $-2x_2^2x_3$ is encoded as \{ \texttt{-}, \texttt{c2}, \texttt{*}, \texttt{x1}, $\wedge$, \texttt{0}, \texttt{*}, \texttt{x2}, $\wedge$, \texttt{2}, \texttt{*}, \texttt{x3}, $\wedge$, \texttt{1}\}. In contrast, the encoding method in \citep{DBLP:conf/casc/ChenJQYZ24} neither explicitly distinguishes between coefficients and exponents nor fully encodes the polynomial term (e.g. variables with zero exponents are omitted). Its encoded result for the same example is \{ \texttt{-}, \texttt{2}, \texttt{*}, \texttt{x2}, $\wedge$, \texttt{2}, \texttt{*}, \texttt{x3}\}.

During model training, sequences shorter than the maximum length within a batch are padded with $\langle \text{pad}\rangle$ tokens appended to their end until they match the batch’s longest sequence, whereas sequences already at the maximum length remain unchanged without any padding.

\begin{example}
{\rm
Next, we illustrate the encoding scheme in detail through tokenizing the polynomial system $sys:=\{-6x_1^3x_2-4x_1x_2x_3^2+2x_2^2x_3+1=0, -5x_3^4+x_3^3-7=0\}$ and its outputs for different tasks presented in Example~\ref{exam:features}. 

The sequence of tokens representing the input system $sys$ is given as \{\texttt{$\langle s\rangle$}, \texttt{-}, \texttt{c6}, \texttt{*}, \texttt{x1}, $\wedge$, \texttt{3}, \texttt{*}, \texttt{x2}, $\wedge$, \texttt{1}, \texttt{*}, \texttt{x3}, $\wedge$, \texttt{0}, \texttt{-}, \texttt{c4}, \texttt{*}, \texttt{x1}, $\wedge$, \texttt{1}, \texttt{*}, \texttt{x2}, $\wedge$, \texttt{1}, \texttt{*}, \texttt{x3}, $\wedge$, \texttt{2}, \texttt{+}, \texttt{c2}, \texttt{*}, \texttt{x1}, $\wedge$, \texttt{0}, \texttt{*}, \texttt{x2}, $\wedge$, \texttt{2}, \texttt{*}, \texttt{x3}, $\wedge$, \texttt{1}, \texttt{+}, \texttt{c1},\texttt{=}, \texttt{c0}, \texttt{,}, \texttt{-}, \texttt{c5}, \texttt{*}, \texttt{x1}, $\wedge$, \texttt{0}, \texttt{*}, \texttt{x2}, $\wedge$, \texttt{0}, \texttt{*}, \texttt{x3}, $\wedge$, \texttt{4}, \texttt{+}, \texttt{x1}, $\wedge$, \texttt{0}, \texttt{*}, \texttt{x2}, $\wedge$, \texttt{0}, \texttt{*}, \texttt{x3}, $\wedge$, \texttt{3}, \texttt{-}, \texttt{c7},\texttt{=}, \texttt{c0}, \texttt{$\langle /s\rangle$}\}. 
The tokenization of an output ordering $[x_2,x_1,x_3]$ is simply 
\{\texttt{$\langle s\rangle$},  \texttt{x2}, \texttt{x1}, \texttt{x3}, \texttt{$\langle /s\rangle$}\}.

%{ Next we present the tokenization of the output of each task introduced in Definition~\ref{def:pre-training}.}

The tokenization of 
$\texttt{feature\_e}$ is given as \{\texttt{$\langle s\rangle$}, \texttt{3}, \texttt{$\langle \text{sep}\rangle$}, \texttt{1}, \texttt{$\langle \text{sep}\rangle$}, \texttt{0}, \texttt{;},\texttt{1}, \texttt{$\langle \text{sep}\rangle$}, \texttt{1}, \texttt{$\langle \text{sep}\rangle$}, \texttt{2}, \texttt{;},\texttt{0}, \texttt{$\langle \text{sep}\rangle$}, \texttt{2}, \texttt{$\langle \text{sep}\rangle$}, \texttt{1}, \texttt{,},\texttt{0}, \texttt{$\langle \text{sep}\rangle$}, \texttt{0}, \texttt{$\langle \text{sep}\rangle$}, \texttt{4}, \texttt{;},\texttt{0}, \texttt{$\langle \text{sep}\rangle$}, \texttt{0}, \texttt{$\langle \text{sep}\rangle$}, \texttt{3}, \texttt{$\langle /s\rangle$}\}. It is important to note that while semicolon ``;" are used to separate features, comma ``\texttt{,}" are employed to distinguish different polynomial systems. As a result, both semicolons and commas may appear simultaneously in this pre-training task. 

The tokenization of the other features are similar. 
Note that ``\texttt{,}"  will not appear in them and integers 
with multiple digits have to be separated into different tokens. 
For instance,  $\texttt{feature\_r}$ is represented as \{\texttt{$\langle s\rangle$}, \texttt{3}, \texttt{1}, \texttt{$\langle \text{sep}\rangle$}, \texttt{1}, \texttt{9}, \texttt{$\langle \text{sep}\rangle$}, \texttt{4}, \texttt{0}, \texttt{;}, \texttt{1}, \texttt{3}, \texttt{$\langle \text{sep}\rangle$}, \texttt{1}, \texttt{1}, \texttt{$\langle \text{sep}\rangle$}, \texttt{2}, \texttt{3}, \texttt{;}, \texttt{6}, \texttt{0}, \texttt{$\langle \text{sep}\rangle$}, \texttt{4}, \texttt{0}, \texttt{$\langle \text{sep}\rangle$}, \texttt{1}, \texttt{4}, \texttt{4}, \texttt{;},\texttt{1}, \texttt{2}, \texttt{$\langle \text{sep}\rangle$}, \texttt{1}, \texttt{2}, \texttt{$\langle \text{sep}\rangle$}, \texttt{4}, \texttt{5}, \texttt{$\langle /s\rangle$}\}.  
}
\end{example}

%%%%%%%%%%%%%%%%%%%%%%%%%%%%%%%%%%%%%%%%%%%%%%%%%%%%%%%%%%%%%%%
%{
\subsection{Put the models into practice}
\label{subsec:practice}
In previous subsections, we have presented in detail
how to train a Transformer model to predict the optimal variable ordering
for CAD.
The datasets for training the models are described in Section~\ref{sec: datasets}. 
Let $n$ be the number of variables.
In this work, a Transformer model is trained for a fixed value of $n=3,4$.
If $n\geq 5$, it will be very difficult
to obtain a large dataset, where each example is labelled with the optimal variable ordering, 
to train machine learning models. 
For this reason, the work~\citep{DBLP:conf/icms/ChenZC20}
proposed an iterative approach to search
better variable orderings than the heuristic methods
and use the searched variable orderings as labels to train classification models. 
The performance of the classification models largely depend on the quality of the searched variable orderings, which are usually not optimal.
The work~\citep{Jing2024} employed reinforcement learning
to search better variable orderings than the heuristic methods
for systems up to $16$ variables. 
However, it relies on the strong assumption that the agent 
has been trained on systems with the same support sets as the targeted system. 
In this subsection, we propose a method 
to  apply the trained models for $n$ variables to handle problems in $m$ variables.
 
Case I: $m=n$, we directly apply the trained models.
Case II: $m<n$. The input system in $m$ variables is also a legal system in $n$
variables and can be safely fed to the trained models (after variable renaming).
Then it suffices to force the model to ignore the other $n-m$ variables at the output layer.
% As all variable tokens in the $m$-dimensional data 
% already appear in the $n$-dimensional model’s vocabulary (up to variable renaming),
% if the model generates a variable token not in the lower-dimensional input, it suffices to discard it.

Case III: $m>n$.  There exist $t=\left( {m \atop n} \right) $ ways to choose $n$ different variables from $m$ ones. 
For each choice, we set the rest $m-n$ variables to $1$ in the input system $F$
and get a sub-system $F_i$, $i=1,\ldots,t$, involving only $n$ variables $X_i=x_{i_1}, \ldots, x_{i_n}$. 
Note that we purposely choose $1$ instead of random numbers to evaluate the variables
to prevent the resulting system from getting longer when tokenized. 
For each $F_i$ in variables $X_i$, we call an $n$-dimensional model to get an ordering $O_i$. 
It then suffices to design a scoring scheme to recover a variable ordering for $F$
from  $O_1,\ldots,O_t$.
Given  $O_i=x_{i_1}>\cdots>x_{i_n}$, 
let ${\rm score}(x_{i_j}, O_i)=n-j$, $j=1,\ldots, n$.
Then the final score of $x_k$, $k=1,\ldots, m$, 
would be ${\rm score}(x_k)=\sum_{i=1}^t {\rm score}(x_k, O_i)$.
Sorting $x_i$ by ${\rm score}(x_k)$ in descending order provides 
the final ordering for $F$.

In practice, the $m$ variables might  already satisfy some block variable ordering $X_1>X_2>\cdots>X_r$,
where the variables in each block $X_i$ is free to take any ordering. 
For instance, block variable ordering naturally arises from real quantifier elimination, which is one of the main applications of cylindrical algebraic decomposition.
If $m\leq n$, we firstly apply the strategies in Case I and Case II to choose a variable ordering $O$ and then force the order $O$ to respect the block ordering. 
If $m>n$, instead of directly following the strategy in Case III, we  take advantage of the block structure. More precisely, let $BF_i$ be the sub-system obtained by assigning variables 
in $X\setminus X_i$ to $1$ in $F$. We apply the previous strategies on $(BF_i, X_i)$ to get an ordering $O_i$. Then the final ordering for $F$ is $O_1>O_2>\cdots >O_r$.

% \subsubsection{Handle higher-dimensional systems}

% \subsubsection{Handle systems in block variable ordering}
%}

\section{Datasets}
\label{sec: datasets}
In this section, we present the datasets used in the experiments. 
In Section~\ref{subsec:CAD-pretrain-dataset}, we firstly recall two public CAD variable ordering datasets DQ-3 and DQ-4 introduced in ~\citep{DBLP:conf/casc/ChenJQYZ24} and then present how to generate pre-training datasets  from them. 
In Section~\ref{subsec:dq4b}, we introduce a new four variables dataset for CAD variable ordering selection. 
This new dataset DQ-4b enhances DQ-4 in Section~\ref{subsec:CAD-pretrain-dataset}  by incorporating more difficult examples to improve its quality. 
In Section~\ref{subsec:smt-data}, we recall an existing three variables  CAD variable ordering dataset  built with SMT examples~\citep{DBLP:conf/casc/RioE22}.

\subsection{Two public random CAD variable ordering datasets and the corresponding pre-training datasets}
\label{subsec:CAD-pretrain-dataset}

The objective of this study is to train machine learning models to improve the accuracy of variable order prediction in CAD. To this end, high-quality CAD random datasets are required to evaluate model performance. The work in \citep{DBLP:conf/casc/ChenJQYZ24} provides two benchmark datasets, DQ-3 and DQ-4, corresponding to three-variables and four-variables settings, respectively. These datasets contain polynomial systems along with the associated CAD running time under different variable orders.

\begin{table}[!ht]
    \caption{Two random CAD order datasets from~\texorpdfstring{\citep{DBLP:conf/casc/ChenJQYZ24}}{DBLP:conf/casc/ChenJQYZ24}.}

    \label{tab:dataset-names}
    \centering
    \begin{tabular}{c l}
        \toprule
        \text{Dataset} & \text{Names of Sub-Datasets} \\
        \midrule
        \multirow{2}{*}{\text{DQ-3}} &
        %\begin{varwidth}[t]{0.5\textwidth}
         %   \RaggedRight
            \texttt{REcMdMn2p0tMv3}, \texttt{REcMeEn2p0rCv3}, \texttt{REcMeEn2tMv3},
            \texttt{REdEn4rCv3}, \texttt{REdEn5rCv3}, \texttt{REdEn6rCv3}\\
            & \texttt{REdHn2v3}, \texttt{REdMn2rCv3}, \texttt{REdMn2v3},
            \texttt{REeEn2tMv3}, \texttt{REeEn3p0rHv3}, \texttt{REeMn2tMv3}
        %\end{varwidth}
        \\
        \midrule
        \multirow{2}{*}{\text{DQ-4}} &
        %\begin{varwidth}[t]{0.7\textwidth}
            \RaggedRight
            \texttt{REdEn2rCv4}, \texttt{REdEn2rHv4}, \texttt{REdEn2rMv4}, \texttt{REdEn2v4},
            \texttt{REdEn3rCv4}, \texttt{REdEn3rHv4},\\ & \texttt{REdEn3rMv4}, \texttt{REdEn3v4}, \texttt{REdEn4rHv4}, \texttt{REeEn2rHv4}
        %\end{varwidth} 
        \\
        \bottomrule
    \end{tabular}
\end{table}

DQ-3 consists of 12 sub-datasets and DQ-4 consists of 10 sub-datasets, with their names listed in Table ~\ref{tab:dataset-names}. The rules for generating these sub-datasets are encoded in their names. 
For instance, the dataset \texttt{REdEn4rCv3} consists of randomly generated systems in three variables (v3). 
The system has 4 constraints (n4) and each constraint in the system is a pure polynomial (rC) of total degree 
in the range of $1..2$ (dE). The other settings, such as the number of terms and the value of coefficients
take the default value (RE), specified in  Table 1 of \citep{DBLP:conf/casc/ChenJQYZ24}.
% In this paper, we adopt DQ-3 and DQ-4 as the original CAD datasets for three- and four-variables cases.
These raw datasets are preprocessed following the rules in Section~\ref{subsec:data}. 
%and subsequently preprocess them into Parquet format for storage. 
The final sizes of the four datasets are summarized in Table ~\ref{tab:data-size}. 
%For simplicity, we refer to them as Train, Valid, Test-Valid, and Test, corresponding to the training, validation, test-validation, and test sets, respectively. 
%Specifically to be specified, 
To be specific, the training set is used for model training, the validation set is used to evaluate the model's performance during the training process, the testing-validation set is used for model prediction evaluation after training completion, and the holdout testing set is used for the final evaluation (used only once and never seen before the final test) of the model. 

Following the same protocol for generating the DQ-3 and DQ-4 sub-datasets, we randomly generated 200,000 unlabeled examples (i.e. polynomial systems) for each sub-dataset. These were then split into training, validation, testing-validation, and testing sets in a 7:1:1:1 ratio.
%By concatenating the corresponding partitions across all sub-datasets, we obtain the complete pre-training datasets. 
%The resulting split sizes are 1,680,000:240,000:240,000:240,000 for the three-variables dataset and 1,400,000:200,000:200,000:200,000 for the four-variables dataset. 
The datasets are preprocessed following the steps described in Section~\ref{subsec:data}. 
%For each pre-training task, we construct task-specific input–output pairs and store them in Parquet format for model training. 
Furthermore, the preprocessing procedure is tailored to the specific features of each task.
In $\texttt{task\_m}$, polynomial systems with excessively long product sequences are pruned by retaining only those with a sequence length $\le$ 512. For $\texttt{task\_p}$ and $\texttt{task\_s}$, a small number of degenerate cases with an empty projection factor set ($pf$) are discarded. 
Furthermore, in the four-variables scenario, a 3-second timeout per example is enforced during feature computation for $\texttt{task\_r}$ and $\texttt{task\_s}$ in Maple to ensure feasibility.
Examples exceeding this limit are discarded. 
The final dataset sizes
%after preprocessing
are presented in Table \ref{tab:data-size}.
%}

\begin{table}[!ht]
    \caption{Data size for each task.}
    \label{tab:data-size}
    \centering
    \begin{tabular}{cc *{4}{c}}
        \toprule
        \text{Dataset} & \text{Task} & \text{Train} & \text{Valid} & \text{Test-Valid} & \text{Test} \\
        \midrule
        \multirow{4}{*}{\text{DQ-3}} 
            & \texttt{task\_c} & 8,724 & 1,238 & 1,214 & 1,215 \\
            & \texttt{task\_e}/\texttt{task\_f}/\texttt{task\_r}  & 1,680,000 & 240,000 & 240,000 & 240,000 \\
            %& \textbf{task\_f} & 1,680,000 & 240,000 & 240,000 & 240,000 \\
            & \texttt{task\_m} & 1,155,452 & 165,020 & 164,935 & 165,068 \\
            & \texttt{task\_p}/\texttt{task\_s} & 1,679,997 & 240,000 & 239,999 & 239,999 \\
            % & task\_r & 1,680,000 & 240,000 & 240,000 & 240,000 \\
            % & \textbf{task\_s} & 1,679,997 & 240,000 & 239,999 & 239,999 \\
        \midrule
        \multirow{6}{*}{\text{DQ-4}} 
            & \texttt{task\_c} & 10,574 & 1,488 & 1,180 & 1,180 \\
            & \texttt{task\_e}/\texttt{task\_f}  & 1,400,000 & 200,000 & 200,000 & 200,000 \\
            %& task\_f & 1,400,000 & 200,000 & 200,000 & 200,000 \\
            & \texttt{task\_m} & 1,168,113 & 166,865 & 166,772 & 167,034 \\
            & \texttt{task\_p} & 1,399,999 & 200,000 & 200,000 & 200,000 \\
            & \texttt{task\_r} & 1,399,998 & 200,000 & 200,000 & 200,000 \\
            & \texttt{task\_s} & 1,399,996 & 200,000 & 200,000 & 200,000 \\
        \bottomrule
    \end{tabular}
\end{table}

It is important to note that, due to practical considerations such as prolonged training time and the substantial computational cost associated with hyperparameter tuning, we do not directly employ the full pre-training datasets in the initial training and evaluation stages. Instead, using a fixed random seed, we randomly sample one-tenth of the data from each pre-training task to serve as the learning corpus. This strategy not only substantially reduces training time and experimental overhead but also enables more efficient exploration of model architectures and hyperparameter configurations within a manageable experimental scale, thereby laying a solid foundation for subsequent large-scale training on the complete datasets.

\subsection{An enhanced \texorpdfstring{$4$}{4}-variable random dataset}

\label{subsec:dq4b}
In this section, we introduce an enhanced four-variables dataset, denoted as DQ-4b. This dataset is constructed by substantially augmenting the number of polynomial instances in the original DQ-4 dataset presented in \citep{DBLP:conf/casc/ChenJQYZ24}. As a result, DQ-4 is a strict subset of DQ-4b.
More precisely, we expand the existing sub-dataset \texttt{REeEn2rHv4} in DQ-4 and introduce two additional sub-datasets, \texttt{REeEn3rHv4} and \texttt{REcMeEn3p0rCv4}. 
In addition, we generate pre-training data for these two new sub-datasets following the same procedure used for the other sub-datasets in DQ-4. 
The newly generated pre-training data are then combined with the existing DQ-4 pre-training data to form the full pre-training corpus for DQ-4b.
Due to constraints on training time and computational resources, we randomly sample one-tenth of this full corpus for use in the DQ-4b pre-training tasks. 
Table~\ref{tab:data-size-4b} reports the amount of data used for each task.
Finally, for consistency in task naming, we designate \texttt{task\_b} as the identifier for the CAD variable-ordering selection task on the DQ-4b dataset.

\begin{table}[!ht]
    \caption{Data size for tasks related to DQ-4b.}
    \label{tab:data-size-4b}
    \centering
    \begin{tabular}{c *{4}{c}}
        \toprule
      \text{Task} & \text{Train} & \text{Valid} & \text{Test-Valid} & \text{Test} \\
        \midrule
            \texttt{task\_b} & 12,366 & 1,741 & 1,431 & 1,431 \\
             \texttt{task\_e}/\texttt{task\_f}/\texttt{task\_r} & 168,000 & 24,000 & 24,000 & 24,000 \\
           % & task\_f & 168,000 & 24,000 & 24,000 & 24,000 \\
             \texttt{task\_m} & 124,789 & 17,815 & 17,702 & 17,796 \\
             \texttt{task\_p} & 167,999 & 24,000 & 24,000 & 24,000 \\
           % & task\_r & 168,000 & 24,000 & 24,000 & 24,000 \\
             \texttt{task\_s} & 167,813 & 23,982 & 23,962 & 23,969 \\
        \bottomrule
    \end{tabular}
\end{table}

\begin{figure}[ht!]
    \centering
    \includegraphics[width=0.7\textwidth]{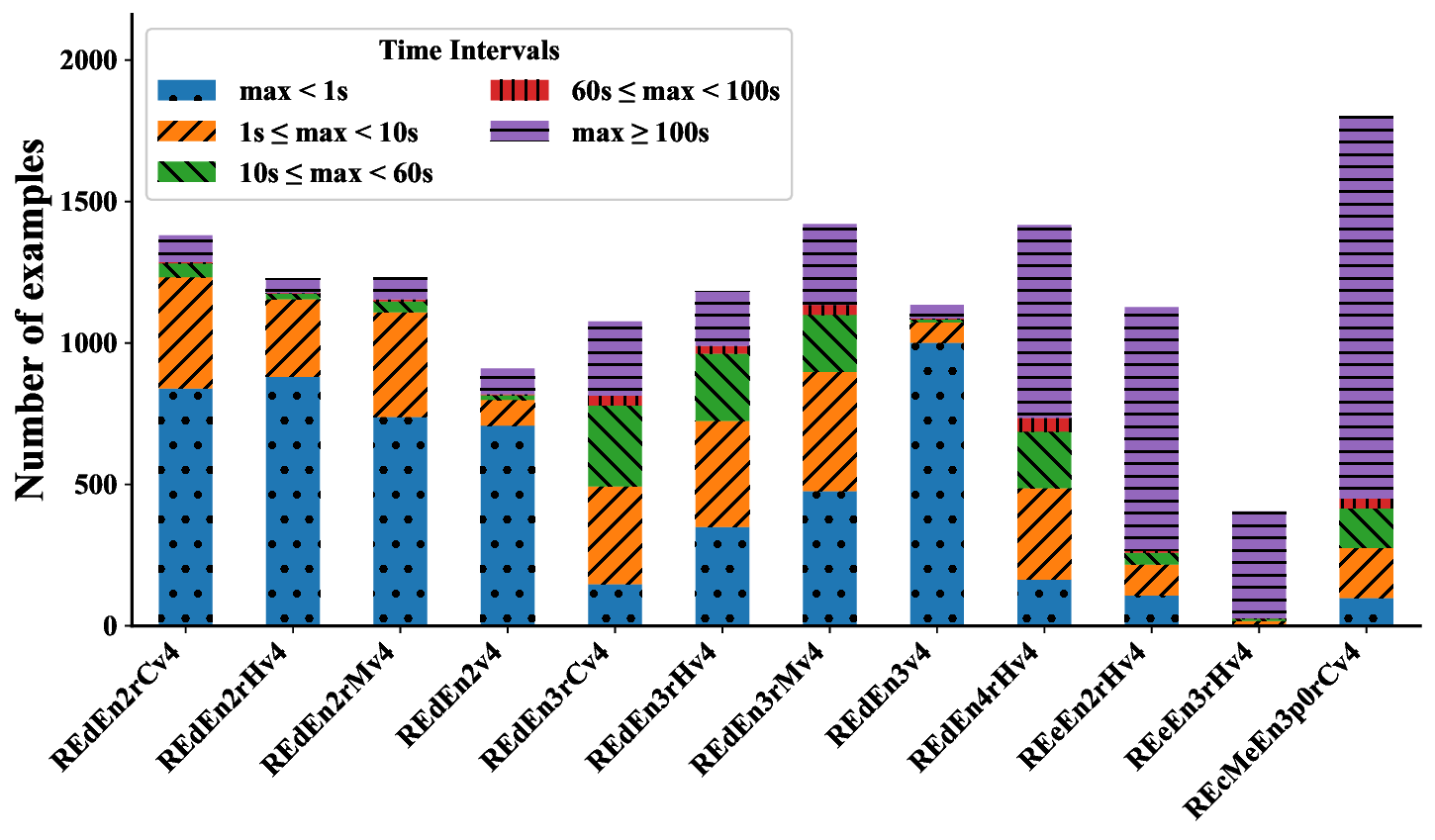}
    %\caption{Distribution statistics of max runtime for RE-Subsets in DQ-4b.}
    \caption{Distribution of the number of examples according to different ranges of 
\texorpdfstring{\texttt{max}}{max} for RE-Subsets in DQ-4b.}

    \label{fig:dq4b-distribution}
\end{figure}

%Furthermore, we performed a statistical analysis of the maximum computation times for each of the 12 RE sub-datasets within the DQ-4b dataset.
Following the methodology described in \citep{DBLP:conf/casc/ChenJQYZ24}, 
for each system we compute the maximum CAD running time, denoted by \texttt{max}, over all 24 possible variable orderings. 
Each sub-dataset is then partitioned into five time intervals according to their corresponding \texttt{max} values, as shown in Figure~\ref{fig:dq4b-distribution}.
From the distribution, we observe that the expanded sub-dataset and the two newly added sub-datasets (corresponding to the three rightmost bars in the figure) contain a substantially higher proportion of difficult instances---those with $\texttt{max} \ge 100\text{s}$---compared to the remaining sub-datasets.

\begin{figure}[ht!]
    \centering
    \includegraphics[width=0.7\textwidth]{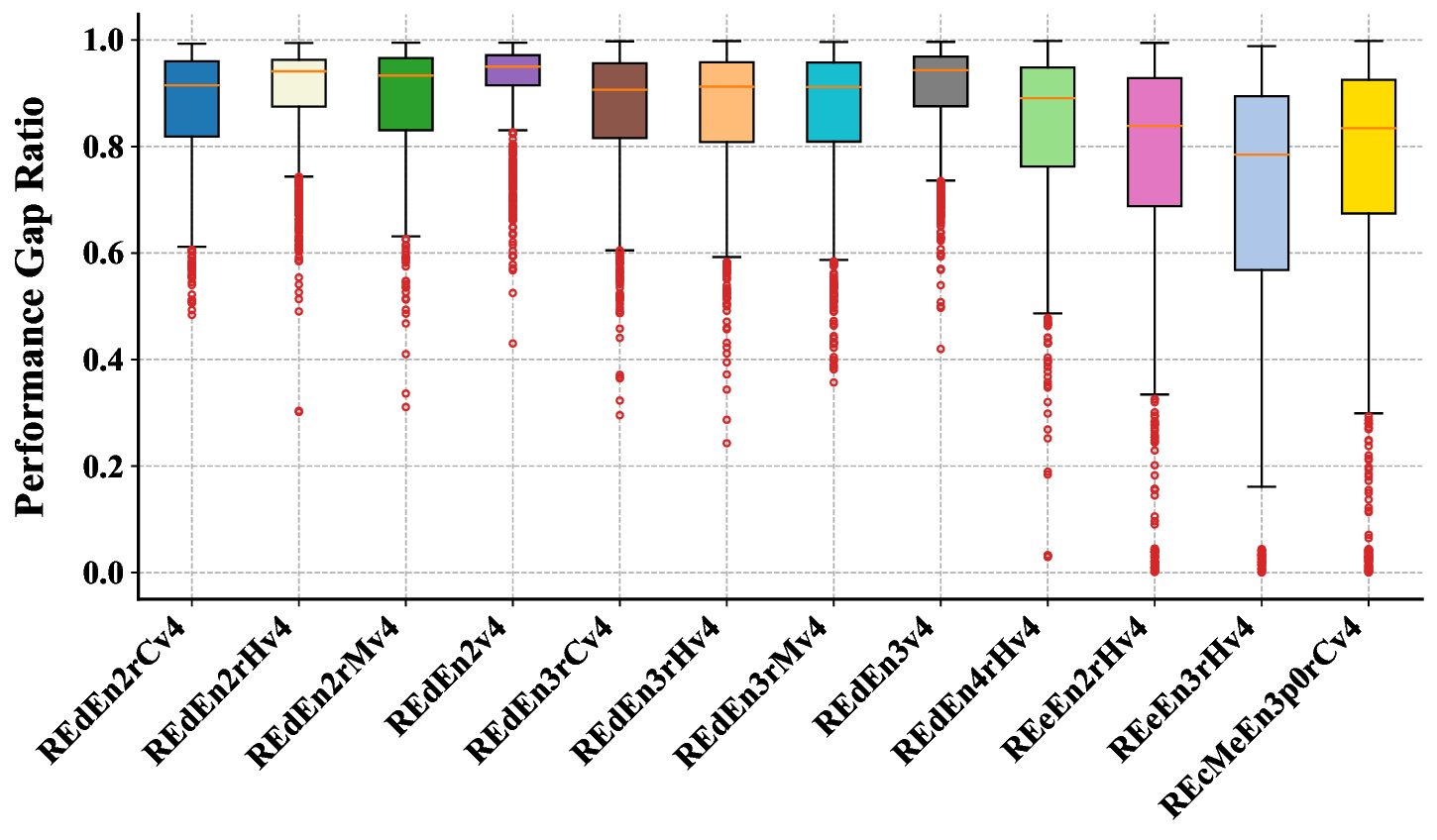}
    \caption{Performance gap ratios for RE-Subsets in DQ-4b.}
    \label{fig:dq4b-histogram}
\end{figure}

For a given system, let $t_{\rm min}$ denote its minimum CAD running time under all variable orderings.
Let $\mathcal{O}_{\mathrm{rel}}$ be  the set of all relative optimal orderings, that is the orderings whose corresponding CAD running times do not exceed $1.03\, t_{\min}$.
Let $\mathcal{O}$ be the set of all variable orderings, and define 
$
\overline{\mathcal{O}} := \mathcal{O} \setminus \mathcal{O}_{\mathrm{rel}} .
$
Let
$
t_{{\max}_{\mathcal{O}_{\mathrm{rel}}}}
$
(resp. $t_{{\min}_{\overline{\mathcal{O}}}}$)
be the maximum (resp. minimum) CAD running time under the orderings in $\mathcal{O}_{\mathrm{rel}}$  (resp. $\overline{\mathcal{O}}$).
We define the \emph{performance gap ratio} as
$
t_{{\max}_{\mathcal{O}_{\mathrm{rel}}}}/t_{{\min}_{\overline{\mathcal{O}}}},
$
which measures the performance advantage of the relative optimal orderings over the rest. Here smaller values indicate better data quality.
The box plot in Figure~\ref{fig:dq4b-histogram} illustrates the distribution of this ratio across all DQ-4b sub-datasets.
The third quartile (and median) values of the three rightmost sub-datasets are noticeably smaller than those of the others, indicating their superior quality.
%}

\subsection{SMT dataset} 
\label{subsec:smt-data}

In this section, we briefly review a  dataset relevant to SMT solving, which is called SMT CAD Order Data in this paper, introduced in~\citep{DBLP:conf/casc/RioE22}.
This dataset consists of 1{,}019 instances, each comprising a set of polynomials in three variables. 
Its purpose in our study is to evaluate whether our model trained on randomly generated data can generalize effectively to practical SMT inputs.
For this purpose, the SMT CAD Order Data is partitioned into training, validation, testing-validation, and testing splits using a 7:1:1:1 ratio. 
The training subset is used to fine-tune both the pre-trained model and the CAD-fine-tuned model. 
%To maintain consistent task naming across our framework, we denote the fine-tuning task on the SMT dataset by $task\_t$.
The experiments taken on this dataset are introduced in Section \ref{subsubsec:perform-on-SMT}.
%}

\begin{figure}[ht!]
  \centering
  \begin{subfigure}{0.46\textwidth}
    \centering
    \includegraphics[width=\textwidth]{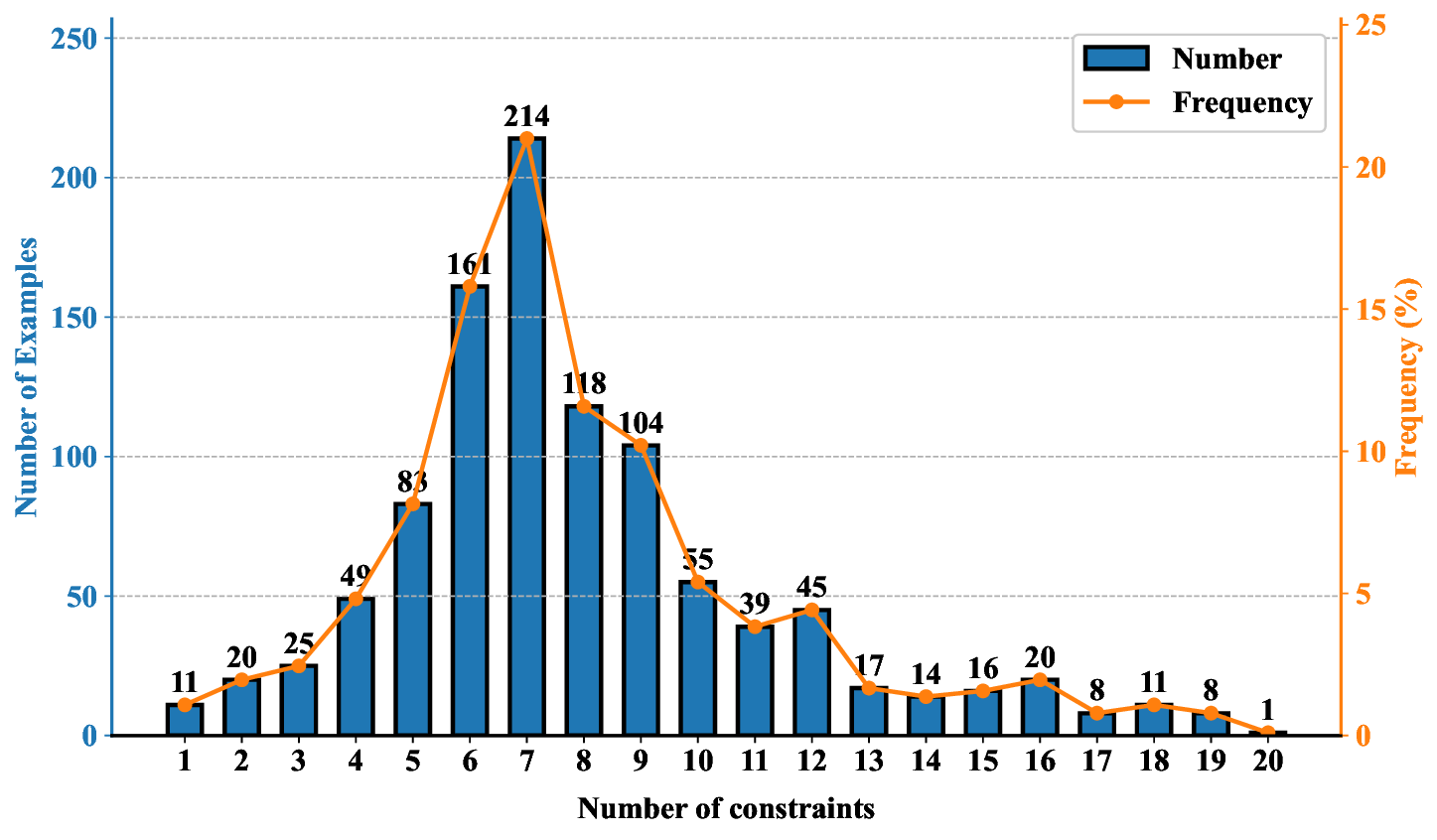}
    % \caption{Constraint number frequency statistics.}
    \caption{Number of constraints}
    \label{fig:smt-constraints}
  \end{subfigure}%\hfill
  \begin{subfigure}{0.46\textwidth}
    \centering
    \includegraphics[width=\textwidth]{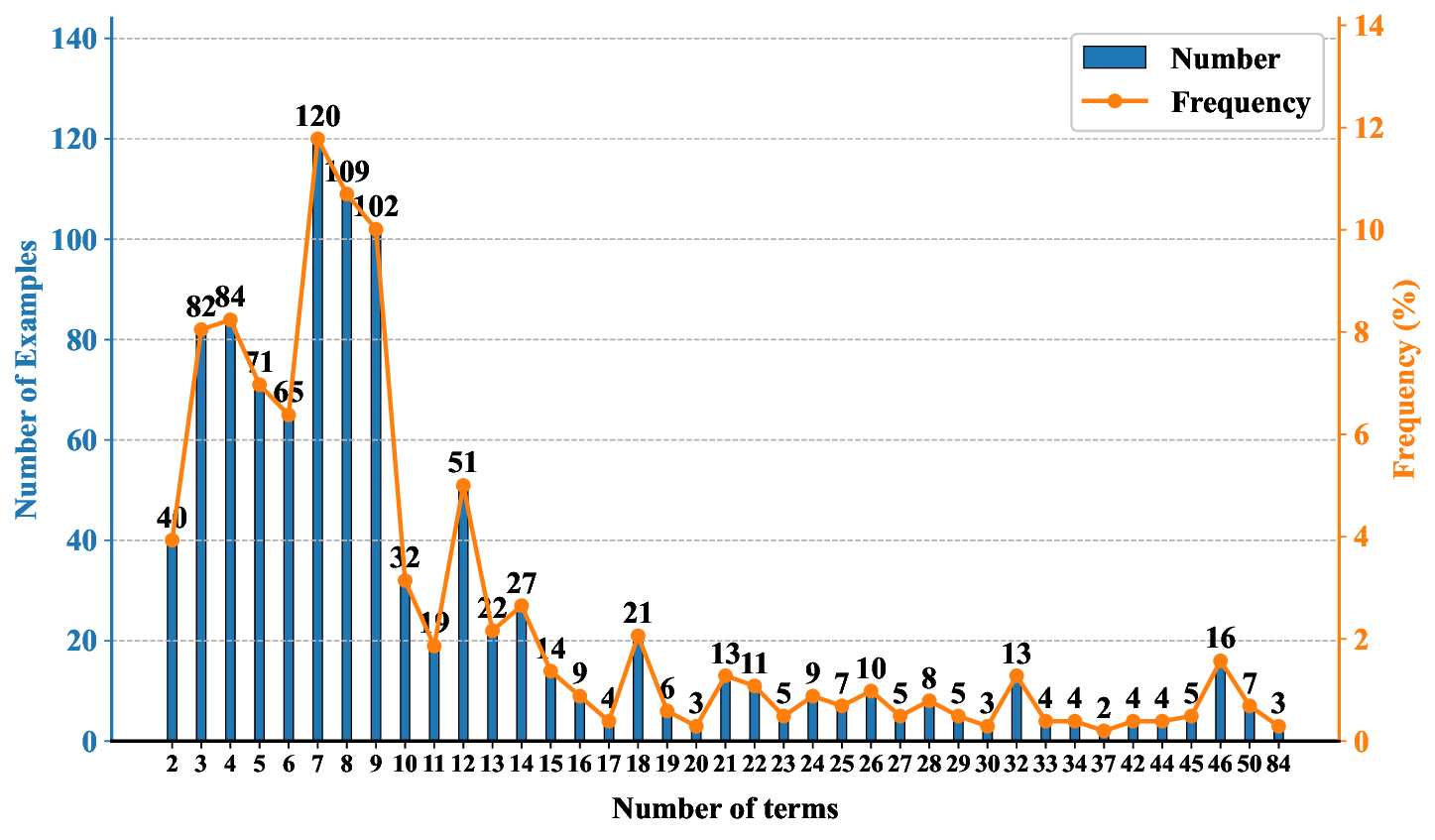}
    %\caption{Term number frequency statistics.}
    \caption{Number of terms}
    \label{fig:smt-terms}
  \end{subfigure}
  
    \begin{subfigure}{0.46\textwidth}
    \centering
    \includegraphics[width=\textwidth]{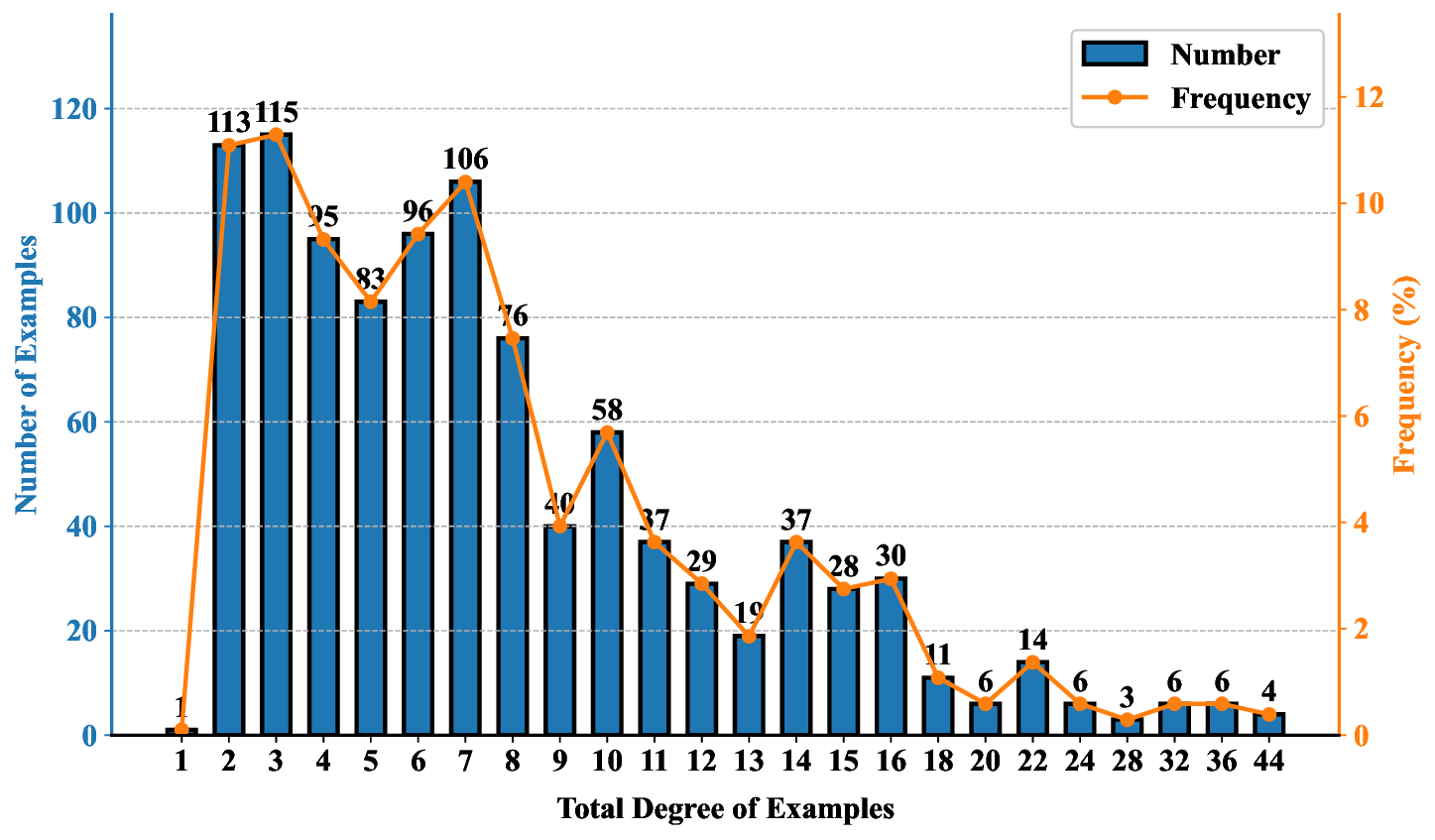}
   % \caption{Exponents number frequency statistics.}
   \caption{Total degree}
    \label{fig:smt-exponents}
  \end{subfigure}
  %\hfill
  \begin{subfigure}{0.46\textwidth}
    \centering
    \includegraphics[width=\textwidth]{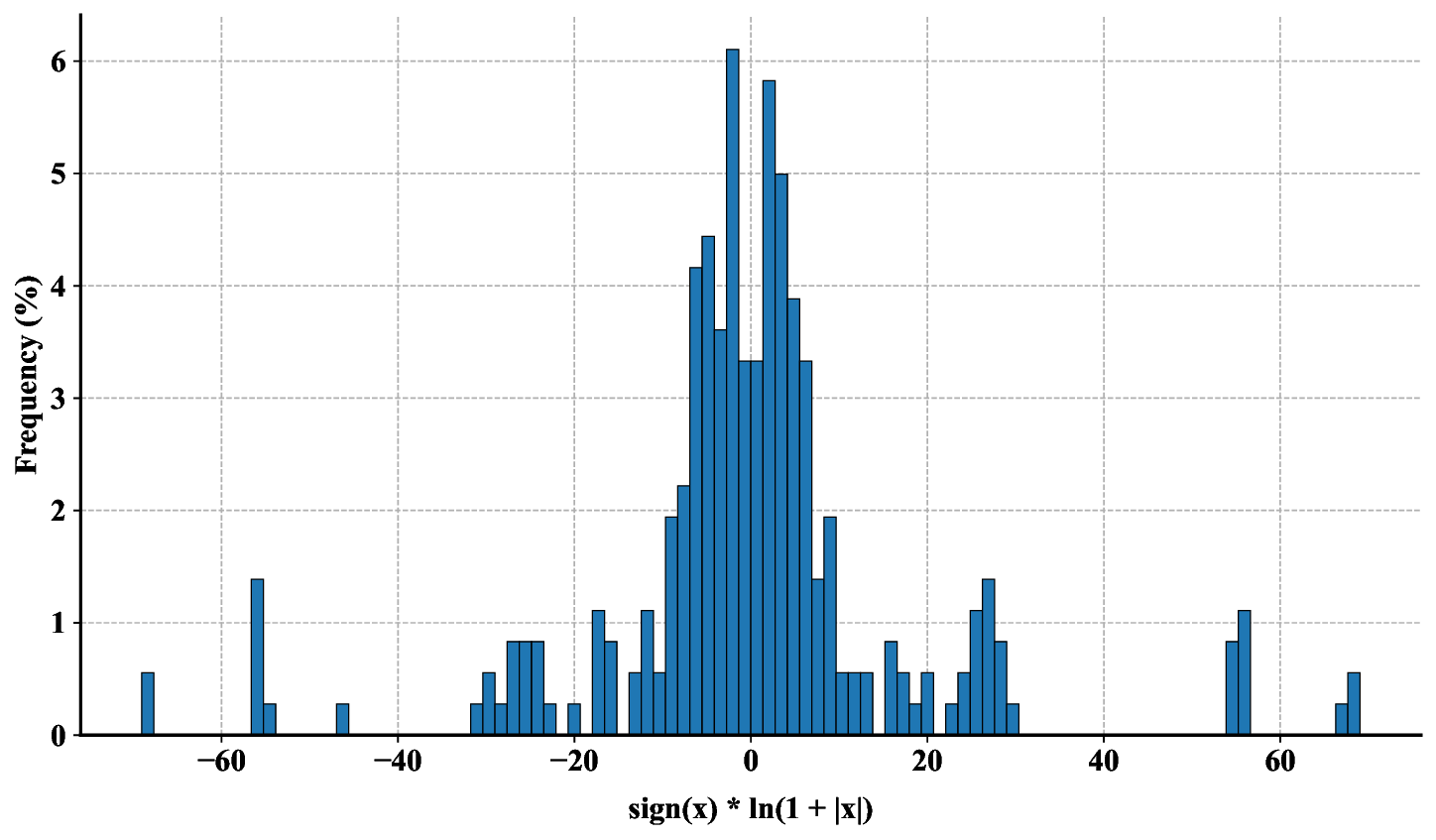}
    %\caption{Coefficient density statistics.}
    \caption{Density of coefficients}
    \label{fig:smt-coefficients}
  \end{subfigure}

    \caption{Statistical information of the SMT CAD order dataset.}
    \label{fig:qualities}
  %\caption{Frequency statistics of constraints and exponents.}
  %\label{fig:smt-constraints-exponents}
\end{figure}

% Figure~\ref{fig:smt-constraints-exponents} and Figure~\ref{fig:smt-terms-coefficients}
Figure~\ref{fig:qualities} presents the distributions of these examples with respect to the number of constraints, the number of terms, the total degrees, and the density of coefficients. 
Figures~\ref{fig:smt-constraints}, \ref{fig:smt-terms}, and \ref{fig:smt-exponents} show that most examples have 5--9 constraints, 3--9 terms, and total degrees ranging from 2 to~8.
For any system, we choose the maximum and the minimum of all its coefficients, and apply the logarithmic transformation, that is
$
    x \rightarrow  \mathrm{sign}(x)\cdot \ln(1 + |x|).
$
Figure~\ref{fig:smt-coefficients} reports the distribution of these values across all systems in the dataset.
% Since the dataset contains $1{,}019$ instances, this yields a total of $1{,}019 \times 2 = 2{,}038$ extracted coefficients. 
% The figure shows, for each possible coefficient value, how frequently it appears as either a maximum or a minimum among all instances.
%%
It shows a clear central peak around~0, with the distribution tapering off symmetrically in both directions.
% { explain (d), For Frequency, change number to percent.}

%%%%%%%%%%%%%%%%%%%%%%%%%%%%%%%%%%%%%%%%%%%%%%%%%%%%%%%%%%%%%%%

\section{Heuristic baselines}
\label{sec: heuristics}
To better evaluate the performance of Transformer models, in this section, we present 11 heuristic methods for CAD variable ordering selection as baselines 
and evaluate their performance on four CAD order datasets. 
Five of them are existing heuristics, including the classical Brown's heuristic method \texttt{svob}~\citep{Brown04}, the command \texttt{SuggestVariableOrder} in the RegularChains library of Maple 
~\citep{DBLP:journals/cca/ChenDLMXXX11} (\texttt{svoc} for short), and three recent ones:  \texttt{gmods} in~\citep{DBLP:conf/casc/RioE22}, 
the pseudo graph feature \texttt{pgf}~\citep{DBLP:conf/icms/ChenZC20},
and \texttt{tone} (T1 in~\citep{DBLP:journals/jsc/PickeringAEC24}). 
The rest are variants of them.
{Note that the chordal approach proposed in~\citep{DBLP:journals/jsc/LiXZZ23}
is not included for comparison, as that method is specially designed for input systems
whose chordal graphs are sparse. The sparsity immediately gets destroyed as long as there is one input polynomial containing all the variables. 
For this reason,~\citep{DBLP:journals/jsc/LiXZZ23} suggests that
the chordal graph approach is ``used in a complementary way with the existing heuristics rather than used competitively". }

\subsection{Introduction of the heuristics}
\label{subsec: introduction-heuristics}

The heuristic methods for suggesting a variable ordering for CAD are based on 
sorting the feature vectors associated with the variables. 
More precisely, let $F\subset\Q[x_1,\ldots,x_n]$. 
For each variable $x_i$, one computes a feature vector ${\bf y}_i$ of a fixed size $s$ and appends $i$ to the end of ${\bf y}_i$ to break the tie. 
One then sorts the vectors ${\bf y}_1, \ldots, {\bf y}_n$ in the lexicographical ascending order into ${\bf y}_{i_1}, \ldots, {\bf y}_{i_n}$. 
Then the corresponding suggested variable ordering is $x_{i_1}>\cdots>x_{i_n}$.

\begin{table}[!ht]
\caption{Heuristic methods for CAD.}
\label{tab:heuristic-description}
\centering
\begin{tabular}{>{\centering\arraybackslash}m{0.08\textwidth}  p{0.35\textwidth}}
\toprule
\text{Heuristic} & \text{Features} \\
\midrule
\texttt{svob}   & $(\text{max\_max}\_{\sf d},\; \text{max\_max}\_{\sf e},\; \text{sum\_sum}\_{\sf a})$ \\
\texttt{svoc}   & $(\text{max\_max}\_{\sf d},\; \text{max}\_{\sf l},\; \text{sum\_max}\_{\sf d})$ \\
\texttt{gmods}  & $(\text{sum\_max}\_{\sf d})$ \\
\texttt{tone}   & $(\text{sum\_max}\_{\sf d},\; \text{avg\_avg}\_{\sf d},\; \text{sum\_sum}\_{\sf d})$ \\
\texttt{slm}    & $(\text{sum\_max}\_{\sf d},\; \text{max}\_{\sf l},\; \text{max\_max}\_{\sf d})$ \\
\texttt{SmSsAa} & $(\text{sum\_max}\_{\sf d},\; \text{sum\_sum}\_{\sf d},\; \text{avg\_avg}\_{\sf d})$ \\
\texttt{SmAaMl} & $(\text{sum\_max}\_{\sf d},\; \text{avg\_avg}\_{\sf d},\; \text{max}\_{\sf l})$ \\
% \texttt{isf}    & $ie_{11}(F)$ \\
\texttt{isf}    & \texttt{feature\_f} \\
% \texttt{psf}    & $ie_{11}\!\left(\cup_{i=1}^n pf(F,x_i)\right)$ \\
\texttt{psf}    & \texttt{feature\_p}\\
\texttt{ipf}    & (\texttt{feature\_f}, \texttt{feature\_p}) \\
\texttt{pgf}    & pseudo graph feature vector of $F$ \\
\bottomrule
\end{tabular}
\end{table}

Different heuristic methods compute different feature vectors. 
Table~\ref{tab:heuristic-description} summarizes the state-of-the-art heuristic methods 
as well as their variants. 
In the table, given $F=\{f_1,\ldots,f_m\}$ and a variable $v$, max\_{\sf l} 
is a number computed in the following way. For each $f_i$, let ${\sf l}(f_i, v)=0$ if $v$ does not appear in $f_i$,
and otherwise let ${\sf l}(f_i, v)$ be the total degree of the leading coefficient of $f_i$ w.r.t. $v$. 
The number max\_{\sf l} is then assigned to be the maximum element of $\{{\sf l}(f_1, v),\ldots,{\sf l}(f_m, v)\}$.
It is interesting to observe that the quantities ${\sf d}$ and ${\sf a}$ are only relevant to a given variable $v$, 
while the quantifiers given by ${\sf e}$, ${\sf l}$ as well as the pseudo graph feature vectors
reveal also the impact of the rest variables on $v$.
The top $8$ heuristics and \texttt{pgf} listed in Table~\ref{tab:heuristic-description} rely only on explicit information from the input polynomials, and thus can be applied on the fly to select a variable ordering. The rest two heuristics involve the computation 
of resultants of input polynomials and thus may consume more time.

\subsection{Evaluation of the heuristics}
\label{subsec: evaluation-heuristics}

Based on the formal definitions of the heuristic methods introduced above, this section evaluates their performance on the CAD order datasets (i.e. SMT, DQ-3,  DQ-4 and DQ-4b) presented in Section~\ref{sec: datasets}. 
% To ensure a comprehensive and rigorous evaluation, we not only examined their performance on the test sets but also analyzed their behavior across the entire dataset. Such a broad evaluation provides a clearer perspective on the practical effectiveness of the heuristic methods, highlighting both their advantages and inherent limitations, thereby offering deeper insights into their overall validity. 
Table ~\ref{tab:heuristic-results} presents the absolute accuracy, the relative accuracy, and the time ratio for the heuristic methods evaluated on the four CAD datasets.
The best value in each column is highlighted in bold. 
Here, for a given example, time ratio measures the ratio of the running time of CAD for this example under the ordering selected by the heuristic over the minimal running time of CAD under all possible variable orderings.
Note that for this evaluation,  the entire data of each dataset is used.

\begin{table}[htbp]
\centering
\setlength{\tabcolsep}{3pt}
\caption{Performance comparison of heuristic methods across CAD datasets.}
\label{tab:heuristic-results}

\begin{tabular}{@{}c c*{12}{c} @{}}
\toprule
\multirow{2.5}{*}{ Heuristics} 
  & \multicolumn{4}{c}{ Abs\_Acc (\%)} 
  & \multicolumn{4}{c}{ Rel\_Acc (\%)} 
  & \multicolumn{4}{c}{ Time\_Ratio} \\
\cmidrule(lr){2-5} \cmidrule(lr){6-9} \cmidrule(lr){10-13}
 & SMT & DQ-3 & DQ-4 & DQ-4b 
 & SMT & DQ-3 & DQ-4 & DQ-4b 
 & SMT & DQ-3 & DQ-4 & DQ-4b \\
\midrule
\texttt{svob}   & 55.45 & 37.86 & 20.41 & 18.06 & 67.32 & 41.07 & 27.69 & 24.38 & 1.25 & 6.86 & 3.85 & 4.50 \\
\texttt{svoc}   & 53.88 & 37.45 & 18.26 & 16.64 & 65.36 & 40.61 & 25.16 & 22.64 & 1.31 & 8.46 & 4.46 & 4.41 \\
\texttt{gmods}  & \bf{56.72} & 44.57 & 15.78 & 15.30 & \bf{68.89} & 48.13 & 22.16 & 20.98 & \bf{1.18} & 5.73 & 3.76 & 3.68 \\
\texttt{tone}   & 54.17 & 46.96 & 18.92 & 17.54 & 66.44 & 50.66 & 26.05 & 23.82 & \bf{1.18} & 5.36 & 3.61 & 3.87 \\
\texttt slm    & 55.64 & 46.20 & 18.35 & 17.45 & 67.91 & 49.81 & 25.27 & 23.60 & \bf{1.18} & 5.65 & 3.84 & 3.69 \\
\texttt{SmSsAa} & 54.96 & 46.90 & 18.81 & 17.48 & 67.12 & 50.51 & 25.93 & 23.76 & 1.19 & \bf{5.28} & 3.62 & 3.84 \\
\texttt{SmAaMl} & 54.27 & 47.75 & 20.31 & 18.66 & 66.63 & 51.40 & 27.89 & 25.30 & \bf{1.18} & 5.37 & \bf{3.57} & 3.86 \\
\texttt{isf}    & 55.25 & \bf{48.53} & 20.71 & 19.20 & 67.81 & 52.32 & 28.15 & 25.75 & 1.19 & \bf{5.28} & 3.60 & 3.79 \\
\texttt{psf}    & 40.63 & 29.94 & \bf{22.15} & \bf{21.04} & 50.74 & 32.55 & \bf{29.78} & \bf{28.04} & 1.55 & 7.65 & 4.23 & \bf{3.41} \\
\texttt{ipf}    & 55.05 & 48.51 & 21.51 & 19.86 & 67.71 & \bf{52.35} & 29.24 & 26.63 & 1.20 & \bf{5.28} & 3.60 & 3.78 \\
\texttt{pgf}    & 53.58 & 47.75 & 19.44 & 18.13 & 66.05 & 51.38 & 26.48 & 24.38 & 1.33 & 5.51 & 3.68 & 3.75 \\
\bottomrule
\end{tabular}
\end{table}

From Table~\ref{tab:heuristic-results}, we have the following observations. 
Among all four datasets,  the SMT dataset is much less challenging than the others  in terms of both accuracy and time ratio.
On the SMT dataset,  \texttt{gmods} outperforms the others in terms of both accuracy and time ratio. 
%The performance differences among heuristics are more evident on the other three random datasets, especially on DQ-3 and DQ-4b. 
On DQ-3, \texttt{isf} and \texttt{ipf} perform the best. 
On DQ-4, \texttt{psf} and \texttt{SmAaMl} outperform the others in terms of accuracy and time ratio respectively. On DQ-4b, \texttt{psf}  outperforms the rest for both accuracy and time ratio. 
Overall, no heuristic methods dominate on all datasets.

%{\color{blue}
Interestingly, only by reordering the features in \texttt{svoc}, the resulting heuristic \texttt{slm} performs much better 
than \texttt{svoc} on all datasets. The top feature in \texttt{slm} is now  the same as in \texttt{gmods} and \texttt{tone}, that is sum\_max\_d. 
This feature reflects the partial degree of the product of input polynomials w.r.t. each variable, 
which plays an important role in measuring the complexity of CAD~\citep{DBLP:conf/casc/RioE22}. 
We remark that \texttt{svoc} was originally not designed as a dedicated variable ordering 
selection method for CAD. Rather it targeted on algorithms solving (mainly) polynomial equations based on triangular decompositions, 
which explains the appearance of the feature max\_max\_d (measuring maximum partial degree of the polynomials) and max\_l (measuring total degree of the leading coefficient). 
The heuristic \texttt{SmAaMl} replaces the last feature sum\_sum\_d in \texttt{tone} by the feature max\_l. 
On all the datasets, it attains higher accuracy than \texttt{tone}.
It also achieves the lowest time ratio on DQ-4.

\begin{figure}[htbp]
  \centering
  \begin{subfigure}{0.48\textwidth}
    \centering
    \includegraphics[width=\textwidth]{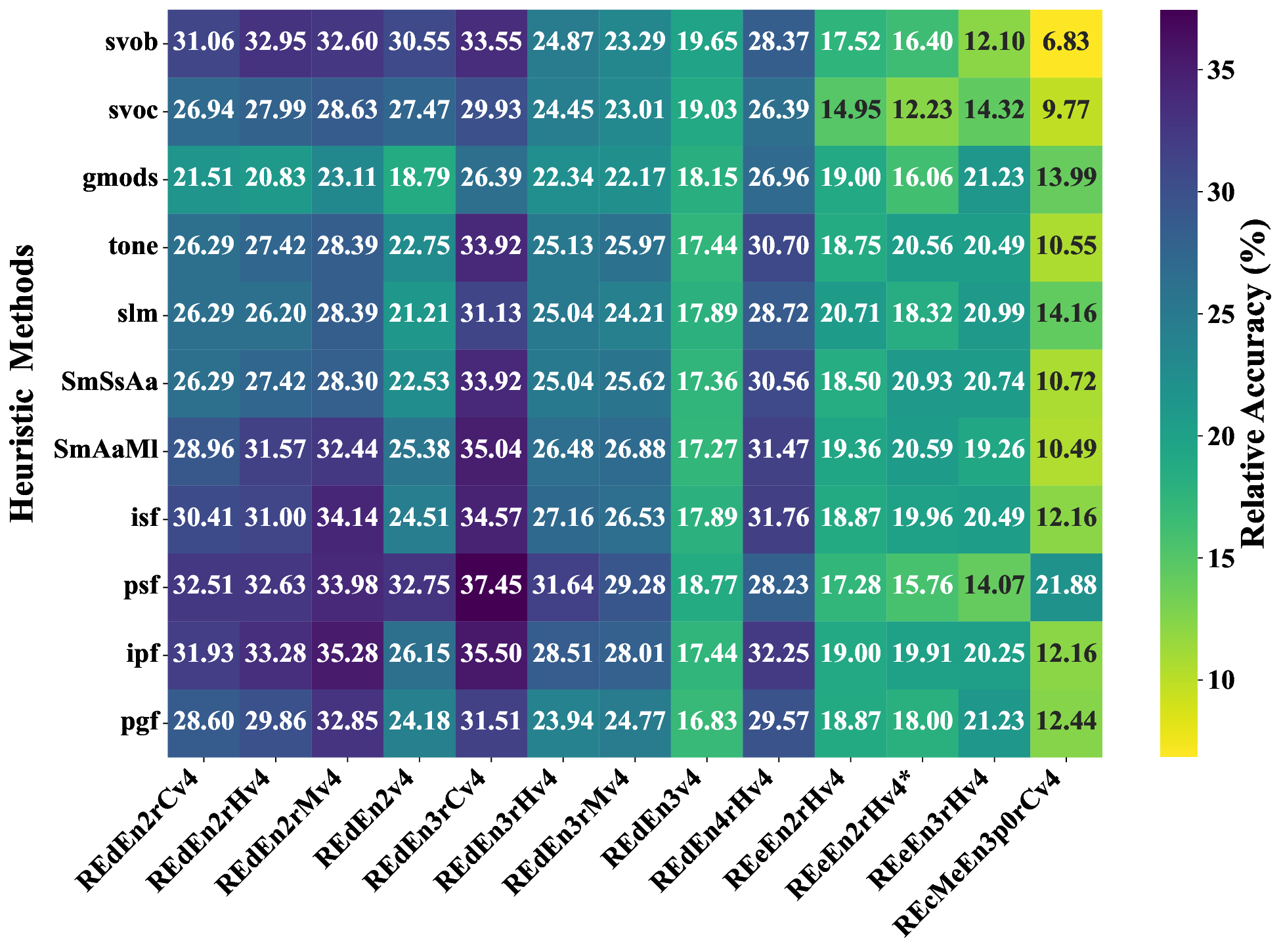}
    \caption{Accuracy}
  \end{subfigure}\hfill
  \begin{subfigure}{0.48\textwidth}
    \centering
    \includegraphics[width=\textwidth]{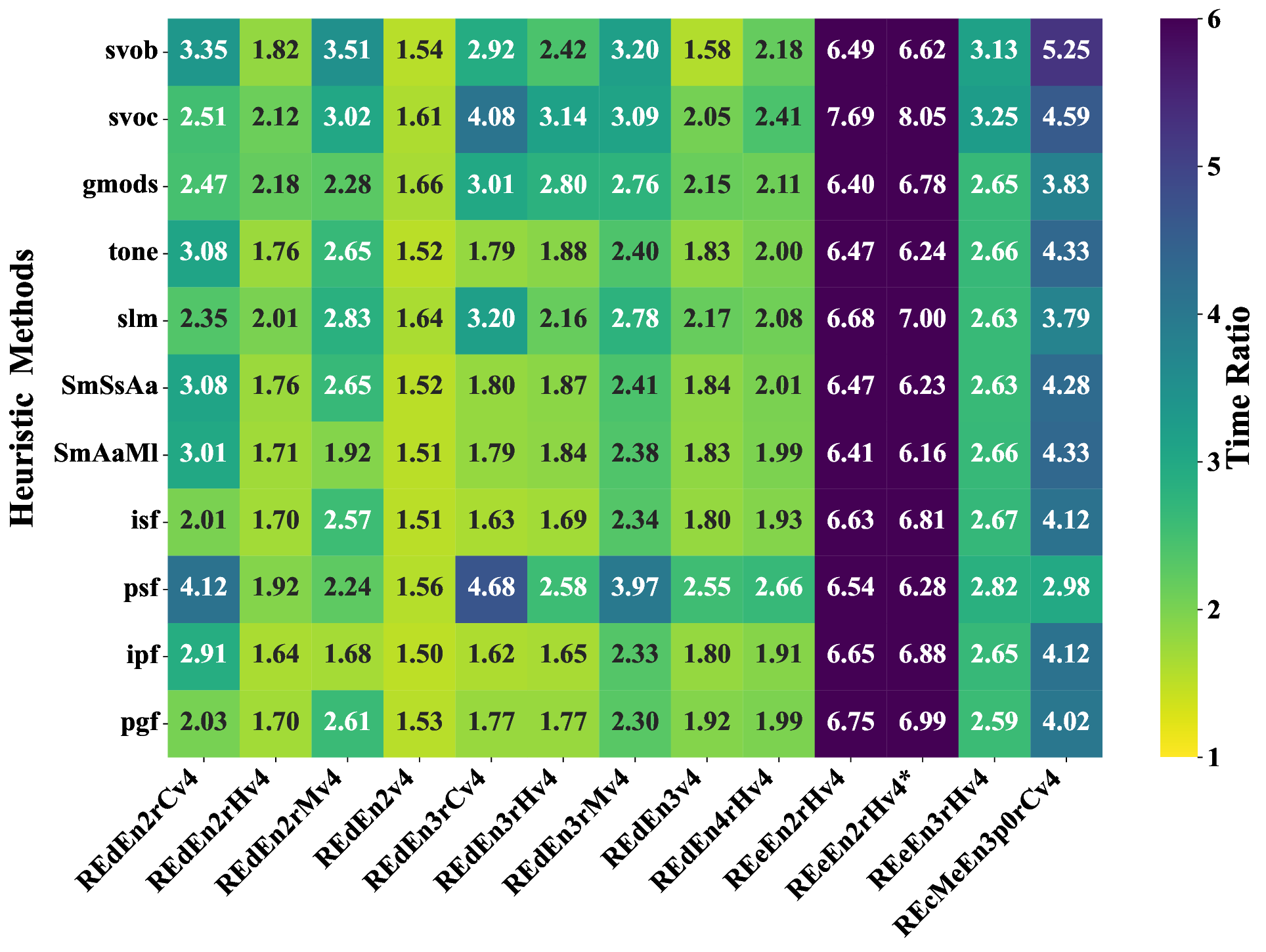}
    \caption{Time Ratio}
  \end{subfigure}
  \caption{Performance of the heuristics on all sub-datasets of DQ-4 and DQ-4b.}
  \label{fig:heuristic-sub}
\end{figure}

From Table ~\ref{tab:heuristic-results}, we also observe that there is certain disagreement 
between the highest accuracy and the lowest time ratio. 
In particular, \texttt{psf} achieves the highest 
accuracy on DQ-4, but its time ratio is worse than most of the other heuristics. 
To further analyze this, we visualize the performance of all heuristics on all sub-datasets of DQ-4 and DQ-4b
in Figure~\ref{fig:heuristic-sub} (the first ten sub-datasets belong to  DQ-4).
From this figure, We notice that on the fifth sub-dataset \texttt{REdEn3rCv4}, \texttt{psf} achieves the highest accuracy 
but also the highest time ratio. On the contrast, although the accuracy of \texttt{ipf} 
is apparently lower than \texttt{psf}, it attains the lowest time ratio. 
To understand such a contradiction between accuracy and time ratio on the sub-dataset  \texttt{REdEn3rCv4}, 
we further split  \texttt{REdEn3rCv4} into two subsets: \textit{normal\_set} and \textit{abnormal\_set}. The \textit{normal\_set} consists of examples where no
timeout occurs for any variable ordering and the \textit{abnormal\_set} consists of the rest of examples. 
Table~\ref{tab:psf-ipf} compares the total CAD running time under the variable orderings predicted by \texttt{psf} and \texttt{ipf} 
on these two sub-datasets of different difficulty. 
From the table, we see that when \texttt{psf} makes a wrong prediction, it is penalized much more than \texttt{ipf} on the more difficult dataset \textit{abnormal\_set}. 
%}

\begin{table}[!ht]
    \caption{Further experiments on the dataset \texorpdfstring{\texttt{REdEn3rCv4}}{REdEn3rCv4}.}

    \label{tab:psf-ipf}
    \centering
    \begin{tabular}{ccccc}
        \toprule
        \text{Category (number)} & \text{Heuristic} & {\text{n\_time}} & {\text{ab\_time}} & {\text{timeout}} \\
        \midrule
        \multirow{2}{*}{Both correctly predicted (236)}
            & \texttt{psf} & 42.08 & 19.55 & 0 \\
            & \texttt{ipf} & 42.05 & 19.54 & 0 \\
        \midrule
        \multirow{2}{*}{Both wrongly predicted (527)}
            & \texttt{psf} & 432.44 & 1429.94 & 1 \\
            & \texttt{ipf} & 300.70 & 215.98 & 0 \\
        \midrule
        \multirow{2}{*}{\texttt{psf} correct but \texttt{ipf} wrong (167)}
            & \texttt{psf} & 36.82 & 53.30 & 0 \\
            & \texttt{ipf} & 91.76 & 76.44 & 0 \\
        \midrule
        \multirow{2}{*}{\texttt{psf} wrong but \texttt{ipf} correct (146)}
            & \texttt{psf} & 97.61 & 300.90 & 0 \\
            & \texttt{ipf} & 46.86 & 42.07 & 0 \\
        \bottomrule
    \end{tabular}
\end{table}

\section{Experiments}

In this section, we present the experimental results. 
Experiments were performed on a workstation with an Intel Xeon Platinum 8272CL CPU @ 2.60\,GHz, 256 GB RAM, and an NVIDIA RTX 4090 GPU, running Ubuntu 22.04.
The software environment includes Python 3.10 and Maple 2024. 
The content is organized as follows. Firstly, we describe the architectures and hyperparameters selected after extensive tuning to achieve optimal performance on the validation sets of $\texttt{task\_c}$.
We have also  evaluated the performance of these models on the testing-validation dataset to 
have a feeling on the generalization ability of these models. 
Secondly, we report the performance of these models on hold-out testing datasets. We emphasize that the testing datasets were never seen by the tested models before the final test.
By default, only the results on the testing datasets are shown.
However, when there is an obvious performing difference between testing-validation and testing datasets, 
we present both experimental results to avoid the bias due to data splitting.
%{ Thirdly, we summarize the different approaches explored in pursuit of such performance. }
Thirdly, we present several (including unsuccessful) attempts made aiming 
to push the limit of pre-training \& fine-tuning for CAD variable ordering selection.

\subsection{Detailed information on the architecture and parameters}
\label{subsec:params-setting}

In this study, all models are built upon a Transformer-based architecture. The architectural specifications and training configurations—including batch size, optimizer settings (Adam with a fixed learning rate of 0.0001), dropout rate, and early stopping criteria—are summarized in Table~\ref{tab:model-parameters}. The experimental workflow begins with a pre-training stage, where six models are trained, each corresponding to one of the six pre-training tasks. For fine-tuning, the pre-trained weights are used for initialization. 
The embedding layer and the first three encoder layers are frozen, while the remaining parameters are updated on the CAD order datasets. This procedure produces task-specific fine-tuned models optimized for CAD variable ordering prediction.

In addition, following the data processing strategy outlined in Section~\ref{subsec:CAD-pretrain-dataset}, we utilized the \textit{HuggingFace Datasets library} ~\citep{lhoest-etal-2021-datasets} for data loading and storage, adopting the Apache Parquet format~\citep{Vohra2016}. This columnar storage format supports efficient compression and parallel processing, significantly improving data-loading performance. Compared with plain text storage, it greatly accelerates data access and facilitates the efficient reuse of training data across multiple tasks, thereby enhancing the overall efficiency of the experimental workflow.

\begin{table}[!ht]
    \caption{Summary of model architecture and training parameters.}
    \label{tab:model-parameters}
    \centering
    \footnotesize

    %\begin{tabularx}{0.75\textwidth}{@{} >{\bfseries}l l >{\RaggedRight\arraybackslash}X @{}}
    \begin{tabular}{lll}
        \toprule
        Category & Parameter & Value/Description \\
        \midrule
        Model Architecture & Embedding Dimension & 256 \\
        & Attention Heads & 8 \\
        & Encoder Layers & 7 \\
        & Decoder Layers & 6 \\  
        \midrule        %\addlinespace[2pt]\hline\addlinespace[2pt]
        Training Configuration & Batch Size & 128 (64 for $\texttt{task\_m}$ on 4-variable data) \\
        & Epoch Definition & Each epoch spans the entire training set \\ 
        \midrule        %\addlinespace[2pt]\hline\addlinespace[2pt]
        Optimization & Optimizers & 2 (independent: one for encoder, one for decoder) \\
        & Optimizer Algorithm & Adam \\
        & Learning Rate & 0.0001 (fixed) \\ 
        \midrule        %\addlinespace[2pt]\hline\addlinespace[2pt]
        Regularization \& Early Stopping & Dropout Rate & 0.1 \\
        & Early Stopping Patience & 30 consecutive epochs (no improvement in validation accuracy) \\
        & Model Selection & Checkpoint with highest validation accuracy \\
        %\addlinespace[2pt]\hline\addlinespace[2pt]
        Pre-training Strategy & Number of Pre-training Models & 6 (one for each pre-training task) \\  
        \midrule       % \addlinespace\hline\addlinespace[2pt]
        Fine-tuning Strategy & Initialization & Pre-trained weights \\ 
        & Frozen Layers & Input embedding layer + lowest 3 encoder layers \\
        & Adapted Parameters & Remaining parameters \\ 
        \midrule
        %\addlinespace[2pt]\hline
        Data Handling & Dataset Storage Format & Parquet (for direct access and faster loading) \\
        & Data Reuse & Cached data reused for tasks sharing identical training data \\
        \bottomrule
        \end{tabular}
   % \end{tabularx}
\end{table}

% { The following remark need to be polished. According to this remark, is it possible 
% to add some previous tries in ablation analysis.}

\begin{remark}
{\rm
Table~\ref{tab:model-parameters} presents the hyperparameter configuration employed for our main experiments. However, this specific parameter set was not arbitrarily chosen; rather, it is the culmination of extensive exploration and careful consideration. Indeed, prior to this, we conducted a substantial number of experimental explorations through systematic hyperparameter search and evaluation.
Recognizing that a full training run for each hyperparameter combination would be computationally prohibitive and time-consuming, we adopted an efficient trial-training strategy: each hyperparameter trial was limited to the first 10 epochs, during which the validation set accuracy's evolution and performance were continuously monitored. This approach enabled us to effectively filter out underperforming parameter combinations within a significantly shorter timeframe. Following extensive trial-training and comparative analysis, 
evaluated mainly on the validation set of DQ-3,
we ultimately determined the optimal hyperparameter set  presented in Table~\ref{tab:model-parameters}.
}
\end{remark}

\subsection{Performance of the Transformer models}
\label{subsec:main}
This section reports the performance of Transformer models 
on both the pre-training tasks and CAD variable ordering prediction tasks.

\subsubsection{Performance on pre-training tasks}
\label{subsubsec:performance-on-pretask}
The performance of models for six pre-training tasks on three random datasets are reported in Table ~\ref{tab:pretrain-valid}. 
Here, ``valid", ``test-valid", ``test" refer to the validation set, the testing-validation, and the testing set, respectively.
Note that, the performance on the validation set is obtained in the teacher-forcing mode
while the performance on the testing-validation and testing sets are obtained in the autoregressive mode. 
The data on the same task show that inference in autoregressive mode does not lead to performance degradation. 
In addition, the similarity between the performance on validation, test-validation and testing sets shows that the models 
have good in-distribution generalization ability.

\begin{table}[!ht]
    \caption{Accuracies of pre-training tasks.}
    \label{tab:pretrain-valid}
    \centering
    % \renewcommand{\arraystretch}{1.1}
   % \begin{tabular}{@{} c c *{6}{c} @{}} 
    \begin{tabular}{cc|ccc|ccc} 
        \toprule
        \text{Dataset} & \text{Category} & {\texttt{task\_e}} & {\texttt{task\_f}} & {\texttt{task\_m}} & {\texttt{task\_p}} & {\texttt{task\_r}} & {\texttt{task\_s}} \\
        \midrule
        \multirow{3}{*}{\text{DQ-3}}
            & valid & 100.00\% & 98.00\% & 93.49\% & 25.29\% & 20.28\% & 9.30\% \\
            & test-valid & 99.99\%  & 98.08\% & 93.65\% & 24.75\% & 19.93\% & 9.22\% \\
            &test & 100.00\% & 98.11\% & 93.78\% & 24.87\% & 19.83\% & 9.02\% \\
        \midrule
        \multirow{3}{*}{\text{DQ-4}}
            & valid & 99.99\% & 99.83\% & 94.03\% & 65.88\% & 49.80\% & 41.01\% \\
            & test-valid & 99.98\%  & 99.85\% & 93.91\% & 65.49\% & 49.44\% & 41.24\% \\
             &test & 100.00\% & 99.79\% & 94.25\% & 65.14\% & 49.36\% & 40.76\% \\
        \midrule
        \multirow{3}{*}{\text{DQ-4b}}
            & valid & 99.98\%  & 99.90\% & 93.08\% & 56.62\% & 41.57\% & 35.19\% \\
            & test-valid & 99.98\%  & 99.90\% & 92.75\% & 56.02\% & 41.31\% & 35.46\% \\
             &test & 100.00\% & 99.99\% & 93.32\% & 56.09\% & 41.03\% & 35.10\% \\
        \bottomrule
    \end{tabular}
\end{table}

Now, we focus on the performance of these models on the test data. 
For the two simple tasks (\texttt{task\_e} and \texttt{task\_f}) only involving extracting the features of input systems, 
the pre-training models achieve nearly perfect accuracies. 
For the slightly more complex polynomial multiplication task (only predicting supports of the product), 
the accuracies are all higher than $90\%$. 
The accuracies drop dramatically for those much more complex pre-training tasks, 
which involve the computation of resultants. 
For these more complex tasks, 
it is interesting to observe that the pre-training models perform better on four-variables datasets than the three-variables dataset.
To understand the performance difference for different tasks on different datasets, 
we introduce the {\bf mean feature-digit length} to measure the difficulty of the output. 
More precisely, recall that the output of every example for each pre-training task 
consists of a set $m$ of features and each feature is a $n$-tuple of integers, where $n$
is the number of variables. 
The mean feature-digit length is defined as the total number of digits across all feature values divided by $mn$.
For every pre-training task, we then average this quantity across all instances in the corresponding test set, yielding the mean feature-digit length  summarized in Table ~\ref{tab:feature-length}.
By comparing the values in this table with the accuracies of three tasks reported in Table~\ref{tab:pretrain-valid},
we see that the mean feature-digit length does characterize the learning difficulty of 
the three tasks across the testing set.
%}

\begin{table}[!ht]
\caption{Comparison of the mean digit length for \texttt{task\_p}, \texttt{task\_r} and \texttt{task\_s} on testing set.}
\label{tab:feature-length}
\centering
\begin{tabular}{cccc}
\toprule
\text{Dataset} & \texttt{task\_p} & \texttt{task\_r} & \texttt{task\_s} \\
\midrule
\text{DQ-3} & 1.83 & 2.12 & 2.75 \\
\text{DQ-4} & 1.34 & 1.60 & 2.20 \\
\text{DQ-4b} & 1.48 & 1.79 & 2.46 \\
\bottomrule

\end{tabular}
\end{table}

\subsubsection{Performance for CAD variable ordering selection on three random datasets}
\label{subsubsec:performance-on-DQ}

\begin{figure}[htbp]
  \centering
  \begin{subfigure}{0.48\textwidth}
    \centering
    \includegraphics[width=\textwidth]{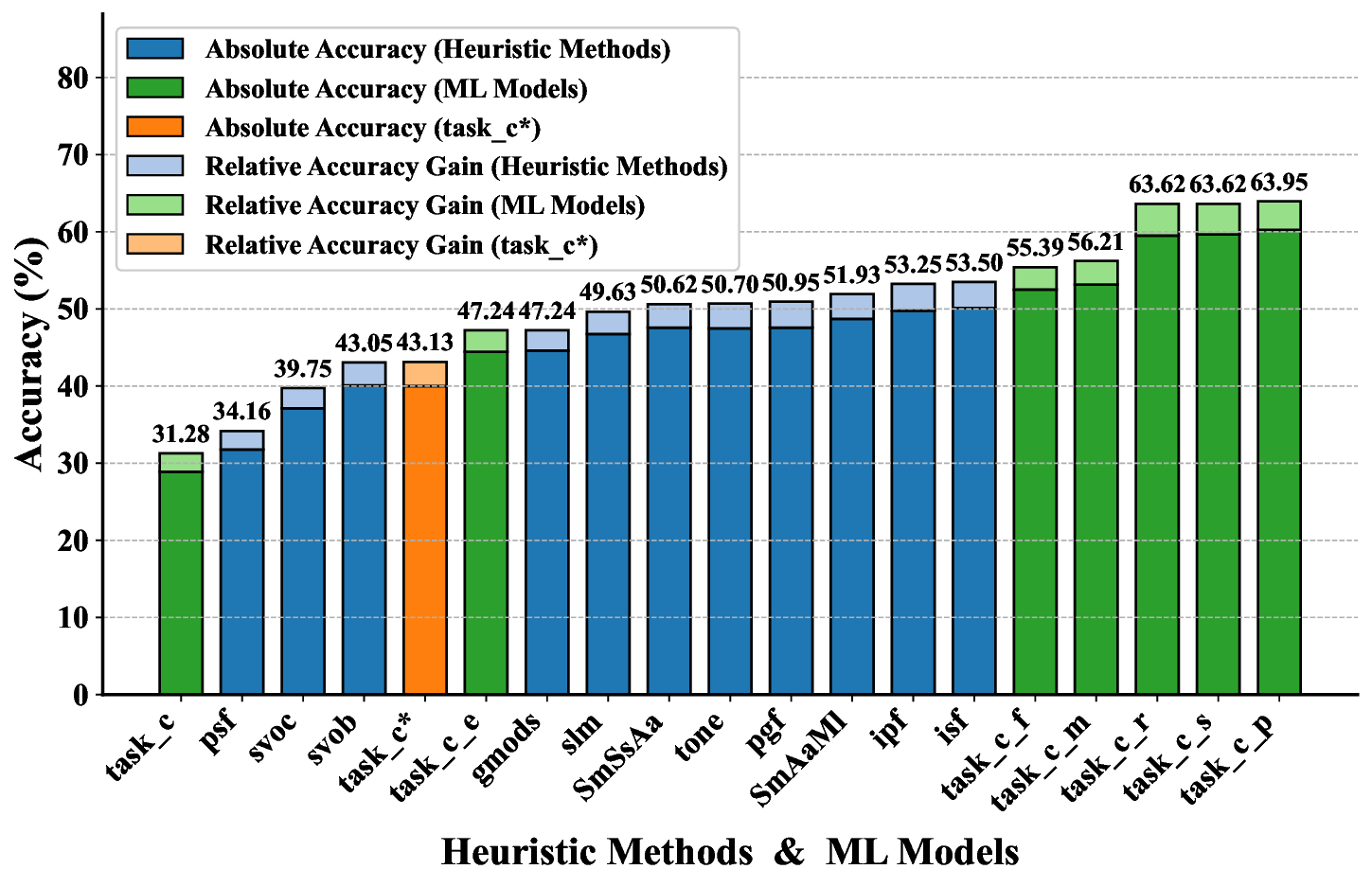}
    \caption{Accuracy results.}
    \label{fig:abs_rel_acc_v3}
  \end{subfigure}\hfill
  \begin{subfigure}{0.48\textwidth}
    \centering
    \includegraphics[width=\textwidth]{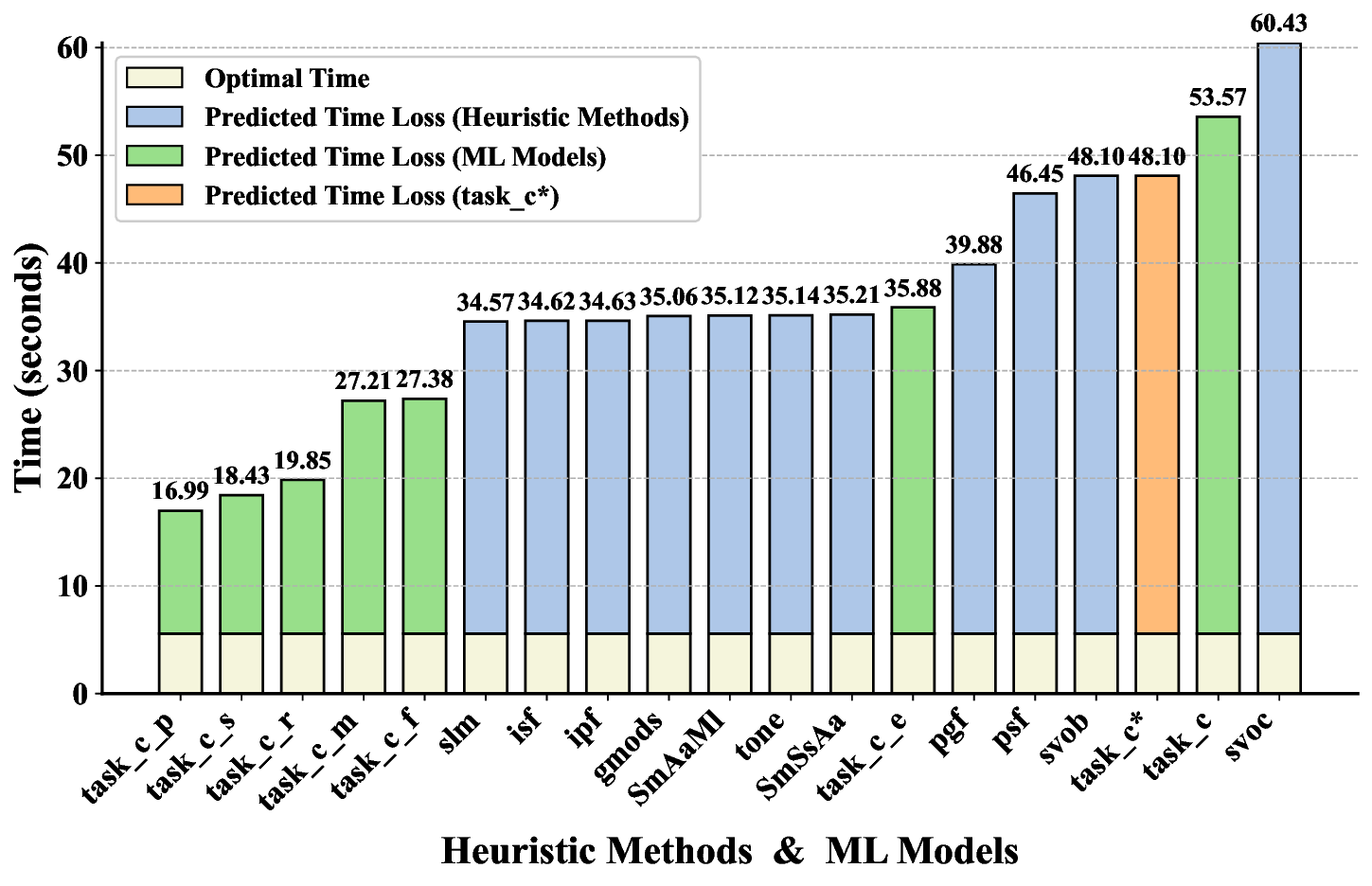}
    \caption{Execution time results.}
    \label{fig:time_v3}
  \end{subfigure}
  \caption{Comparison of heuristic methods and ML models on the DQ-3 testing set.}
  \label{fig:dq3_comparison}
\end{figure}

Figure~\ref{fig:dq3_comparison} compares the performance of ML models and heuristic methods for CAD variable ordering selection 
in terms of both (relative) accuracy and CAD running time on the three variable random dataset DQ-3. 
In the figure, the ML models start with the word ``task". 
We have the following observations: $(i)$ The pre-training \& fine-tuning models involve ``CAD projection" (\texttt{task\_c\_p}, \texttt{task\_c\_s}, \texttt{task\_c\_r}), thus learning deep features,  achieve the highest accuracy 
and the lowest CAD running time. Running CAD under the variable ordering selected by the best pre-training model achieves more than ${\mathbf 2\times}$ {\bf speedup} over running CAD under the one selected by the best heuristic method;
$(ii)$ The pre-training \& fine-tuning models involve ``light computation" (\texttt{task\_c\_m}, \texttt{task\_c\_f}), learning shallow features of the input system,  also outperform the best heuristic, 
although less promising than those involving ``CAD projection"; $(iii)$ Direct Transformer models (\texttt{task\_c}, \texttt{task\_c}$^\ast$) underperform most of the heuristics, 
which further justify the importance of pre-training and fine-tuning; $(iv)$ The naive pre-training model \texttt{task\_c\_e}, predicting only the exponent vectors of input system, 
still performs better than direct Transformer models, although underperforms most of the heuristics.

In Figure~\ref{fig:dq3_comparison}, 
% the only difference between task\_c$^\ast$  and task\_c and is that the former uses the tokenization scheme introduced in \citep{DBLP:conf/casc/ChenJQYZ24} (specifically, the non-complete scheme for polynomial terms described in Section~\ref{subsec:encode}). 
the only difference between \texttt{task\_c}$^\ast$  and \texttt{task\_c}  is that they use different tokenization schemes. The latter used the default scheme introduced 
in Section~\ref{subsec:encode} while the former used 
the alternative scheme detailed in Section~\ref{subsubsec: inpact_encode}.
As shown in Figure~\ref{fig:dq3_comparison}, for direct Transformer models, the former scheme performs better. 
However, the tokenization scheme has different impacts on different tasks. A comprehensive evaluation is provided in Section~\ref{subsubsec: inpact_encode}.

\begin{figure}[htbp]
  \centering
  \label{fig:abs_rel_acc_time_v4}
  \begin{subfigure}{0.46\textwidth}
    \centering
    \includegraphics[width=\textwidth]{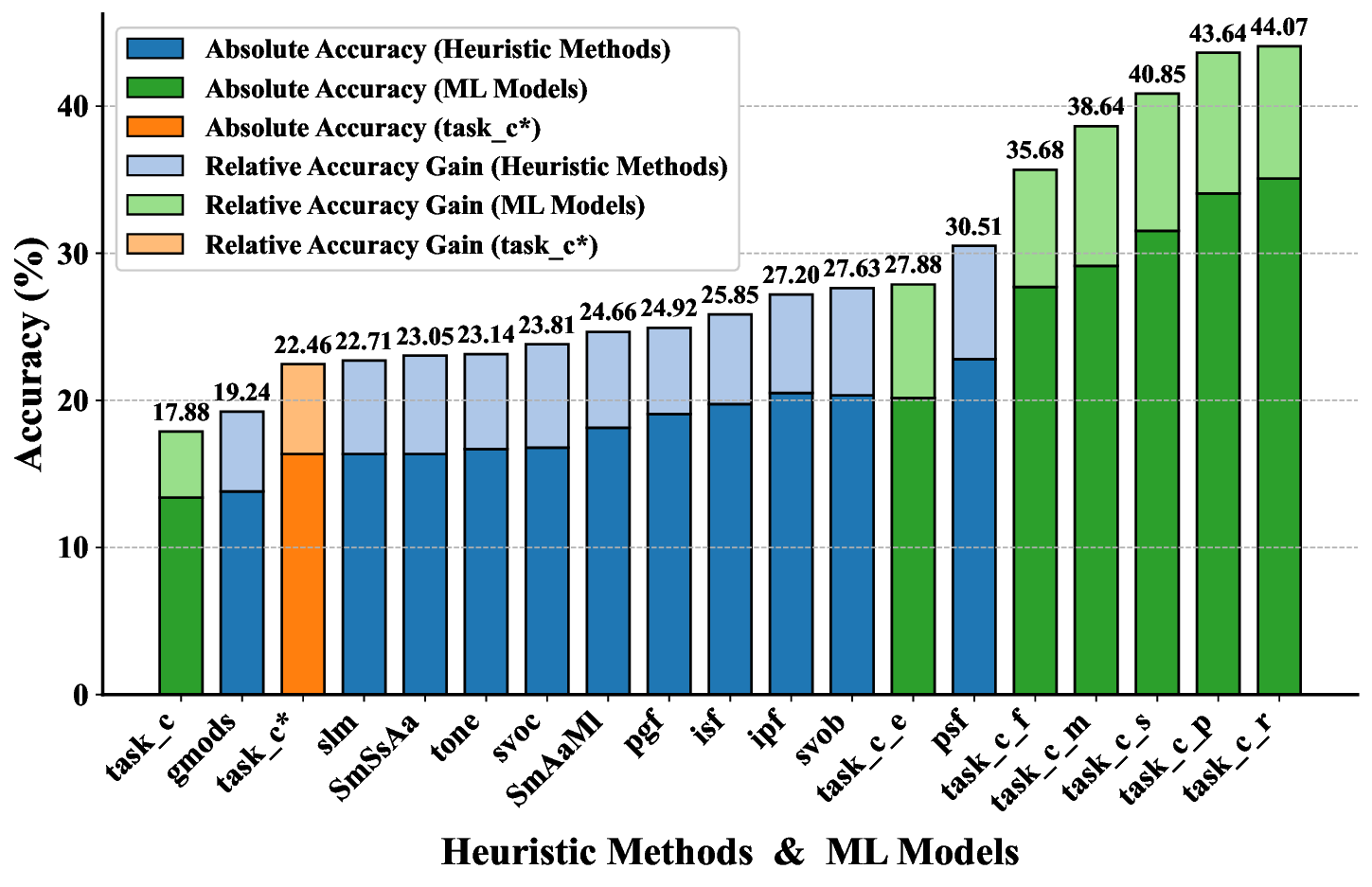}
    \caption{Accuracy results on the testing-set.}
  \end{subfigure}\hfill
  \begin{subfigure}{0.46\textwidth}
    \centering
    \includegraphics[width=\textwidth]{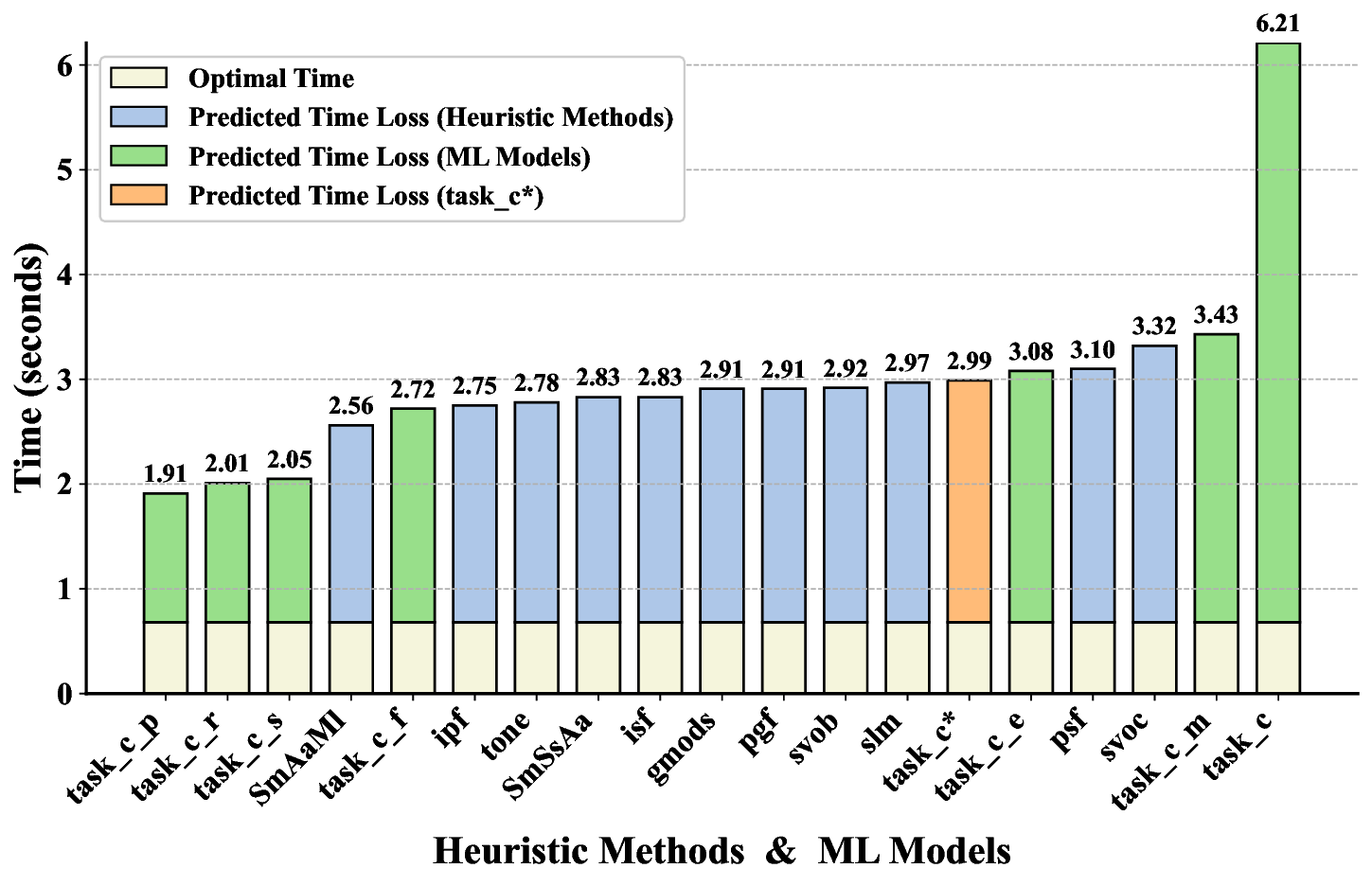}
    \caption{Execution time results on the testing-set.}
    % \label{fig:time_v4}
  \end{subfigure}
  
   \begin{subfigure}{0.46\textwidth}
    \centering
    \includegraphics[width=\textwidth]{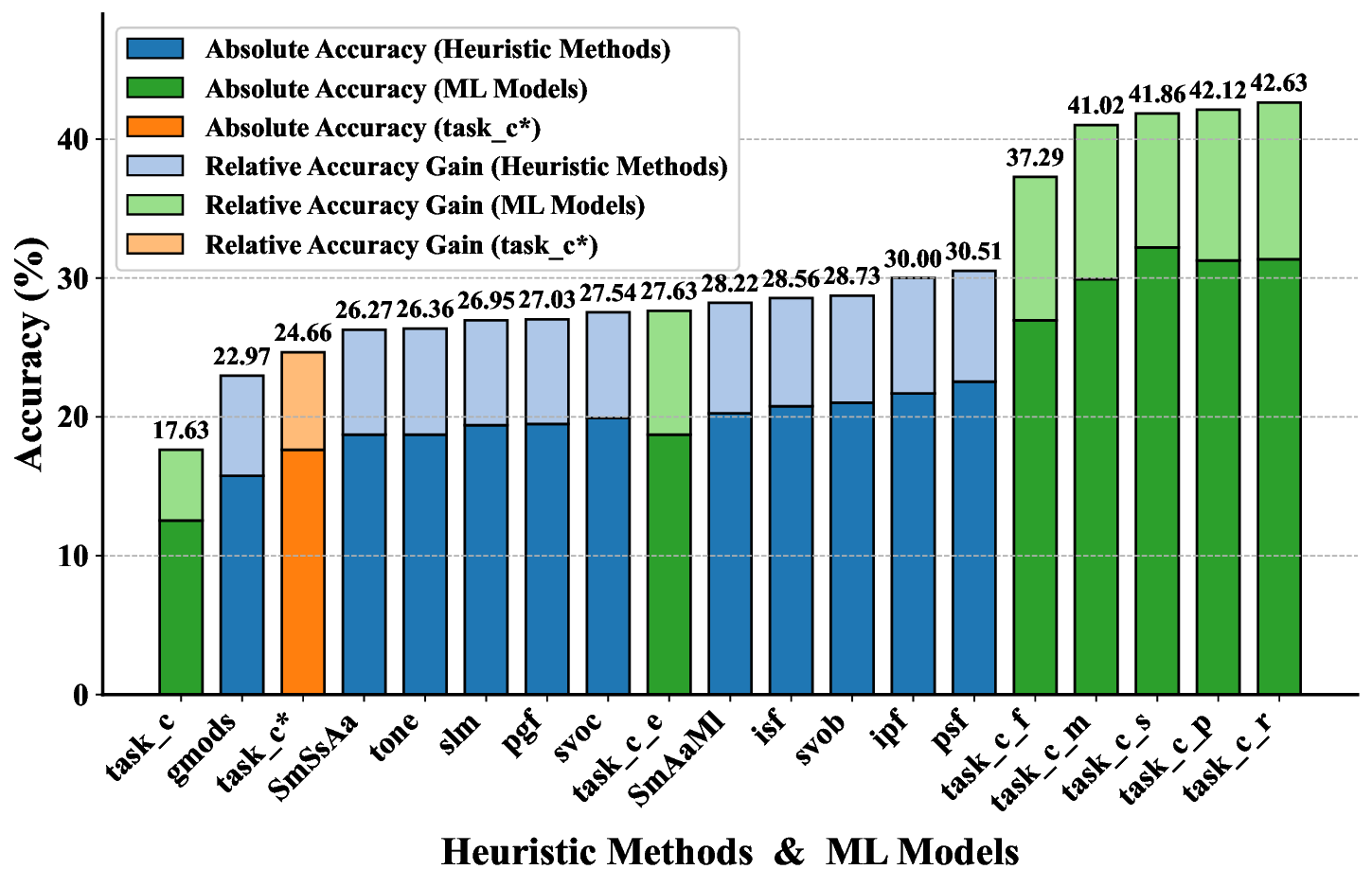}
    \caption{Accuracy results on the testing-validation set.}
  \end{subfigure}\hfill
  \begin{subfigure}{0.46\textwidth}
    \centering
    \includegraphics[width=\textwidth]{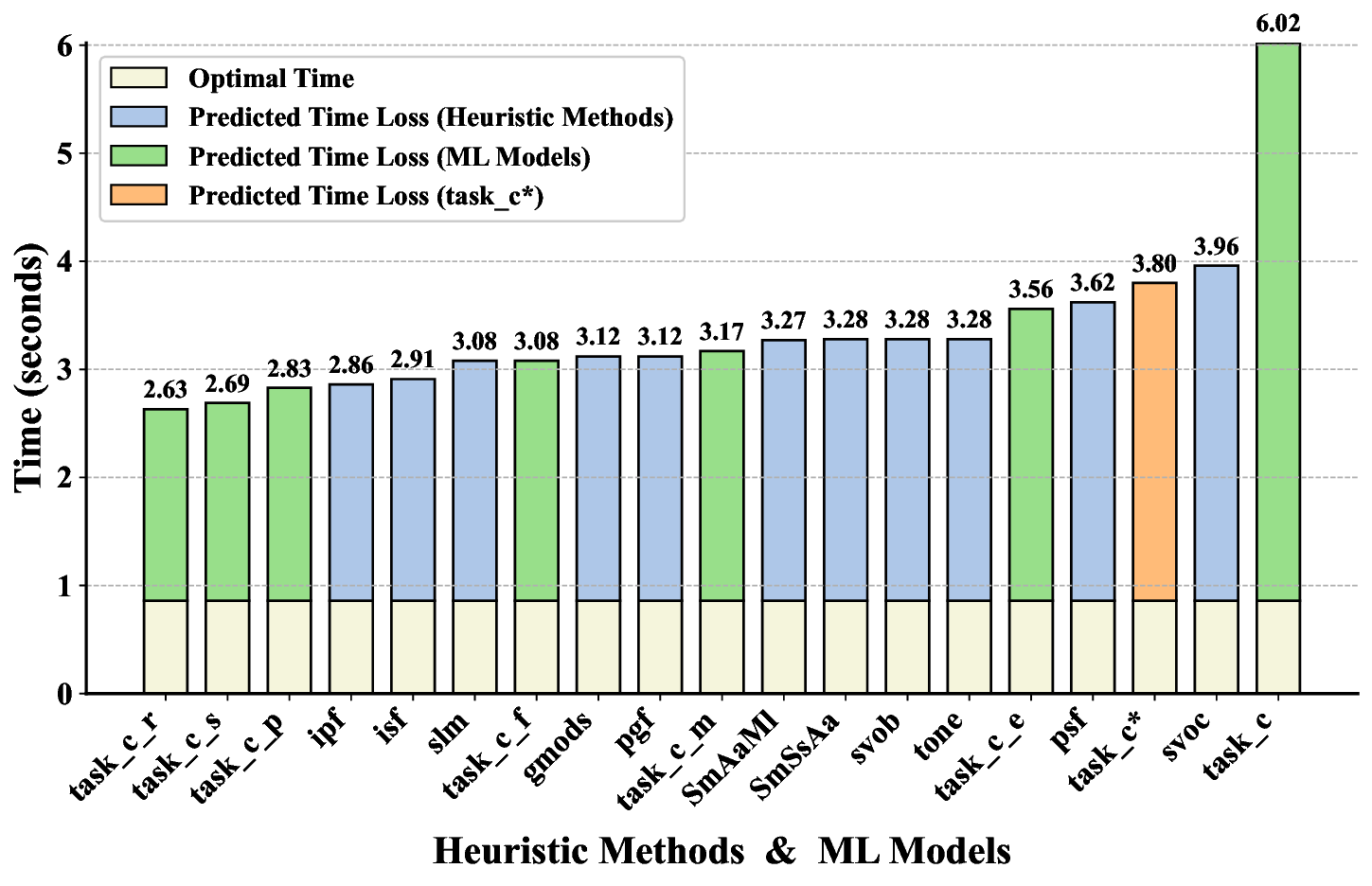}
    \caption{Execution time results on the testing-validation set.}
    % \label{fig:time_v4}
  \end{subfigure}
  \caption{Comparison of heuristic methods and ML models on the DQ-4 testing set and testing-validation set.}
  \label{fig:dq4_comparison}
\end{figure}

Figure~\ref{fig:dq4_comparison} compares 
the performance of ML models with heuristic methods
on testing and testing-valid set of the random four variables dataset DQ-4.
The performance of some of the models on these two sets are different. 
So we include both in order to draw a more solid conclusion. 
Firstly, for both figures, the trend observed in $(i)$ for DQ-3 still holds here. 
The three pre-training models learning deep features still perform the best after fine-tuning,
although the speedup over the best heuristic is less significant ($\mathbf{1.3\times}$ on testing set).
Secondly, the pre-training model $\texttt{task\_c\_f}$ learning the shallow features 
still outperforms most of the heuristics as well as the direct Transformer models after fine-tuning. The other pre-training model $\texttt{task\_c\_m}$, however, 
does not perform as well as in the three variable case.

\begin{figure}[!htbp]
  \centering
  \begin{subfigure}{0.48\textwidth}
    \centering
    \includegraphics[width=\textwidth]{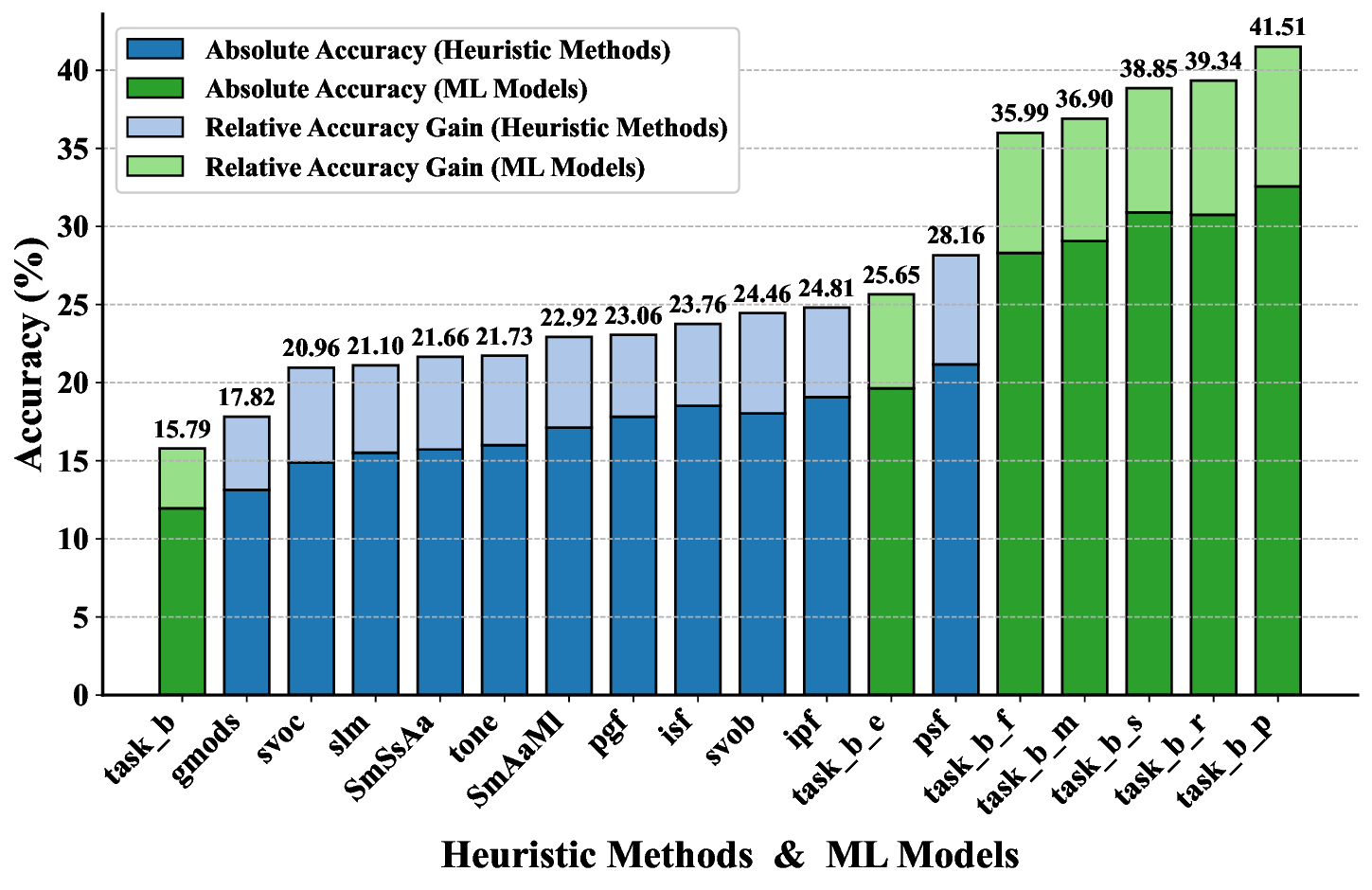}
    \caption{Accuracy results.}
    \label{fig:abs_rel_acc_v4b}
  \end{subfigure}\hfill
  \begin{subfigure}{0.48\textwidth}
    \centering
    \includegraphics[width=\textwidth]{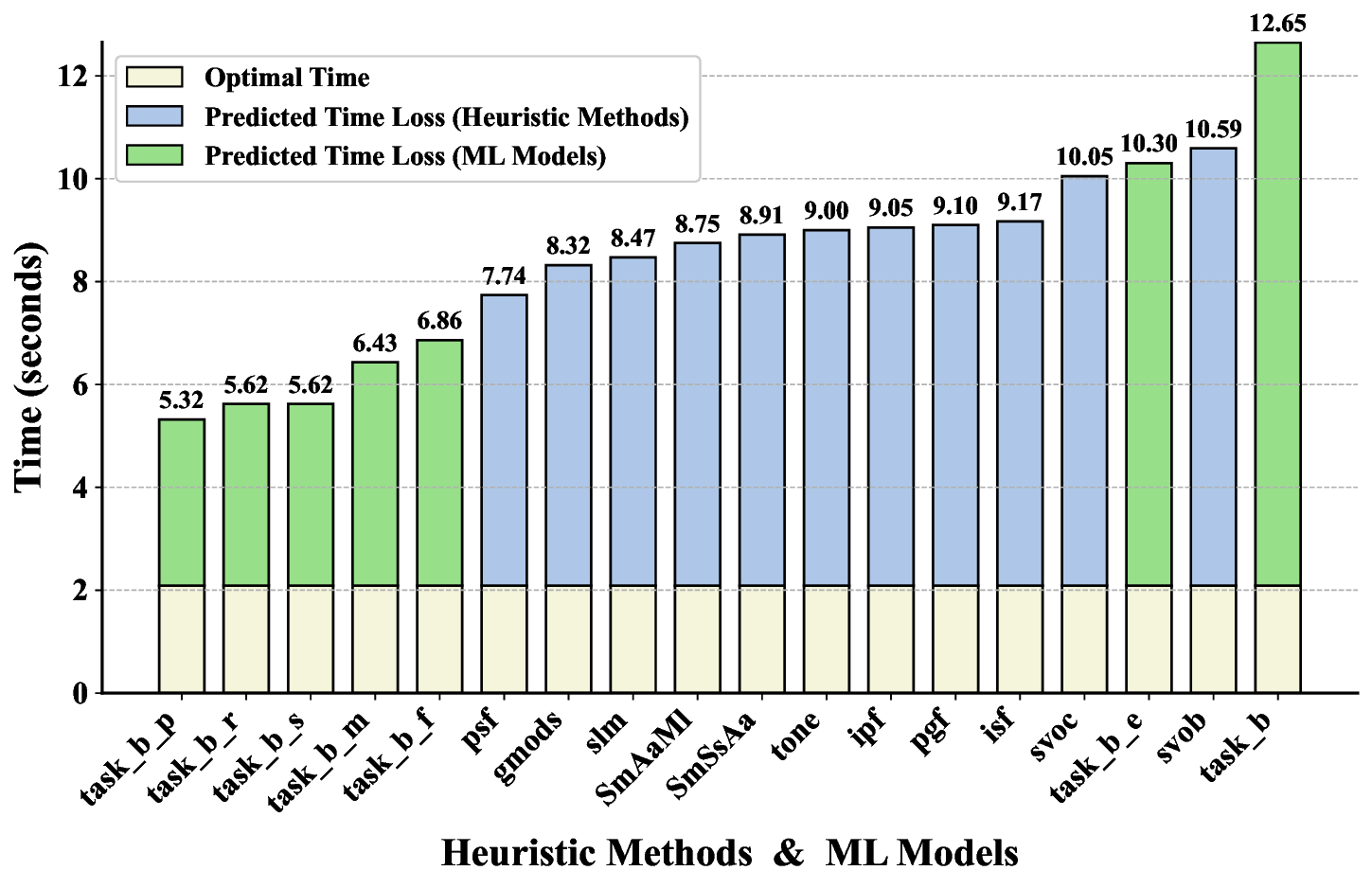}
    \caption{Execution time results.}
    \label{fig:time_v4b}
  \end{subfigure}
  \caption{Comparison of heuristic methods and ML models on the DQ-4b testing set.}
  \label{fig:dq4_dq4b_comparison}
\end{figure}

Next, we report the performance on DQ-4b, which includes DQ-4 as a subset and enhances 
DQ-4 with more difficult examples. 
The results are shown in Figure~\ref{fig:dq4_dq4b_comparison}.
All four observations  $(i)-(iv)$ made for DQ-3 apply here as well, 
except that the speedup of the best pre-training \& fine-tuning model over the best heuristic is less significant ($\mathbf{1.4\times}$ on testing set). We notice that, the speedup over the classical heuristic method \texttt{svob}
is close to ${\mathbf 2}\times$.

\subsubsection{Performance on the SMT dataset}
\label{subsubsec:perform-on-SMT}
In this section, we present the performance of various models on the SMT dataset. 
The models under consideration are categorized into two groups : ``Random Models" and ``SMT Models", illustrated in Figure~\ref{fig:smt}:

\begin{figure}[!htbp]
    \centering
    \includegraphics[width=0.6\textwidth]{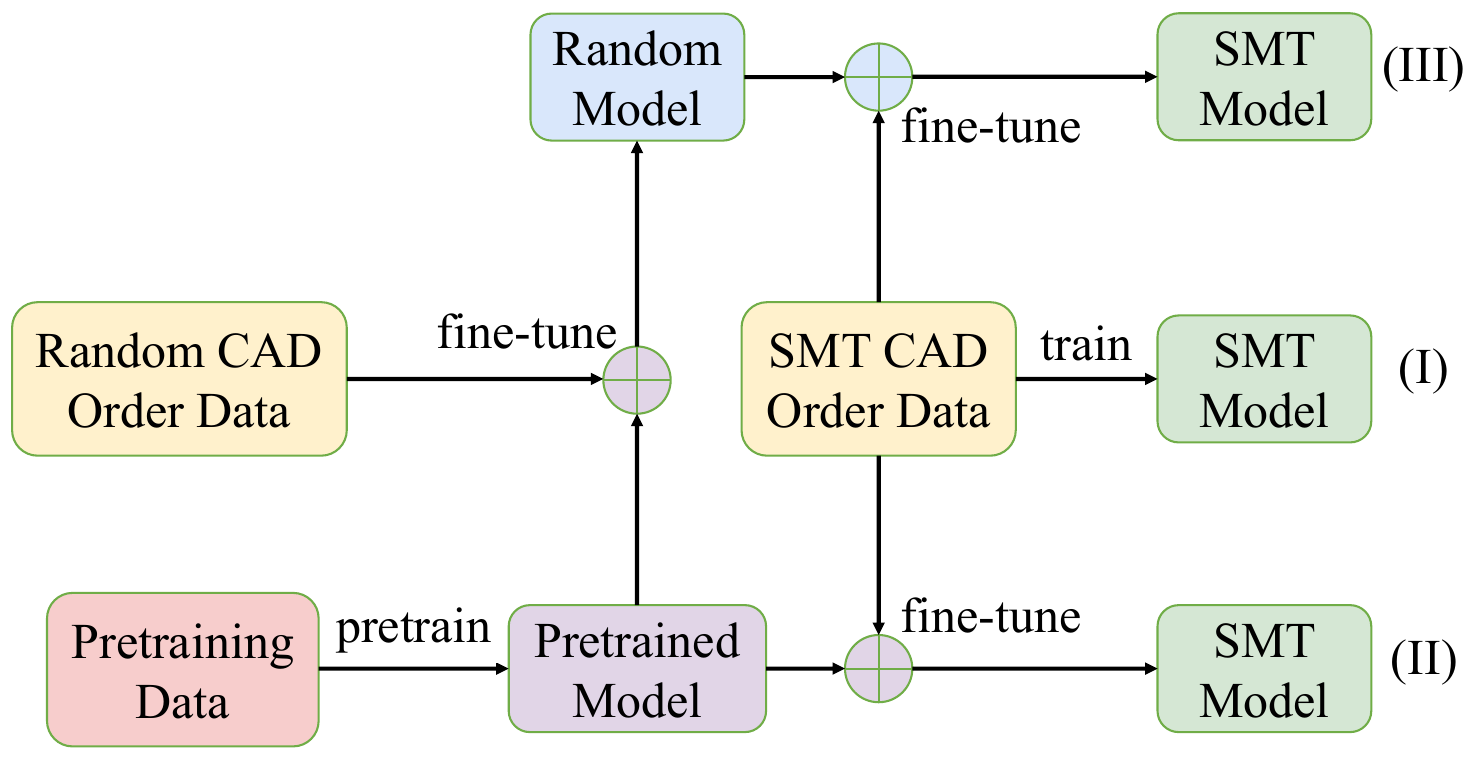}
    \caption{An overview of comparison for CAD model and SMT model.}
    \label{fig:smt}
\end{figure}

\begin{enumerate}
[label=(\arabic*)]
\item \textbf{Random Models}
These are the models that we have tested on the DQ-3 dataset. 
That is, they are either the Transformer models directly trained 
on DQ-3 or pre-trained model but fine-tuned 
on DQ-3. These models are not fine-tuned on the SMT dataset.
 \item \textbf{SMT Models} The models in this category are further divided into three subcategories. 
 The model in the first subcategory is obtained 
 by directly training the Transformer model with the training set 
 of the SMT dataset, denoted by \texttt{task\_t}.  
 The models in the second subcategory are obtained 
 by directly fine-tuning those pre-trained models 
 on the training set of the SMT dataset, named as 
  $\texttt{task\_t\_e}$, $\texttt{task\_t\_f}$, $\texttt{task\_t\_m}$, $\texttt{task\_t\_p}$, $\texttt{task\_t\_r}$, and $\texttt{task\_t\_s}$.
 The models in the third subcategory are obtained 
 by further fine-tuning those pre-trained models, 
 which have already been fine-tuned on the CAD order dataset DQ-3, on the  the training set of the SMT dataset, 
 named as $\texttt{task\_t\_c\_e}$, $\texttt{task\_t\_c\_f}$, $\texttt{task\_t\_c\_m}$, $\texttt{task\_t\_c\_p}$, $\texttt{task\_t\_c\_r}$, and $\texttt{task\_t\_c\_s}$.

 \end{enumerate}

\begin{figure}[!htbp]
    \centering
    \includegraphics[width=0.95\textwidth]{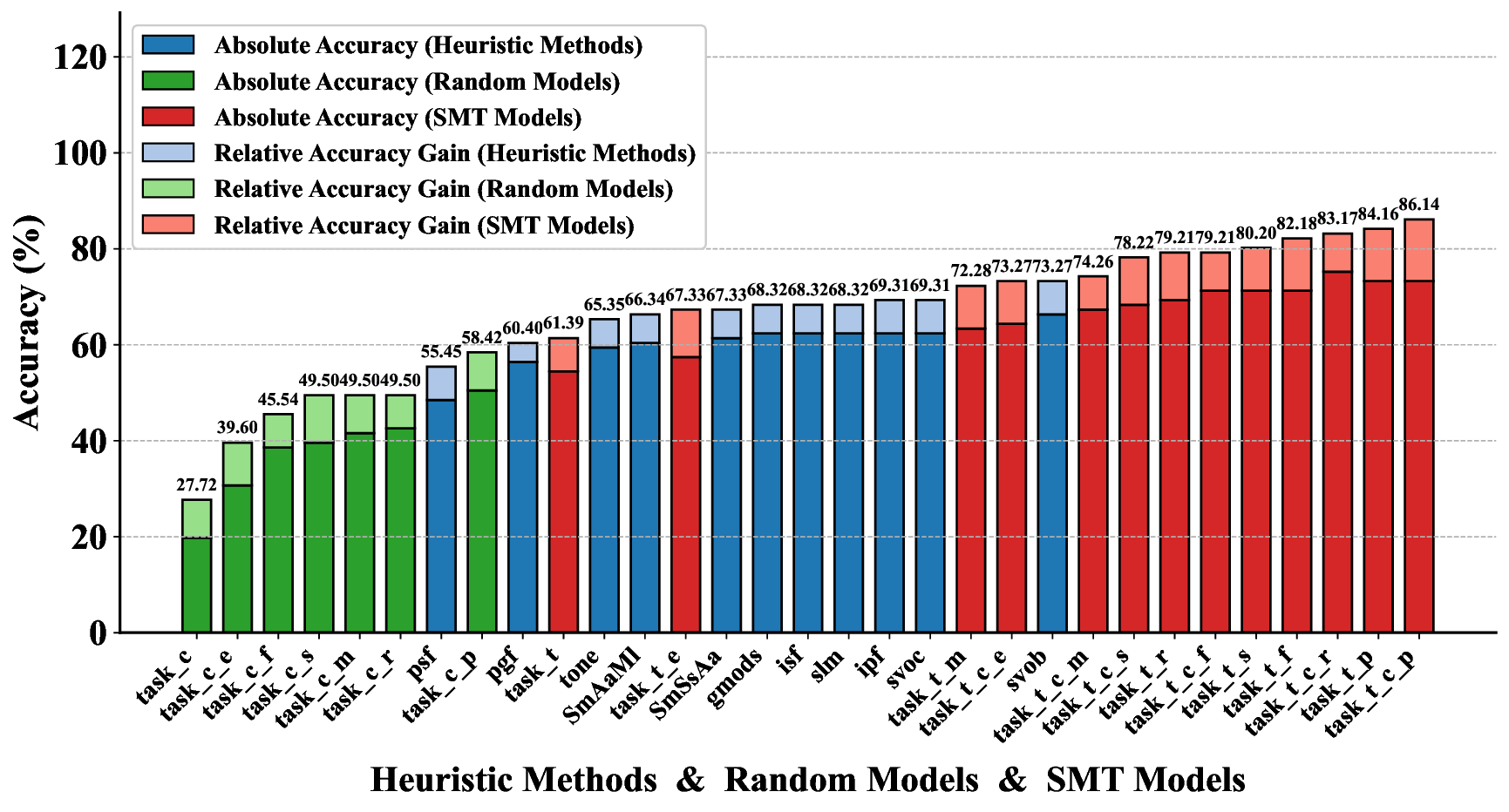}
    \caption{Accuracy comparison of heuristic methods and ML models on the SMT testing dataset.}
    \label{fig:with_finetune_smt_cad_acc_v3}
\end{figure}

\begin{figure}[!htbp]
    \centering
    \includegraphics[width=0.95\textwidth]{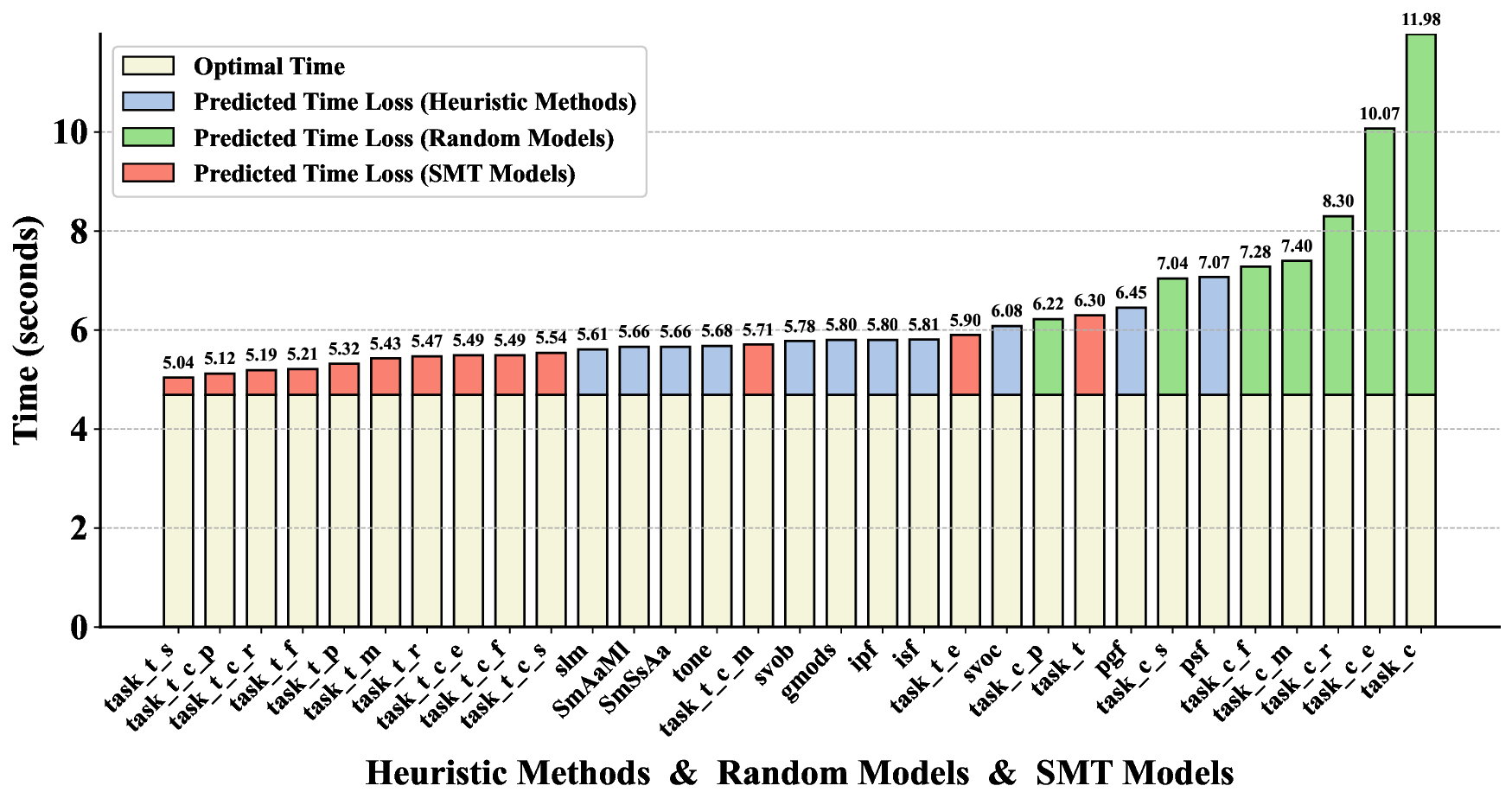}
    \caption{Runtime comparison of heuristic methods and ML models on the SMT testing dataset.}
    \label{fig:with_finetune_smt_cad_time_v3}
\end{figure}

Comparisons of 13 SMT models, 7 random models, and 11 heuristic methods are provided in Figures~\ref{fig:with_finetune_smt_cad_acc_v3} and \ref{fig:with_finetune_smt_cad_time_v3}, measuring respectively 
the accuracy and CAD running time. 
We have the following observations: (i) The pre-trained models
fine-tuned further on the SMT dataset perform the best. 
Most of these models (actually all pre-trained model exploiting deep features) outperform the best heuristic method; 
(ii) If the models are finally fine-tuned with the SMT dataset, then firstly fine-tuning on the random CAD order set does not help much and sometimes even reduce the performance; 
(iii) The pre-trained models not fine-tuned on the SMT dataset 
show limited generalization ability and they underperform 
most of the heuristic methods and the Transform model 
directly trained on the SMT dataset, although still outperform 
the one directly trained on the random dataset DQ-3.

\subsection{Some attempts to push the limit of pre-training \& fine-tuning models}
From the evaluations on both random and real datasets in last subsection, we have the following
common observations: $(i)$ Pre-training \& fine-tuning can substantially improve the performance of Transformer models on the CAD variable ordering selection task though firstly learning to predict some features relevant to the projection computation of CAD; $(ii)$ Pre-training \& fine-tuning brings such an improvement by creating 
labelled pre-training datasets $10\times$  larger, 
 but with much lower cost, than the variable ordering dataset. 

These positive results encouraged us to explore in different directions aiming 
to further improve the performance. 
More precisely, this section is designed to further address the following research questions:
\begin{itemize}
    \item $\mathbf{(RQ_1)}$ Can further performance improvement be achieved if the size of the pre-training dataset increases by another magnitude? 
    \item $\mathbf{(RQ_2)}$ Can further performance improvement be achieved if we increase the capacity of Transformer models? 
    \item $\mathbf{(RQ_3)}$ Instead of relying on pre-training to learn important features for CAD variable ordering selection, what if we first compute these features and directly use these features to train Transformer models?
    \item $\mathbf{(RQ_4)}$ As the performance of pre-training models is critical to the CAD order task, can we adopt 
    a multi-stage pre-training scheme to further lift the performance?
    \item $\mathbf{(RQ_5)}$ What is the impact of different tokenization schemes?
\end{itemize}
% The evaluations for addressing these questions were mainly made on the testing-validation sets of DQ-3. 
% To testify the solidness of the conclusion, the testing-validation sets of DQ-4 were also used for some evaluations.
The evaluations for addressing these questions were made on the testing-validation sets of DQ-3.

\subsubsection{The impact of dataset size on pre-training tasks and CAD order tasks}
\label{subsubsec:size}
In Section ~\ref{subsec:CAD-pretrain-dataset}, we successfully generated  pre-training datasets comprising millions of instances for  DQ-3. 
In the experimental results reported in Section~\ref{subsec:main}, however, we only used $1/10$ of the pre-training data for training the models.
It is natural to ask what if all the pre-training data are used. 

\begin{figure}[htbp]
  \centering
    \includegraphics[width=0.5\textwidth]{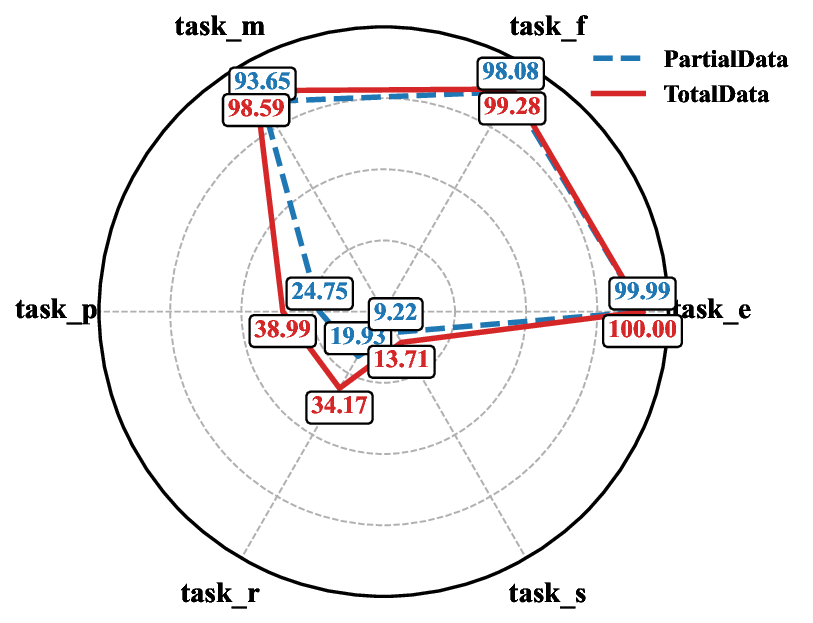}
  \caption{Performance comparison of pre-trained models with partial data and total data.}
  \label{fig:pretrained_total_data}
\end{figure}

The impact of data augmentation strategies on the accuracy of pre-training models for three-variables and four-variables problems is illustrated in Figure ~\ref{fig:pretrained_total_data}. This chart compares the accuracies of six pre-training tasks under partial-data and total-data settings. 
%, underscoring the benefits of data augmentation.
On one hand, more data do boost the performance of models on difficult pre-training tasks, namely \texttt{task\_p}, \texttt{task\_r}
and \texttt{task\_s}. On the other hand, in the best case, the accuracy does not double even the size of the dataset 
gets $10\times$ larger.

\begin{figure}[htbp]
  \centering
    \includegraphics[width=0.8\textwidth]{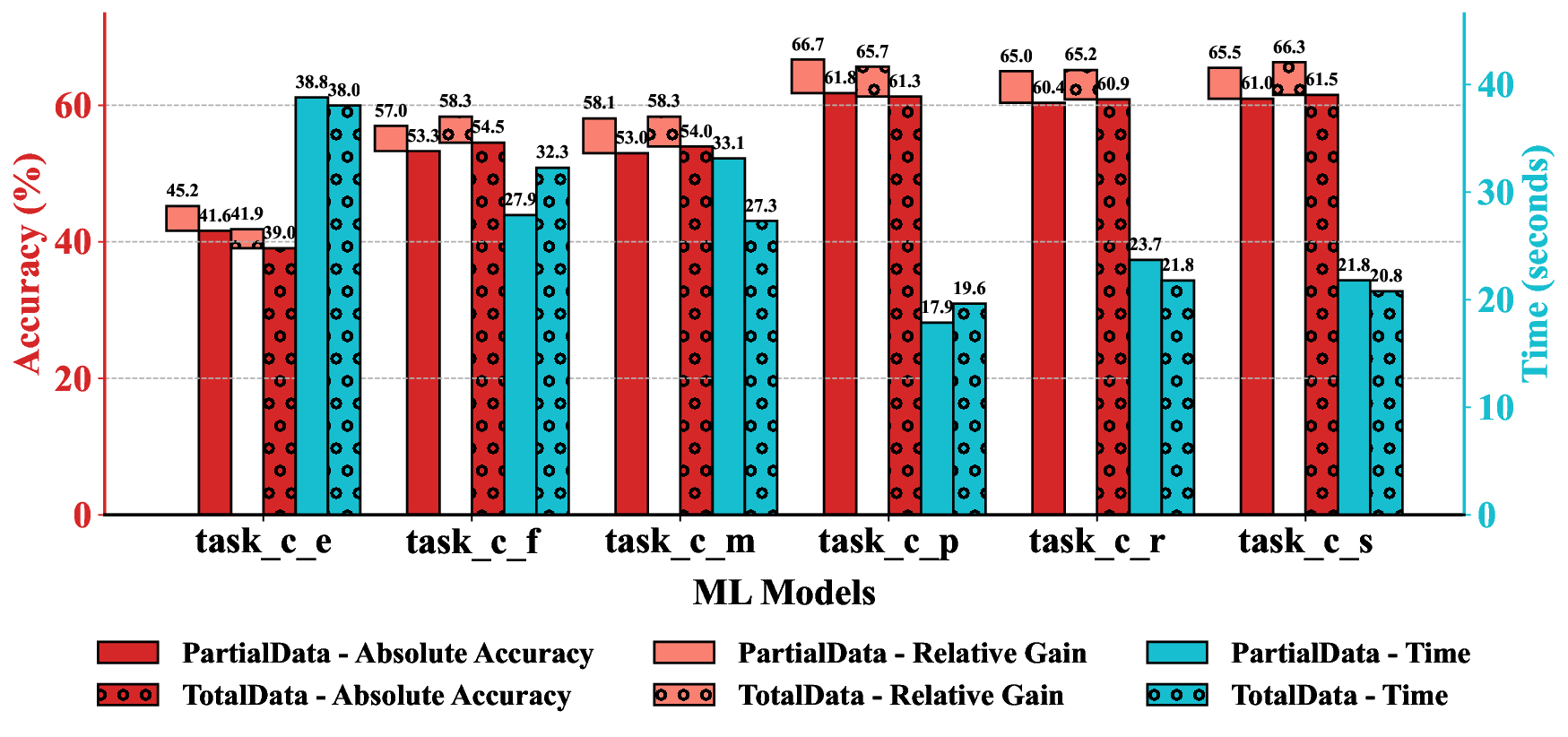}
   \caption{Performance comparison of models pre-trained with partial data and total data after fine-tuning 
  on CAD order task.}
  \label{fig:total_experiments_cad}
\end{figure}

Still, given the previous experimental results revealing that limited accuracy on pre-training task greatly improve 
the performance on CAD ordering selection task, we expect that the pre-trained models trained with much more data 
can further lift such a performance. The experimental results are reported in Figure~\ref{fig:total_experiments_cad}.
Regarding the accuracy, the difference between the total-data and partial-data models are negligible on most of the tasks. 
For CAD running time, the total-data model slightly outperforms the partial-data model on four tasks but 
underperforms the total-data model on two tasks.
As an answer to $(\mathbf{RQ_1})$, overall, the benefit of further increasing 
the dataset size of pre-training datasets for pre-training tasks is significant, but the benefit of transferring to CAD order task is negligible.

\subsubsection{The impact of model's capacity on pre-training tasks and CAD order tasks}
In this section, we use \texttt{task\_s} as an example to understand the impact of model's capacity on the performance. 
In Section~\ref{subsubsec:size}, we have observed that increasing the dataset size can substantially  improve the performance 
of pre-training models. As the dataset size has increased, it is natural to increase the model's capacity accordingly. 
 Table~\ref{tab:model-parameters-elegant} shows the changes. 
Note that limited to the hardware limit of our GPU, the batch size is reduced. 
We also reduce the number of encoders by one, but double the number of decoders. 
This adjustment is based on the observation from our previous experimental results, 
namely the accuracies of models on tasks for learning input features (namely \texttt{task\_e} and \texttt{task\_f}) 
are nearly perfect, but the accuracies of models on generating deep features for CAD ordering are relatively low.
Thus, we slightly reduce the number of encoders, but significantly increase the number of decoders.

\begin{table}[!ht]
    \caption{Comparison on model architecture and training parameters.}
    \label{tab:model-parameters-elegant}
    \centering
    \begin{tabular}{ccccccc}
    \toprule
       {Task}        & Embedding\_Dimension & Heads & Encoders/Decoders & Batch\_Size & Epochs \\ 
       \midrule
       \texttt{task\_s}  & 256              & 8        & 7/6       & 128        & 100   \\
       \texttt{task\_ss} & 512              & 16       & 6/12       & 32         & 100  \\
       \bottomrule
    \end{tabular}
\end{table}

For clarity and distinction, we designate the version of \texttt{task\_s} trained with the new set of hyperparameters as \texttt{task\_ss}. As shown in Table ~\ref{tab:other-experiments}, the pre-training accuracy of \texttt{task\_ss} significantly improved from \textbf{13.71\%} for \texttt{task\_s} to \textbf{18.57\%}. This result strongly suggests that even for a challenging task like \texttt{task\_s}, increasing the model's architectural capacity (through higher embedding dimensions, more attention heads, and deeper layers) leads to a substantial improvement in pre-training performance. However, when transferred to CAD tasks, \texttt{task\_ss} did not benefit from the more complex model architecture and enhanced hyperparameters. 
This provides an answer to $(\mathbf{RQ_2})$.

\begin{table}[!ht]
\caption{Performance variation after increasing the capacity of a pre-training model.}
\label{tab:other-experiments}
    \centering
    \begin{tabular}{cc|ccccc}
    \toprule
     \text{Pre-training\_Task}     & acc & \text{Order\_Task} & abs\_acc & rel\_acc & predict\_time & optimal\_time\\ 
     \midrule
    \texttt{task\_s}                & 13.71\%  & \texttt{task\_c\_s}   &  61.53\% &  66.31\%  &  20.77 & 6.28\\ 
     \texttt{task\_ss}              &  18.57\% & \texttt{task\_c\_ss} &  61.29\%  &  65.65\%  & 22.02 & 6.28\\ 
     \bottomrule
    \end{tabular}
\end{table}

\subsubsection{Employing pre-training to learn features versus directly using features}
In Section ~\ref{subsec:pretask}, we elaborated on the design of polynomial system pre-training tasks, where the core idea involves taking raw polynomial systems as input and generating a set of carefully designed, representative polynomial system features as output. Given the objective of generating these features, a natural and critical question arises: Do these features, extracted from the pre-training tasks, possess sufficient discriminative power to be directly applied to the CAD variable ordering task? More specifically, can we forgo the original polynomial system and instead use these features as direct model inputs, training a dedicated model with the CAD optimal ordering of the polynomial system as the target output?

To address this, for each pre-training task $\texttt{task\_x}$ presented in Definition~\ref{def:pre-training} of Section~\ref{subsec:pretask}, 
we name its targeted learning object as $\texttt{exp\_x}$. 
we compare two types of models: \textbf{RandomModel} and \textbf{FeatureModel}.
The former model is the previously introduced pre-training \& fine-tuning model, namely a  model is trained 
to learn $\texttt{exp\_x}$ and then gets fine-tuned to learning the best CAD variable ordering.
In contrast, the latter model is trained to predict the best variable ordering by  taking directly $\texttt{exp\_x}$ as input.

\begin{figure}[htbp]
  \centering
    \includegraphics[width=0.8\textwidth]{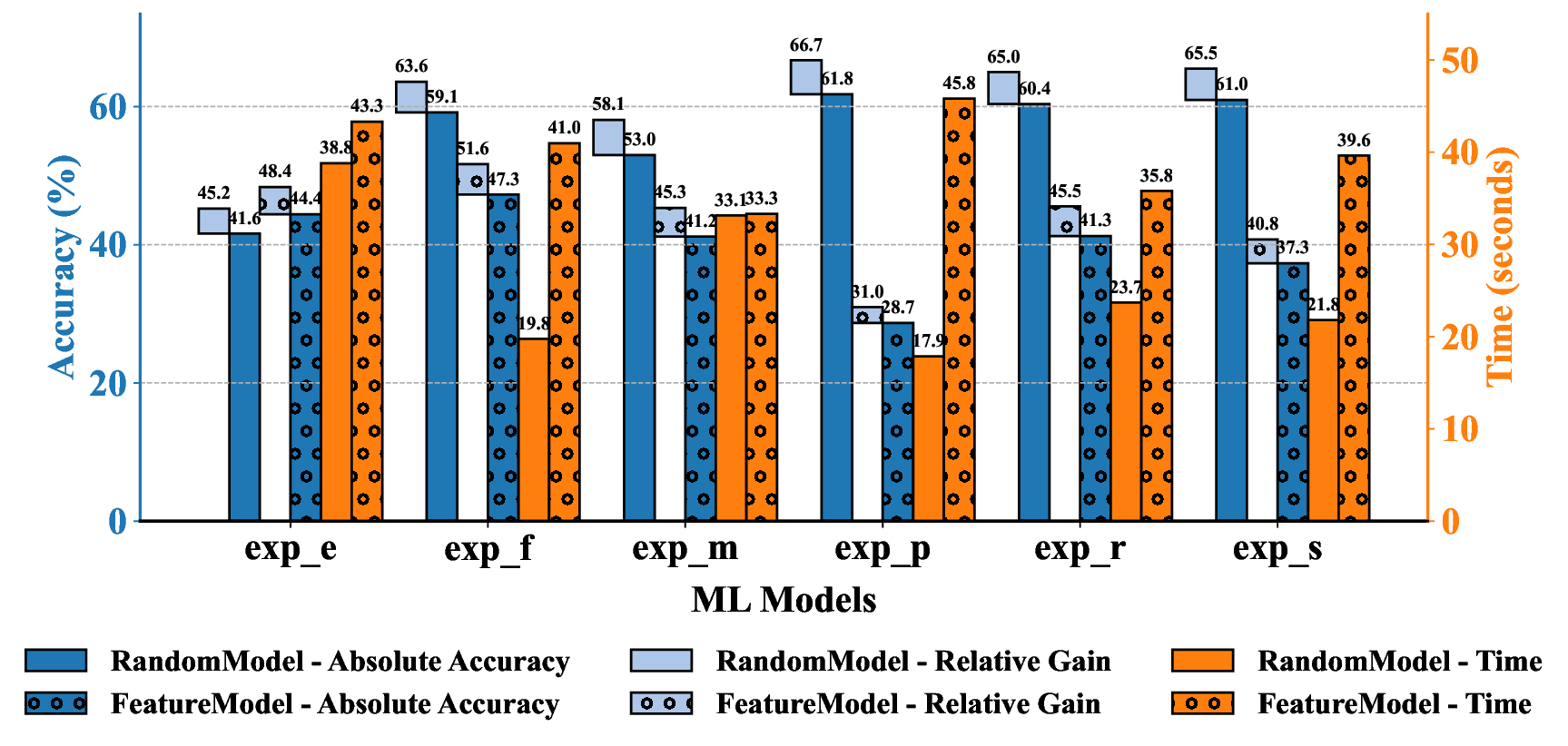}
  \caption{Performance comparison of RandomModel and FeatureModel on CAD order task.}
  \label{fig:with_feature_learning}
\end{figure}

A systematic comparison between RandomModel and FeatureModel across six pre-training tasks is presented in Figure~\ref{fig:with_feature_learning}. 
Overall, RandomModel consistently outperforms FeatureModel in the majority of tasks. 
The only exception occurs in $\texttt{task\_e}$:  FeatureModel attains slightly higher accuracy.
But even in this case,  FeatureModel does not show superiority on CAD running time. 
This provides an answer to $(\mathbf{RQ_3})$: directly using pre-training derived features as model inputs is not optimal for downstream performance. Instead, employing such features as auxiliary tools during pre-training, facilitating knowledge transfer and representation learning, proves more effective. In this way, features help capture deep structures and latent distributional patterns, providing a robust initialization for subsequent adaptation.

\subsubsection{The impact of multi-level pre-training}
In Section~\ref{subsec:pretask}, we designed several pre-training tasks aimed at effectively extracting features of polynomial systems and applying them to the CAD ordering task to enhance model performance. Building upon this, a natural question arises: if an individual pre-training task can successfully transfer to the CAD ordering task and improve model performance, could multiple sequential transfers between different pre-training tasks further boost the model’s capability by integrating and leveraging a broader range of pre-trained features?

To answer this question, we conducted a series of experiments. Here, we only show a representative example. As we know, $\texttt{task\_f}$ is designed to extract 11 core features of the input polynomials, whereas $\texttt{task\_p}$ focuses on capturing 11 key features related to the projection factor sets of input polynomials. In principle, the features learned in $\texttt{task\_f}$ should provide valuable assistance during the training process of $\texttt{task\_p}$. Therefore, we first transferred the model pre-trained on $\texttt{task\_f}$ to the training phase of $\texttt{task\_p}$. This sequential transfer task is denoted as $\texttt{task\_p\_f}$, whose main objective is to enhance the model’s learning capability for $\texttt{task\_p}$. Subsequently, the model trained on $\texttt{task\_p\_f}$ was further applied to the CAD ordering task, designated as $\texttt{task\_c\_p\_f}$. 
Table~\ref{tab:other-experiments-2} compares the performance between them. 
 As shown in the results, the multi-transfer models did not outperform the single-transfer models. 
 This provides an answer to $(\mathbf{RQ_4})$.

\begin{table}[!ht]
\caption{Comparison of one-step and multi-step pre-training models.}
\label{tab:other-experiments-2}
    \centering
    \begin{tabular}{cc|cccc}
    \toprule
     \text{Pre-training\_Task}     & acc & \text{Order\_Task} & abs\_acc & rel\_acc & time\_ratio \\ 
     \midrule
      \texttt{task\_f} & 98.08\%   & \texttt{task\_c\_f} &   53.29\% &  57.00\% & 4.43 \\ 
      \texttt{task\_p} & 24.75\%   & \texttt{task\_c\_p} &   61.78\% &  66.72\% & 2.84 \\ 
      \texttt{task\_p\_f} & 22.48\%  & \texttt{task\_c\_p\_f} &  59.14\% & 63.59\% & 3.15 \\ 
      { \texttt{task\_fp}} & { 24.77\%}  & { \texttt{task\_c\_fp}} &  { 58.73\%} & { 62.93\%} & { 3.87} \\ 
      \bottomrule
    \end{tabular}
\end{table}

We attribute this performance degradation to the nature of the sequential transfer process. In multi-transfer settings, the final pre-trained model inherits the lower-level parameters from a simpler pre-training task, in this case, from $\texttt{task\_f}$. Due to the fine-tuning strategy that freezes part of the model’s lower layers, the resulting representation space is dominated by features learned from the simpler task. However, such features are less effective for fine-tuning on complex downstream tasks, leading to suboptimal performance compared to models pre-trained directly on more complex tasks.

Moreover, it is important to note that the pre-training \& fine-tuning paradigm is primarily designed for scenarios where labeled data for the target task is scarce. In such cases, pre-training on related labeled tasks can provide valuable prior knowledge for transfer. However, in our experimental setup, all pre-training tasks already contain sufficient labeled data, allowing the model to learn directly from large-scale supervision without requiring additional transfer steps. Consequently, performing multiple sequential transfers in this context does not yield further benefits and may even hinder the learning of task-specific representations.
{Instead of sequentially using $\texttt{task\_f}$
and $\texttt{task\_p}$, we also tried to merge them into one, named as $\texttt{task\_fp}$. From Table~\ref{tab:other-experiments-2}, we see that it slightly underperforms $\texttt{task\_p\_f}$.}

\subsubsection{The impact of tokenization}
\label{subsubsec: inpact_encode}

\begin{figure}[htbp]
  \centering
    \includegraphics[width=0.6\textwidth]{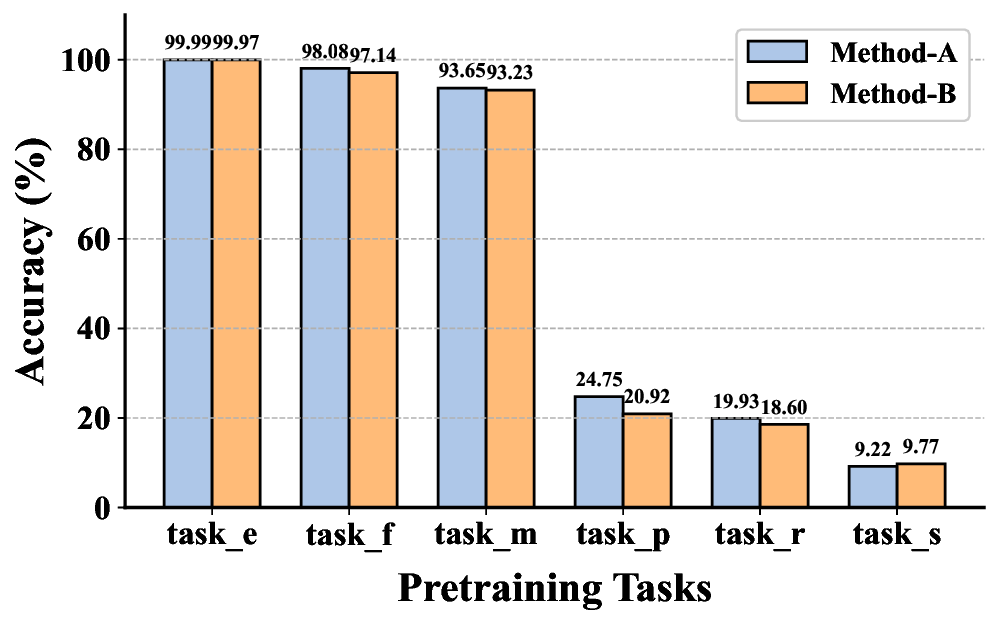}
  \caption{Comparison of two different tokenization schemes on pre-training tasks.}
  \label{fig:pretrained_terms}
\end{figure}

This section is designed to answer ($\mathbf{RQ_5}$).
In Section~\ref{subsec:encode}, we have introduced a tokenization scheme for polynomials, hereafter referred to as Method-A, which completes omitted exponents ($0$ and $1$) in polynomial expressions and distinguishes digits in coefficients and exponents w.r.t. the straightforward scheme  proposed in~\citep{DBLP:conf/casc/ChenJQYZ24}. 
However, we have already observed that Method-A underperforms the straightforward one for \texttt{task\_c}. 
A preliminary investigation shows that if we only distinguish digits in coefficients and exponents but 
do not complete missing exponents w.r.t. the straightforward scheme,  the resulting method, hereafter referred to as Method-B, outperforms both Method-A and the straightforward one for \texttt{task\_c}. 
It is thus worth further investigating the impact of both Method-A and Method-B on the performance of other tasks. 
To illustrate Method-B, we consider the polynomial system discussed in ~\ref{exam:features}. The encoding result produced by Method-A has already been presented in Section ~\ref{subsec:encode}, while the corresponding result under Method-B is given as \{\texttt{$\langle s\rangle$}, \texttt{-}, \texttt{c6}, \texttt{*}, \texttt{x1}, $\wedge$, \texttt{3}, \texttt{*}, \texttt{x2}, \texttt{-}, \texttt{c4}, \texttt{*}, \texttt{x1}, \texttt{*}, \texttt{x2}, \texttt{*}, \texttt{x3}, $\wedge$, \texttt{2}, \texttt{+}, \texttt{c2}, \texttt{*}, \texttt{x2}, $\wedge$, \texttt{2}, \texttt{*}, \texttt{x3}, \texttt{+}, \texttt{c1},\texttt{=}, \texttt{c0}, \texttt{,}, \texttt{-}, \texttt{c5}, \texttt{*}, \texttt{x3}, $\wedge$, \texttt{4}, \texttt{+}, \texttt{x3}, $\wedge$, \texttt{3}, \texttt{-}, \texttt{c7},\texttt{=}, \texttt{c0}, \texttt{$\langle /s\rangle$}\}. 

% Building on these two encoding schemes, this section provides a systematic comparison and analysis of their performance in both the pre-training task and the transfer task to CAD, thereby offering clearer insights into how different encoding strategies affect model effectiveness.

\begin{figure}[htbp]
  \centering
    \includegraphics[width=0.8\textwidth]{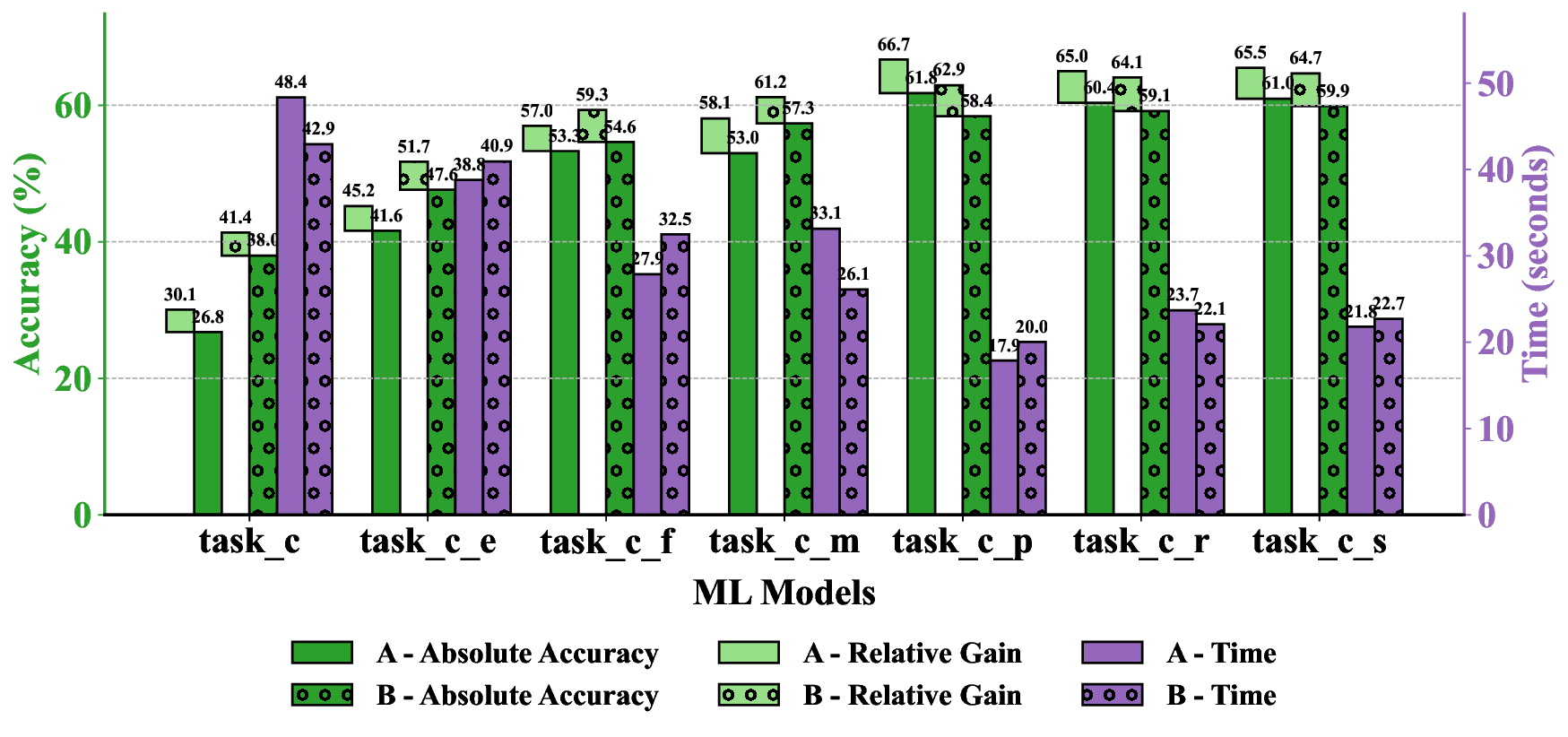}
  \caption{Performance comparison of Method-A and Method-B on seven transferred tasks under the CAD setting.}
  \label{fig:with_simple_terms_cad}
\end{figure}

Figure~\ref{fig:pretrained_terms} compares the performance of Method-A and Method-B on six pre-training tasks. We observe that Method-A (resp. Method-B) has a slight advantage over Method-B (resp. Method-A) on \texttt{task\_p} and \texttt{task\_r} (resp. \texttt{task\_s}). 
Figure~\ref{fig:with_simple_terms_cad} further compares the performance 
of the two encoding schemes on pre-training \& fine-tuning tasks for CAD variable ordering selection.
On the best-performed tasks for ordering selection, the difference between 
the two is not significant. Method-A has a slight advantage over Method-B.

{
\subsection{Further experimental results on putting the models into practice}
In this section, we provide experimental results on the method presented in 
Section~\ref{subsec:practice}, which allows the models trained for a fixed dimension $n=3,4$ to handle problems 
in different dimensions.

\subsubsection{Cross-dimension evaluation}
\label{subsubsec:cross}
The method presented in Section~\ref{subsec:practice} 
allows adpation of the models trained on the three-variable dataset DQ-3 
to predict variable orderings for systems in the four-variable dataset DQ-4b. 
It is interesting to know how these adapted models perform on DQ-4b, 
in particular comparing with the separately trained model for the four-variable case.
Similarly, we want to know how a model trained with four-variable datasets 
performs on three-variable datasets.
The experimental results are reported in Table~\ref{tab:cross-predict}.
On DQ-4b, the adapted three-variable models (starting with \texttt{task\_c}) all have lower accuracies w.r.t. the corresponding separately trained four-variable models (starting with \texttt{task\_b}), which is expected.
On the other hand, regarding CAD running time (lower time\_ratio is better), some of the adapted models outperform the corresponding separately trained models.
However, the best model is still the separately trained one (\texttt{task\_b\_p}).
On DQ-3,  for both the accuracy and the CAD running time, the separately trained three-variable models (starting with \texttt{task\_c}) always outperform the corresponding adapted four-variable models (starting with \texttt{task\_b}).

\begin{table}[!htbp]
    %\caption{Comparison of adapted cross-dimension models and separately trained models without weight.}
    \caption{Comparison of adapted cross-dimension models and separately trained models on testing datasets.}
    \label{tab:cross-predict}
    \centering
    % \renewcommand{\arraystretch}{1.1}
   % \begin{tabular}{@{} c c *{6}{c} @{}} 
    \begin{tabular}{cc|ccccccc} 
        \toprule
         \text{Dataset} & \text{Category} & \texttt{task\_b} & {\texttt{task\_b\_e}} & {\texttt{task\_b\_f}} & {\texttt{task\_b\_m}} & {\texttt{task\_b\_p}} & {\texttt{task\_b\_r}} & {\texttt{task\_b\_s}} \\
        \midrule
        \multirow{8}{*}{\text{DQ-4b}}
            & abs\_acc & 11.95\%  & 19.64\% & 28.30\% & 29.07\% & {\bf 32.56\%} & 30.75\% & 30.89\% \\
            & rel\_acc & 15.79\% & 25.65\% & 35.99\% & 36.90\% & {\bf 41.51\%} & 39.34\% & 38.85\% \\
            & time\_ratio & 6.05 & 4.92 & 3.28 & 3.07 & {\bf 2.54} & 2.69 & 2.69 \\
         \cmidrule(lr){2-9}
         & \text{Category} & \texttt{task\_c} & {\texttt{task\_c\_e}} & {\texttt{task\_c\_f}} & {\texttt{task\_c\_m}} & {\texttt{task\_c\_p}} & {\texttt{task\_c\_r}} & {\texttt{task\_c\_s}} \\
       \cmidrule(lr){2-9}
    
            & abs\_acc & 8.60\%  & 15.44\% & 19.50\% & 21.80\% & 23.06\% & 23.13\% & {\bf 24.32\%} \\
            & rel\_acc & 12.51\% & 21.80\% & 25.79\% & 28.65\% & 29.63\% & 29.84\% & {\bf 31.94\%} \\
            & time\_ratio & 5.97 & 4.87 & 3.50 & 3.04 & 2.75 & 2.73 & {\bf 2.65} \\
        \midrule
      
        \text{Dataset} & \text{Category} & \texttt{task\_c} & {\texttt{task\_c\_e}} & {\texttt{task\_c\_f}} & {\texttt{task\_c\_m}} & {\texttt{task\_c\_p}} & {\texttt{task\_c\_r}} & {\texttt{task\_c\_s}} \\
        \midrule
        \multirow{8}{*}{\text{DQ-3}}
            & abs\_acc & 28.89\%  & 44.44\% & 52.51\% & 53.17\% & {\bf 60.25\%} & 59.51\% & 59.67\% \\
            & rel\_acc & 31.28\% & 47.24\% & 55.39\% & 56.21\% & {\bf 63.95\%} & 63.62\% & 63.62\% \\
            & time\_ratio & 9.61 & 6.44 & 4.91 & 4.88 & {\bf 3.05} & 3.56 & 3.31 \\
         \cmidrule(lr){2-9}
         & \text{Category} & \texttt{task\_b} & {\texttt{task\_b\_e}} & {\texttt{task\_b\_f}} & {\texttt{task\_b\_m}} & {\texttt{task\_b\_p}} & {\texttt{task\_b\_r}} & {\texttt{task\_b\_s}} \\
         \cmidrule(lr){2-9}
       
            & abs\_acc & 20.08\%  & 22.22\% & 30.04\% & 42.72\% & 43.29\% & {\bf 47.74\%} & 33.83\% \\
            & rel\_acc & 21.56\% & 24.28\% & 32.10\% & 45.84\% & 46.83\% & {\bf 50.86\%} & 36.79\% \\
            & time\_ratio & 12.52 & 10.64 & 8.44 & 6.29 & {\bf 5.56} & 6.33 & 7.62 \\
       
        \bottomrule
    \end{tabular}
\end{table}

\subsubsection{Model recommendation}
\label{subsec:model}

In Section~\ref{subsubsec:cross}, we have compared cross-dimension models and separately trained models. 
The conclusion is that for $n=3$ (resp. $n=4$), the best performed model in terms of both the accuracy and the CAD running time is the separately trained model \texttt{task\_c\_p} (resp. \texttt{task\_b\_p}). 
On the other hand, we also notice that for $n=3$ (resp. $n=4$), the performance difference between \texttt{task\_c\_p}, \texttt{task\_c\_r} and \texttt{task\_c\_s} 
(resp. \texttt{task\_b\_p}, \texttt{task\_b\_r} and \texttt{task\_b\_s})
is not significant. 
In particular, the CAD running time of \texttt{task\_b\_r} and \texttt{task\_b\_s} is the same. 
Thus, we may also need to take the cost of training these models
and the inference time of these models into account. 

Here, we take $n=4$ as an example and report the training and the inference cost in Table~\ref{tab:computational cost}. 
We notice that the inference time, 
that is the time for predicting the variable ordering, is negligible and is almost the same for three models, 
which is expected as the input, the output, 
as well as the size of these three models are the same. 
However, for the pre-training time, 
there is a substantial difference. 
From the table, we see that training \texttt{task\_s} takes both the most number of epochs and the longest time per epoch. It also takes more epochs to get fine-tuned than the other two models.
On the contrast,  the training of \texttt{task\_r} takes the least number of epochs and the shortest time per epoch. 
The training cost of \texttt{task\_p} sits in the middle. 
Note that in the table, 
the epoch count is not a manually set hyperparameter but is determined by early stopping based on the highest validation accuracy. 
This efficiency gap can be largely attributed to the label length difference in their training datasets,
which are respectively 108.09, 43.65, 151.33 for \texttt{task\_p}, \texttt{task\_r} and \texttt{task\_s}.

Combining the results in both Table~\ref{tab:cross-predict}
and Table~\ref{tab:computational cost}, we have the following recommendation on the models. 
For $n=4$, our first choice is  the separately trained model \texttt{task\_b\_p} for the following reasons: 
firstly this model performs the best in terms of both accuracy and CAD running time; secondly, its inference is as efficient as that of other models. Although the training time of this model is not the shortest, it is only a one-time price to pay. 
Regarding the second choice, \texttt{task\_b\_r}
is preferred to \texttt{task\_b\_s} as the former has a slight advantage 
over the latter while meantime the training time of the former is 
substantially shorter than the latter. 
For the same reason, when $n=3$, the order of recommendation is: \texttt{task\_c\_p}>\texttt{task\_c\_s}>\texttt{task\_c\_r}.

\begin{table}[!ht]
\caption{Comparison of the costs of pre-training, training and inference.}
\label{tab:computational cost}
    \centering
    \begin{tabular}{ccc|ccccc}
    \toprule
      \text{Pre-training}   & \multirow{2}{*}{\text{epochs}}  & \text{training\_time}   & \text{Order}  & \multirow{2}{*}{\text{epochs}}  & \text{training\_time} & \multicolumn{2}{c}{\text{inference\_time}} \\ 
     \text{Task} & & \text{per\_epoch}& \text{Task} &  &\text{per\_epoch} &\text{per\_epoch} &  \text{per\_example}\\ 
     \midrule
     \texttt{task\_p} & 183 & 507 & \texttt{task\_b\_p} & 11 & 26 & 30 & 0.0210 \\ 
     
    \texttt{task\_r} & 134 & 400 & \texttt{task\_b\_r} & 9 & 28 & 31 & 0.0217 \\ 
    
     \texttt{task\_s} & 249 & 615 & \texttt{task\_b\_s} & 16 & 27 & 40 & 0.0280 \\ \bottomrule
    \end{tabular}
\end{table}

\subsubsection{Application to real quantifier elimination problems}
In this section, we apply the Transformer models to predict variable orderings
for some real quantifier elimination (RQE) problems arising in practice following the strategy presented 
in~\ref{subsec:practice}. All these problems have at least five variables.
To simplify the notation, in this section,  the fine-tuned CAD variable ordering models
whose corresponding pre-training models are \texttt{task\_p}, \texttt{task\_r}, \texttt{task\_s}
will be respectively renamed as $M_p$, $M_r$ and $M_s$.
For all these problems, we have also tried all the heuristic approaches presented in Section~\ref{subsec: introduction-heuristics}.
Let $X_1,\ldots,X_r$ be a partition of the set of variables $\{x_1,\ldots, x_n\}$.
Given a prenex formula $PF=(Q_{r-1}X_{r-1}\cdots Q_1 X_1) FF(x_1,\ldots,x_n)$, 
where $Q_i\in\{\forall,\exists\}$ and $FF$ is a quantifier free formula over polynomial constraints,
let $F$ be the set of distinct polynomials appearing in $PF$.
To select a variable ordering for $PF$, similar to the Transformer-based approach, we make use of the heuristic approaches in two different manners: $(i)$ we firstly choose a variable ordering $O$ and then force the order $O$ to respect the block ordering $X_1>\cdots > X_r$;
$(ii)$ let $BF_i$ be the sub-system obtained by assigning variables 
in $X\setminus X_i$ to $1$ in $F$. We apply the previous strategies on $(BF_i, X_i)$ to get an ordering $O_i$. Then the final ordering for $PF$ is $O_1>O_2>\cdots >O_r$.
Unlike the Transformer-based approach, in both $(i)$ and $(ii)$, no further evaluation is needed for heuristic approaches.
For all the tested examples, we use the command {\sf QuantifierElimination}~\citep{DBLP:conf/icms/ChenM14CAD} (with the option {\sf output=rootof})  inside the RegularChains library of Maple 2025 for performing quantifier elimination. 

\begin{example}
  Consider the following RQE problem~\citep{BH91}: 
$$
\exists \{y, x\}, x^2 + y^2 - 1 = 0 \land b^2(x - c)^2 + a^2y^2 - a^2b^2 = 0 \land 0 < a \land a < 1 \land 0 < b \land b < 1 \land 0 \le c \land c < 1.
$$
The heuristic approaches recommend three variable orderings: $x>y>a>c>b$, $x>y>a>b>c$, and $x>y>c>a>b$. 
The first two can be used to solve the problem in $16.947$ and $40.532$ seconds, respectively, while the third completes the computation in just $6.335$ seconds. 
The order predicted by $M_p$ and $M_s$ coincides with the best heuristic ordering, $x>y>c>a>b$, achieving the same runtime of $6.335$ seconds.
The model $M_r$ predicts the order $x>y>a>c>b$, solving the problem in $16.947$ seconds.

\end{example}

%% test 50
\begin{example}
Consider the following RQE problem~\citep{Sturm1997}: 
$$
\exists \{u, v, w\},\ 
x (f - w) = f u \land 
y (f - w) = f v \land 
(u - a)^2 + (v - b)^2 + (w - c)^2 = 1.
$$
The heuristic approaches recommend three variable orderings: 
$u>v>w>x>y>f>a>b>c$, $u>v>w>x>y>a>b>c>f$, and $w>u>v>x>y>a>b>c>f$. 
None of them is able to be used to solve the problem within $100$ seconds. 
In contrast, all three machine learning models produce effective orderings. 
The model $M_p$ predicts $v>w>u>b>c>a>f>y>x$, solving the problem in $8.779$ seconds. 
The model $M_r$ predicts $w>v>u>a>b>c>x>f>y$, solving it in $16.435$ seconds. 
The model $M_s$ predicts $v>u>w>c>f>b>a>y>x$, solving it in $10.633$ seconds. All predicted orderings successfully help solving the problem within reasonable time.

\end{example}

%% test 55
\begin{example}
  Consider the following RQE problem~\citep{Sturm1997}: 
$$
\exists\{u,v,s\}, v = u^2 \land (x - u)^2 + (y - v)^2 + z^2 = r^2 \land 2su = x - u \land (-s = y - v).
$$
The heuristic approaches recommend three variable orderings: 
$s> v> u> r> z> x> y$, $s> v> u> r> x> y> z$, and $s> v> u> r> z> y> x$.
None of them is able to be used to solve the problem within $100$ seconds.  
The model $M_p$ predicts the order $s>u>v>r>y>z>x$, failing to solve the problem within $100$ seconds.
In contrast, the model $M_r$ predicts the order $v>s>u>r>z>y>x$, solving the problem in $17.391$ seconds.
The model $M_s$ predicts the order $v>u>s>z>r>y>x$, solving the problem in $62.557$ seconds.
\end{example}

\subsection{Interpreting the success and failure of Transformer models}
\label{subsec:analysis}
In this section, we try to employ two interpretability techniques, namely cross-attention weights analysis~\citep{bahdanau2015iclr-neural} and linear probing for representation analysis~\citep{alain2017iclr-understanding}, 
to interpret the success and the failure of Transformer models 
on predicting the CAD variable ordering for  concrete examples.

\subsubsection{Cross-attention analysis}

% ============================================
Let $\mathbf{X}\in \mathbb{R}^{\ell \times d}$ denote the output of the encoder,
where $\ell$ is the input sequence length and $d$ is the embedding dimension.
Let $\mathbf{h}^{(t)}\in \mathbb{R}^{d}$ denote the decoder hidden state at step 
$t$, which serves as the query vector for cross-attention.
Let $H$ be the number of attention heads and $a = d / H$ be the per-head dimension.
For head $h \in \{1, \dots, H\}$,
let $\mathbf{q}_t^{(h)} = \mathbf{h}^{(t)} \mathbf{W}_Q^{(h)}\in \mathbb{R}^{a}$
and $\mathbf{K}^{(h)} = \mathbf{X} \mathbf{W}_K^{(h)} \in \mathbb{R}^{\ell \times a}$.
Then the cross-attention weights at step $t$ are:

\begin{equation*}
\bm{\alpha}_t^{(h)} = \text{softmax}\left( \frac{\mathbf{q}_t^{(h)} (\mathbf{K}^{(h)})^\top}{\sqrt{a}} \right) \in \mathbb{R}^{\ell}.
\end{equation*}
The softmax is applied row-wise, ensuring $\sum_{j=1}^{\ell} \alpha_{t,j}^{(h)} = 1$ for all $h, t$.
By averaging the attention weights across all heads, we obtain the final attention vector at step $t$:
$\bar{\bm{\alpha}}_t = \frac{1}{H} \sum_{h=1}^{H} \bm{\alpha}_t^{(h)} \in \mathbb{R}^{\ell}$.
For the full output sequence of length $m$, the cross-attention weight matrix is $\mathbf{A}=[\bar{\bm{\alpha}}_1;\cdots;\bar{\bm{\alpha}}_m]\in \mathbb{R}^{m \times n}$.
We visualize $\mathbf{A}$ as a heatmap, where the
$x$-axis corresponds to input tokens, the $y$-axis to output tokens,
and the color intensity encodes $\mathbf{A}[i, j]$.

Figure~\ref{c3cs} and~\ref{w3cs} show the cross attention weights 
of the model \texttt{task\_c\_s} when respectively making a correct and wrong 
prediction on the CAD  variable ordering of a three-variable example.
Note that for the example in Figure~\ref{c3cs}, there happen to be two optimal variable orderings. 
In both figures, each row corresponds to the attention weights paid 
to the tokens at the x-axis when predicting the output variable on the left.
From Figure~\ref{c3cs}, we observe that all the exponent tokens are paid substantially
more attention than the rest of tokens when selecting the first variable, 
which is critical for obtaining the right ordering. 
For this example, when $x_1$ is selected, the ordering of the other two variables does not matter. 
Correspondingly, it is observed that less attention are paid to the exponents when predicting $x_2$ and $x_0$.
Similarly, for the unsuccessful example in Figure~\ref{w3cs}, only a small fraction of exponents are paid enough attention
when predicting the first variable.
Moreover, even less attentions are paid to the exponents when predicting the rest two variables.

\begin{figure}[htbp]
    \centering
    \includegraphics[width=\textwidth]{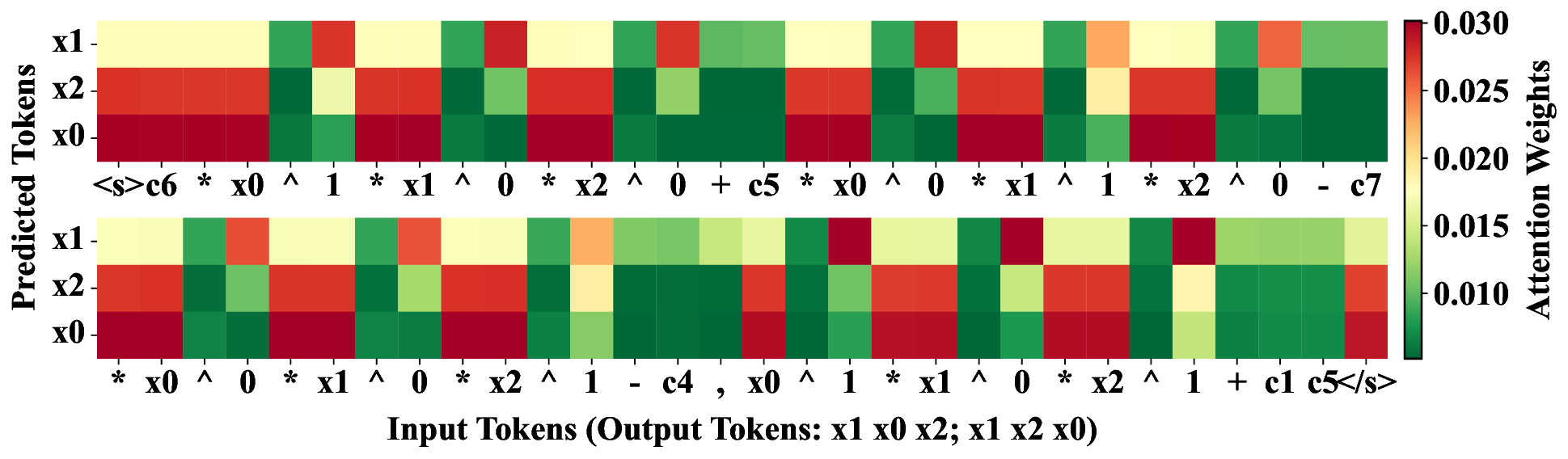}
    \caption{Attention weights for correctly predicted CAD variable ordering for example REdMn2rCv3\_ex-1490 with model \texttt{task\_c\_s}.}
    \label{c3cs}
\end{figure}    

\begin{figure}[htbp]
    \centering
    \includegraphics[width=\textwidth]{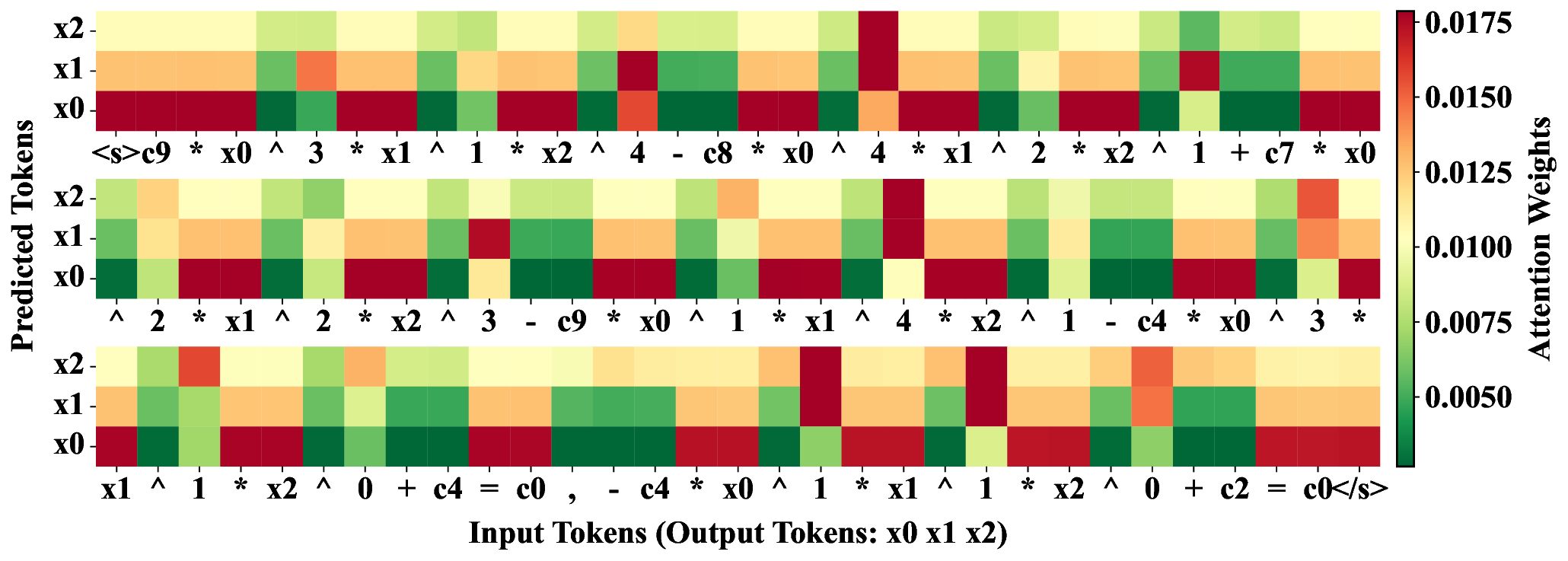}
    \caption{Attention weights for wrongly predicted CAD variable ordering for example REeMn2tMv3\_ex-508 with model \texttt{task\_c\_s}.}
    \label{w3cs}
\end{figure}

Figure~\ref{c3cp} and~\ref{w3cp} further show the cross attention weights 
of the model \texttt{task\_c\_p} when respectively making a correct and wrong 
prediction on the CAD  variable ordering of a three-variable example.
Like \texttt{task\_c\_s}, when making the correct prediction on the first variable $x_1$, more attention is paid on the exponents. 
What is more subtle here is that the exponents of $x_0$ and $x_2$ are paid little attention.
Instead, most of the attention is paid to the powers of $x_1$. 
For the example in Figure~\ref{w3cp},  when making the wrong prediction for the first variable, the exponents in the first polynomial are barely paid attention. 
Instead, considerable attention is paid to the coefficients and operators. 
This  phenomenon is more evident when predicting the rest two variables.

\begin{figure}[htbp]
    \centering
    \includegraphics[width=\textwidth]{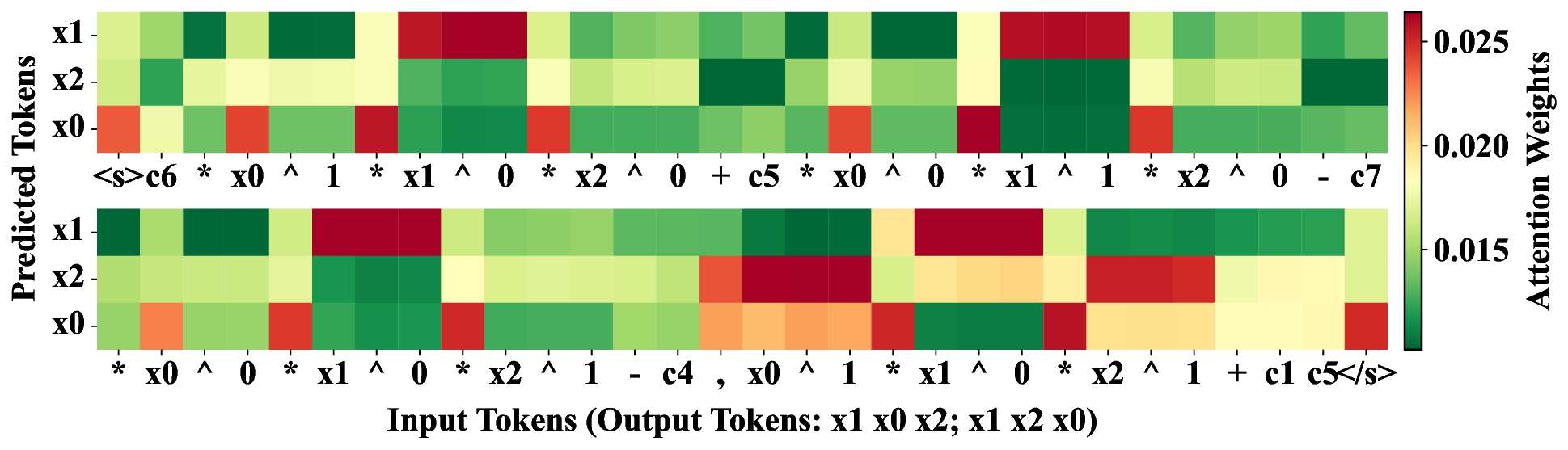}
    \caption{Attention weights for correctly predicted CAD variable ordering for example REdMn2rCv3\_ex-1490 with model \texttt{task\_c\_p}.}
    \label{c3cp}
\end{figure}

\begin{figure}[htbp]
    \centering
    \includegraphics[width=\textwidth]{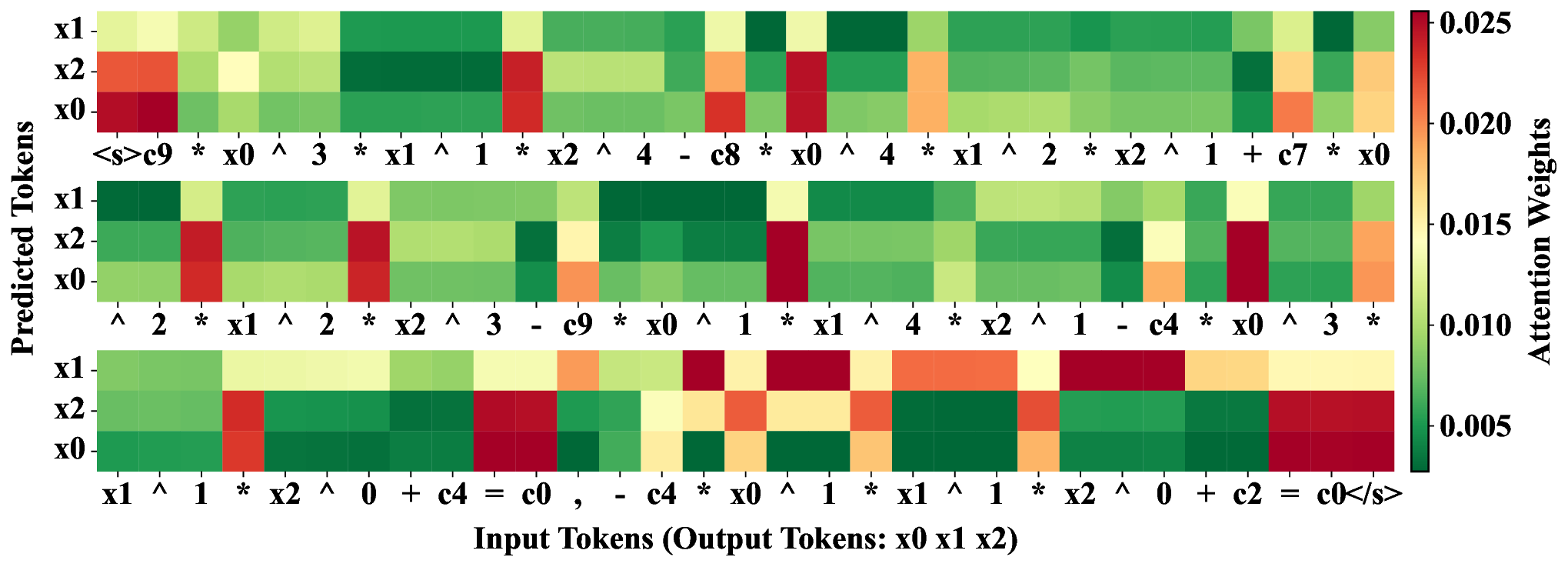}
    \caption{Attention weights for wrongly predicted CAD variable ordering for example REeMn2tMv3\_ex-508 with model \texttt{task\_c\_p}.}
    \label{w3cp}
\end{figure}

In summary, the above cross-attention weights analysis shows
that when the Transformer models succeed on predicting the variable ordering for CAD, more attention is paid to the exponent tokens in the input system. On the contrast, when the Transformer models fail, less attention is paid to the exponent tokens. This makes sense, since the support set captures most of the information in the input polynomials.

\subsubsection{Linear probing}

To further understand the internal mechanisms of Transformers, we design
a probing task for both the model \texttt{task\_c\_s} and the model \texttt{task\_c\_p}, evaluated on the previous two examples.

For a given input sequence $\mathbf{x}$ of length $\ell$, let $\mathbf{X} \in \mathbb{R}^{\ell \times d}$ denote the output 
of the frozen encoder.
By mean-pooling $\mathbf{h} = \frac{1}{\ell} \sum_{i=1}^{\ell} \mathbf{X}[i, :]$,
we obtain a feature vector $\mathbf{h}$ for the input.
We employ a {\em linear probe} consisting of a single fully-connected 
layer with no activation function and no hidden layers $\hat{\mathbf{y}} = \mathbf{W} \mathbf{h} + \mathbf{b}$, where $\mathbf{W} \in \mathbb{R}^{k \times d}$ is the weight matrix, 
$\mathbf{b} \in \mathbb{R}^{k}$ is the bias vector, and $\hat{\mathbf{y}} \in \mathbb{R}^{k}$ 
is the predicted feature vector aiming to regress the features in pre-training tasks.
The probe is trained to minimize the mean squared error (MSE) between 
predicted and ground-truth feature values $\mathcal{L} = \frac{1}{N} \sum_{i=1}^{N} \| \hat{\mathbf{y}}_i - \mathbf{y}_i \|^2$, where $N$ is the number of 
training examples, while all parameters of the pretrained frozen encoders remain unchanged.
The rationale behind the linear probe is that since a linear transformation has limited capacity, the probe primarily reads out information already present in the encoder representations, rather than computing novel features itself.
In this experiment, during the training phase, we randomly sampled $N=5,000$ instances from the existing pre-training dataset as the training set. 
For the regressed feature, we choose the feature \emph{{\rm sum}\_{\rm max}\_{\sf d}}, 
which is the first one in $\texttt{feature\_f}$.
This feature is also used by the heuristic approach \texttt{gmods}.

We selected the pre-trained models \texttt{task\_p} and \texttt{task\_s} on three variables for probing experiments.To avoid ambiguity, we denote the probe model corresponding to the pretraining task $\texttt{task\_id}$ as $\texttt{probe\_id}$.
The results  are shown in Table~\ref{tab:comparison_ps}. 
According to the table, despite the inherent difficulty of exact label recovery via linear probing, the probe achieves high-fidelity approximation of feature values. This capacity demonstrates that the model has acquired meaningful structural representations of the underlying attributes.

\begin{table}[htbp]
\centering
\caption{Output of the probing task.}
\label{tab:comparison_ps}
\centering
\begin{tabular}{cccc}
\toprule
\text{Example} & \text{Metric} & \texttt{probe\_p} & \texttt{probe\_s} \\
\midrule
\multirow{2}{*}{\text{REdMn2rCv3\_ex-1490}}
& label & [2, 1, 2] & [2, 1, 2] \\
& output & [1.9770, 1.8595, 1.8570] & [1.6654, 1.6002, 1.7781] \\
% & dist & 0.8716 & 0.7221 \\
\midrule
\multirow{2}{*}{\text{REeMn2tMv3\_ex-508}}
& label & [5, 5, 4] & [5, 5, 4] \\
& output & [4.5846, 4.4885, 4.3604] & [4.8229, 4.5072, 4.6669] \\
% & dist & 0.7510 & 0.8479 \\
\bottomrule
\end{tabular}
\end{table}

}

%%%%%%%%%%%%%%%%%%%%%%%%%%%%%%%%%%%%%%%%%%%%%%%%%%%%%%%%%%%%%%%

\section{Conclusion and future Work}

In this work, we proposed to leverage pre-training \& fine-tuning Transformer models to 
address the inherent data scarcity problem when employing deep learning to predict the best variable ordering for cylindrical algebraic decomposition. 
By conducting intensive experiments 
on two publicly available random datasets, an enhanced random dataset and a real SMT dataset, 
we have demonstrated that the proposed approach substantially outperforms the state-of-the-art heuristic methods
as well as Transformer models not employing the pre-training technique.
On the  dataset DQ-3, the best pre-training \& fine-tuning model brings $2\times$ speedup for CAD over the best heuristic 
approach, $3\times$ speedup over the classical Brown's approach \texttt{svob}. 
On the dataset DQ-4b, the speedup over \texttt{svob} is near $2\times$.
On a real SMT dataset,  the model brings near-optimal performance (only $1.07\times$ the optimal CAD running time). 
Moreover, we observe that pre-training tasks that exploit the deep features of projection polynomials (\texttt{feature\_p}, \texttt{feature\_s}, \texttt{feature\_r}) consistently outperform the ones that utilize only the shallow features
of input systems (\texttt{feature\_e}, \texttt{feature\_f}, \texttt{feature\_m}) after fine-tuning.

We have also explored several different directions, such as increasing the pre-training dataset size and 
increasing the model capacity, aiming to push the limit of current work. 
We notice that, although the pre-training models have greatly helped the downstream CAD variable ordering selection task, 
the pre-training models for learning information of resultants still have a relatively low accuracy. 
For the future, much work remains to be done for learning with these basic symbolic operations, 
whose success may help with learning more advanced symbolic algorithms via the pre-training \& fine-tuning paradigm.

%%%%%%%%%%%%%%%%%%%%%%%%%%%%%%%%%%%%%%%%%%%%%%%%%%%%%%%%%%%%%%%

\section*{Acknowledgements}
The authors would like to thank anonymous referees for helpful comments and suggestions. 
The authors would also like to thank J\"urgen Gerhard for helping us getting Maple licence.

%%%%%%%%%%%%%%%%%%%%%%%%%%%%%%%%%%%%%%%%%%%%%%%%%%%%%%%%%%%%%%%

\bibliographystyle{elsarticle-harv} 
\bibliography{reference}

\end{document}